\documentclass[journal=jacsat,manuscript=article]{achemso}
\setkeys{acs}{maxauthors=10,etalmode=truncate,chaptertitle =true,articletitle=true}

\usepackage{amssymb}
\usepackage{amsmath}
\usepackage[usenames,dvipsnames]{xcolor}
\usepackage{graphicx}
\usepackage{float}
\usepackage[normalem]{ulem}
\usepackage{caption}
\usepackage{subcaption}
\usepackage{natbib}
\usepackage{multirow}
\usepackage{bm}
\usepackage{array}
\usepackage[version=4]{mhchem} % formatting chemical equations

\SectionNumbersOn

\title{Extending GPU-Accelerated Gaussian Integrals in the TeraChem Software Package to f Type Orbitals: Implementation and Applications.}

\author{Yuanheng Wang}
\affiliation
{Department of Chemistry and The PULSE Institute, Stanford University, Stanford, California 94305, United States}
\alsoaffiliation{SLAC National Accelerator Laboratory, Menlo Park, California 94024, United States}
\author{Diptarka Hait}
\affiliation
{Department of Chemistry and The PULSE Institute, Stanford University, Stanford, California 94305, United States}
\alsoaffiliation{SLAC National Accelerator Laboratory, Menlo Park, California 94024, United States}
\author{K. Grace Johnson}
\affiliation
{Department of Chemistry and The PULSE Institute, Stanford University, Stanford, California 94305, United States}
\alsoaffiliation{SLAC National Accelerator Laboratory, Menlo Park, California 94024, United States}
\author{O. Jonathan Fajen}
\affiliation
{Department of Chemistry and The PULSE Institute, Stanford University, Stanford, California 94305, United States}
\alsoaffiliation{SLAC National Accelerator Laboratory, Menlo Park, California 94024, United States}
\author{Juncheng Harry Zhang}
\affiliation
{Department of Chemistry and The PULSE Institute, Stanford University, Stanford, California 94305, United States}
\alsoaffiliation{SLAC National Accelerator Laboratory, Menlo Park, California 94024, United States}
\author{Rub{\'e}n D. Guerrero}
\affiliation
{Department of Chemistry and The PULSE Institute, Stanford University, Stanford, California 94305, United States}
\alsoaffiliation{SLAC National Accelerator Laboratory, Menlo Park, California 94024, United States}
\author{Todd J. Mart{\'i}nez}
\email{todd.martinez@stanford.edu; toddjmartinez@gmail.com}
\affiliation
{Department of Chemistry and The PULSE Institute, Stanford University, Stanford, California 94305, United States}
\alsoaffiliation{SLAC National Accelerator Laboratory, Menlo Park, California 94024, United States}

\begin{document}
\maketitle

\begin{abstract}
The increasing availability of GPUs for scientific computing has prompted interest in accelerating quantum chemical calculations through their use. The complexity of integral kernels for high angular momentum basis functions however often limits the utility of GPU implementations with large basis sets or for metal containing systems. In this work, we report implementation of $f$ function support in the GPU-accelerated TeraChem software package through the development of efficient kernels for the evaluation of Hamiltonian integrals. The high efficiency of the resulting code is demonstrated through density functional theory (DFT) calculations on increasingly large organic molecules and transition metal complexes, as well as coupled cluster singles and doubles (CCSD) calculations on water clusters. Preliminary investigations into Ni(I) catalysis with DFT and the photochemistry of MnH(CH$_3$) with complete active space self-consistent field (CASSCF) are also carried out. Overall, our GPU-accelerated software appears to be well-suited for fast simulation of large transition metal containing systems, as well as organic molecules. 
\end{abstract}

\section{Introduction}

The increasing availability of computing resources over time has led to a corresponding growth in the use of quantum mechanical methods for modeling chemical problems \cite{general_dft_review, computational_catalysis_review, computational_organic_review, computational_chemistry_review}. The combinatorially scaling computational demand of exact quantum mechanics limits it to small gas-phase systems \cite{mountaineer_excited_state, benzene_ground}, and even practically useful approximations like density functional theory (DFT) \cite{dft_original, kohn_sham_original, general_dft_review} have resource requirements that formally scale quite steeply [$O(N^{3})$ to $O(N^{4})$] with the size $N$ of the system under investigation \cite{computational_chemistry_review_complicated}. \textit{Ab initio} modeling for chemical and material science applications consequently represents a large fraction of the total volume of scientific computing (up to $\sim 30\%$ for some supercomputing clusters like NERSC \cite{nersc}), and the demand is only expected to increase \cite{computational_chemistry_review, computational_organic_review} as research interests shift towards studying the properties and dynamics of complex condensed phase systems. It is therefore quite desirable to develop fast quantum chemistry software that is optimally suited for modern hardware. 

Graphics processing units (GPUs) are increasingly popular for computational applications beyond their original use for video-game graphics, and constitute a large fraction of resources available in modern supercomputing clusters. The highly data parallel structure of GPUs is well suited for embarassingly parallel tasks like the evaluation of Hamiltonian integrals that are required for quantum chemical calculations \cite{gaussian_program, terachem_gpu_1}. This led to the TeraChem software package \cite{terachem_gpu_1, terachem_gpu_2, terachem_gpu_3, terachem_2013, terachem_2021} which pioneered the use of GPUs for Hartree-Fock \cite{terachem_gpu_1}, DFT \cite{terachem_gpu_2,terachem_gpu_3}, coupled cluster \cite{terachem_ccsd_gpu, terachem_rank_reduce_cc_1, terachem_rank_reduce_cc_2, terachem_rank_reduce_cc_3, terachem_rank_reduce_ccsd}, perturbation theory \cite{song_sosmp2_part1,song_sosmp2_part2,song_mp2_gradient}, and multireference \cite{terachem_casscf, terachem_fomo_casci, terachem_direct_ci_1, terachem_direct_ci_2, terachem_direct_ci_3, terachem_rank_reduced_ci, song_caspt2, song_caspt2_gradient, song_xmspt2} calculations. Speedups of one to two orders of magnitude were achieved compared to CPU based programs, stemming from both the computational power of GPUs and the development of GPU optimized algorithms. TeraChem's efficiency permitted the modeling of large chemical systems, such as \textit{ab initio} molecular dynamics simulations on proteins \cite{protein_gpu_1, protein_gpu_2, protein_gpu_3}. Several traditionally CPU based software packages for modeling molecular systems have also started to provide GPU support \cite{gaussian_program, gaussian_program_dft, gaussian_program_high_angular, gamess_rys_quadrature, gamess_sp, gamess_spd, gamess_f, qchem_5_or_higher, brianqc, pyscf_gpu, turbomole, libintx_up_to_iiii, libintx_gpu_2}, highlighting the general interest in the computational chemistry community for GPU acceleration.

Writing GPU-optimized programs is however an involved task, as algorithms must be designed to take advantage of the highly parallel structure of GPUs while minimizing the need for data transfer between CPUs and GPUs. Indeed, it is quite nontrivial to generate kernels for the evaluation of Hamiltonian integrals with high angular momentum basis functions. Until now, TeraChem only supported $s,p$ and $d$ type Gaussian basis functions \cite{terachem_2013}. Although this is sufficient for modeling organic species with polarized double zeta basis sets
\cite{cc_pvxz, red_book}, there are several reasons to desire GPU-accelerated calculations with $f$ type orbitals. Reducing basis set incompleteness errors in DFT calculations below intrinsic functional error is typically believed to require basis sets of at least polarized triple zeta quality
\cite{dft_best_practice}, which usually contain $f$ type functions for elements in the second period and beyond. The dynamic electron correlation recovered by wavefunction based methods like coupled cluster (CC) is also quite sensitive to the size of the basis \cite{kutzelnigg1992rates, helgaker1997basis}, and such calculations therefore benefit from the use of larger basis sets.  Additionally, the modeling of transition metal chemistry and catalysis potentially benefits from $f$ type functions to appropriately polarize $d$ orbitals on metals. 

We have therefore implemented support for utilizing $f$ type orbitals in TeraChem for efficiently performing self-consistent field (SCF) calculations like Hartree-Fock (HF) \cite{blue_book}, Kohn-Sham DFT \cite{kohn_sham_original}, coupled cluster singles and doubles (CCSD) \cite{purvis_full_1982, cc_review} and complete active space SCF (CASSCF) \cite{casscf_original_1, casscf_original_2, casscf_original_3, casscf_original_4} calculations on molecular systems. This is described in the present work, which is organized as follows. In Sec. \ref{sec:theory} we present a brief overview of Gaussian basis sets for molecular systems and the evaluation of the needed integrals in such bases. We subsequently detail the GPU-accelerated implementation for the evaluation of these integrals in Sec. \ref{sec:implementation}, focusing on equation generation utilizing common subexpression elimination, data structures, and kernel design. We characterize the performance of the resulting code via DFT calculations on increasingly large organic and transition metal containing species, as well as compare the performance to the GPU-based BrianQC backend \cite{brianqc} of the Q-Chem software package \cite{qchem_5_or_higher}. The performance for CCSD calculations on water clusters of increasing size is also reported. Finally, two model applications are used to demonstrate potential applicability towards studying transition metal catalysis and photochemistry. 

\section{Theory} \label{sec:theory}

Support for higher angular momentum atomic orbitals requires extension of all existing integral routines, including overlap, electron kinetic energy, nuclear-electron attraction, DFT local exchange-correlation, and electron replusion integrals (ERI). In this section we first present the notation we use for basis functions and pairs, followed by the integral formula for all integrals mentioned above. There exist several excellent reviews of integral generation for atom-centered Gaussian basis sets \cite{head_gordon_pople_flops, helgaker_taylor_book, purple_book, integral_fundamental}, and we encourage the reader to consult these. Nevertheless, we explicitly provide many of the equations needed in a fully functional quantum chemistry code both for completeness and uniformity of notation. 

\subsection{Cartesian Gaussian basis functions}

In TeraChem, contracted atom-centered Cartesian Gaussian type orbitals (GTO) are used as basis functions, which are linear combinations of primitive Gaussian type orbitals (pGTOs):
\begin{align}
\phi(\vec{r}) = \sum_m^{N_{contraction}} C_m^{contraction} \mu_m(\vec{r})
\label{eq:contracted_basis}
\end{align}
The pGTOs $\mu(\vec{r})$ are of the form:
\begin{align}
\mu(\vec{r};\vec{i},a,\vec{A}) = C^{normalization} (x - A_x)^{i_x} (y - A_y)^{i_y} (z - A_z)^{i_z} e^{-a \left| \vec{r} - \vec{A} \right|^2}
\label{eq:primtive_basis}
\end{align}
where $\vec{A}$ and $a$ represent the center location and exponent of the primitive Gaussian function, respectively. The angular momentum index vector $\vec{i}$ (with components $i_x$, $i_y$ and $i_z$) determines the shape of the atomic orbital (for example, if $i_x = i_y = i_z = 0$, then the function represents a $s$ orbital; $i_x = 1, i_y = i_z = 0$ represents a $p_x$ orbital, etc.).

The desired integrals in the contracted GTO basis are just summations over integrals in terms of pGTOs. We therefore will only consider pGTOs from here on, and also condense all constant prefactors (contraction, normalization etc.) into a single $C_\mu$ for the $\mu$-th pGTO.

GTOs appear in pairs for most integrals. For a pair of pGTOs $\mu(\vec{r}; \vec{i},a,\vec{A})$ and $\nu(\vec{r}; \vec{j},b,\vec{B})$, centered at $\vec{A}$ and $\vec{B}$, with coefficients $C_\mu$ and $C_\nu$, exponents $a$ and $b$, and angular momentum indices $\vec{i}$ and $\vec{j}$, respectively, the product (also called pair/overlap distribution) is:
\begin{align}
\mu(\vec{r}; \vec{i},a,\vec{A})\nu(\vec{r}; \vec{j},b,\vec{B}) &= C_\mu C_\nu \left( \prod_{\tau \in \{x,y,z\}} (\tau - A_\tau)^{i_\tau} (\tau - B_\tau)^{j_\tau} \right) e^{-a \left| \vec{r} - \vec{A} \right|^2} e^{-b \left| \vec{r} - \vec{B} \right|^2} \notag \\
    &= C_\mu C_\nu \left( \prod_{\tau \in \{x,y,z\}} (\tau - A_\tau)^{i_\tau} (\tau - B_\tau)^{j_\tau} \right) e^{-\frac{ab}{a+b} \left| \vec{A} - \vec{B} \right|^2} e^{-p \left| \vec{r} - \vec{P} \right|^2}
\label{eq:pair_definition}
\end{align}
where the second equation follows from simplifying the Gaussian exponent. The resulting $\vec{P} = \frac{a\vec{A} + b\vec{B}}{a+b}$ and $p = a+b$ are the new center and exponent of the pair distribution. This pair distribution is often called a ``charge density". \cite{n_electron_integral} From here on, we will drop the parameters for each GTO function and only keep the argument $\vec{r}$ whenever applicable, for notational simplicity.

In the McMurchie-Davidson algorithm \cite{McMurchie_Davidson_original}, the pair distribution along each dimension is rewritten as a sum of Hermite Gaussians ($t$-th derivative of a Gaussian function):
\begin{align}
(\tau - A_\tau)^{i_\tau} (\tau - B_\tau)^{i_\tau} e^{-\frac{ab}{a+b} (A_\tau - B_\tau)^2} e^{-p (\tau - P_\tau)^2} = \sum_{t_\tau = 0}^{i_\tau + j_\tau} E^{i_\tau, j_\tau}_{t_\tau, \tau} \left( \frac{\partial}{\partial P_\tau} \right)^{t_\tau} e^{-p (\tau - P_\tau)^2}
\label{eq:cartesian_to_hermite}
\end{align}
The Cartesian Gaussian to Hermite Gaussian transformation coefficients $E^{i_\tau, j_\tau}_{t_\tau, \tau}$ can be obtained from the McMurchie-Davidson recurrence
relationship:
\begin{align}
E^{i_\tau + 1, j_\tau}_{t_\tau, \tau} &= \frac{1}{2p} E^{i_\tau, j_\tau}_{t_\tau - 1, \tau} + (P_\tau - A_\tau) E^{i_\tau, j_\tau}_{t_\tau, \tau} + (t_\tau + 1) E^{i_\tau, j_\tau}_{t_\tau + 1, \tau} \\
E^{i_\tau, j_\tau + 1}_{t_\tau, \tau} &= \frac{1}{2p} E^{i_\tau, j_\tau}_{t_\tau - 1, \tau} + (P_\tau - B_\tau) E^{i_\tau, j_\tau}_{t_\tau, \tau} + (t_\tau + 1) E^{i_\tau, j_\tau}_{t_\tau + 1, \tau} \\
E^{0,0}_{0,\tau} &= e^{-\frac{ab}{a+b} (A_\tau - B_\tau)^2} \\
E^{i_\tau, j_\tau}_{t_\tau, \tau} &= 0 \qquad \text{ if } t_\tau < 0 \text{ or } t_\tau > i_\tau + j_\tau
\label{eq:E_recursion}
\end{align}

The pair distribution $\mu(\vec{r})\nu(\vec{r})$ can now be rewritten in the Hermite Gaussian form as:
\begin{align}
\mu(\vec{r})\nu(\vec{r}) = C_\mu C_\nu \left( \prod_{\tau \in \{x,y,z\}} \sum^{i_\tau + j_\tau}_{t_\tau = 0} E^{i_\tau, j_\tau}_{t_\tau, \tau} \left( \frac{\partial}{\partial P_\tau} \right)^{t_\tau} \right) e^{-p \left| \vec{r} - \vec{P} \right|^2}
\label{eq:pair_hermite}
\end{align}
We will use this form for our integral evaluations.

\subsection{Overlap integral}

As the simplest type of quantum chemical integral, the overlap integral is defined as
\begin{align}
S_{\mu\nu} = \iiint_{\infty} d\vec{r} \ \mu(\vec{r}; \vec{i},a,\vec{A})\nu(\vec{r}; \vec{j},b,\vec{B})
\label{eq:overlap_definition}
\end{align}
Given that the Gaussian integral $\iiint_{\infty} d\vec{r} \ e^{-p \left| \vec{r} - \vec{P} \right|^2}$ evaluates to $\left( \frac{\pi}{p} \right)^{3/2}$, and is independent of $\vec{P}$, the overlap integral expression can be simplified to:
\begin{align}
S_{\mu\nu} = C_\mu C_\nu E^{i_x, j_x}_{0, x} E^{i_y, j_y}_{0, y} E^{i_z, j_z}_{0, z} \left( \frac{\pi}{p} \right)^{3/2}
\label{eq:overlap_formula}
\end{align}

\subsection{Electron kinetic energy integral}

The electron kinetic energy integral is defined as: 
\begin{align}
T_{\mu\nu} = -\frac{1}{2} \iiint_{\infty} d\vec{r} \ \mu(\vec{r}; \vec{i},a,\vec{A}) \left( \frac{\partial^2}{\partial x^2} + \frac{\partial^2}{\partial y^2} + \frac{\partial^2}{\partial z^2} \right) \nu(\vec{r}; \vec{j},b,\vec{B})
\label{eq:kinetic_definition}
\end{align}
Since derivatives of pGTOs are just sums of (different) pGTOs, closed-form expressions for the electron kinetic energy integrals can be obtained in the same manner as overlap integrals:
\begin{align}
T_{\mu\nu} = C_\mu C_\nu & \left( \left( -\frac{j_x (j_x-1)}{2} E^{i_x, j_x-2}_{0,x} + (2j_x + 1) b E^{i_x, j_x}_{0,x} - 2b^2 E^{i_x, j_x+2}_{0,x} \right) E^{i_y, j_y}_{0,y} E^{i_z, j_z}_{0,z} \right. \notag \\
    & + E^{i_x, j_x}_{0,x} \left( -\frac{j_y (j_y-1)}{2} E^{i_y, j_y-2}_{0,y} + (2j_y + 1) b E^{i_y, j_y}_{0,y} - 2b^2 E^{i_y, j_y+2}_{0,y} \right) E^{i_z, j_z}_{0,z} \notag \\
    & \left. + E^{i_x, j_x}_{0,x} E^{i_y, j_y}_{0,y} \left( -\frac{j_z (j_z-1)}{2} E^{i_z, j_z-2}_{0,z} + (2j_z + 1) b E^{i_z, j_z}_{0,z} - 2b^2 E^{i_z, j_z+2}_{0,z} \right) \right) \left( \frac{\pi}{p} \right)^{3/2}
\label{eq:kinetic_formula}
\end{align}

\subsection{Nuclear attraction integral}

The nuclear-electron attraction operator accounts for the interaction between electrons and nuclei (modeled as point charges). The interaction between electrons and other point charges, such as from classical force-fields in QM/MM calculations, can also be included in this operator. The corresponding matrix element is given by:
\begin{align}
V_{\mu\nu} = \iiint_{\infty} d\vec{r} \ \mu(\vec{r}; \vec{i},a,\vec{A})\nu(\vec{r}; \vec{j},b,\vec{B}) \sum_{C}^{N_{point-charge}} \frac{q_C}{\left| \vec{r} - \vec{C} \right|}
\label{eq:v1e_definition}
\end{align}
where $q_C$ and $\vec{C}$ are the charge and position of each point charge. We simplify the notation by defining:
\begin{align}
V_{\mu\nu C} = \iiint_{\infty} d\vec{r} \ \mu(\vec{r})\nu(\vec{r}) \frac{q_C}{\left| \vec{r} - \vec{C} \right|} \label{eq:v1e_definition2} \\
V_{\mu\nu} = \sum_{C}^{N_{point-charge}}V_{\mu\nu C}
\label{eq:v1e_summation_over_C}
\end{align}

The base case, where both $\mu(\vec{r})$ and $\nu(\vec{r})$ are $s$-orbitals ($\vec{i} = \vec{j} = \vec{0}$), is:
\begin{align}
V_{ss C} &= q_C C_\mu C_\nu e^{-\frac{ab}{a+b} \left| \vec{A} - \vec{B} \right|^2} \iiint_{\infty} d\vec{r} \ e^{-p \left| \vec{r} - \vec{P} \right|^2} \frac{1}{\left| \vec{r} - \vec{C} \right|} \notag \\
    &= q_C C_\mu C_\nu e^{-\frac{ab}{a+b} \left| \vec{A} - \vec{B} \right|^2} \frac{2\pi}{p} F_0\left( p\left| \vec{P} - \vec{C} \right|^2 \right)
\label{eq:v1e_base_case}
\end{align}
The three-dimensional integral over $\vec{r}$ is reduced to the Boys function $F_m(x)$ form,\cite{boys_original} defined as a simple one-dimensional integral:
\begin{align}
F_m(x) = \int_0^1 dt \ t^{2m} e^{-xt^2}
\label{eq:boys_function}
\end{align}
Boys functions are evaluated numerically using interpolation and downward recursion.\cite{boys_interpolation, boys_gpu}

For the general case with $\mu(\vec{r})$ and/or $\nu(\vec{r})$ of arbitrary angular momentum indices, the Hermite Gaussian form of the pair (equation \ref{eq:pair_hermite}) is applied, and equation \ref{eq:v1e_definition2} reduces to
\begin{align}
V_{\mu\nu C} = q_C C_\mu C_\nu \left( \prod_{\tau \in \{x,y,z\}} \sum^{i_\tau + j_\tau}_{t_\tau = 0} E^{i_\tau, j_\tau}_{t_\tau, \tau} \left( \frac{\partial}{\partial P_\tau} \right)^{t_\tau} \right) \frac{2\pi}{p} F_0\left( p\left| \vec{P} - \vec{C} \right|^2 \right)
\label{eq:v1e_general_step1}
\end{align}

In the McMurchie-Davidson formalism, in order to carry out the derivatives, one introduces auxiliary integrals $R_{t_x,t_y,t_z}^m$ as derivatives of Boys functions:
\begin{align}
R_{t_x,t_y,t_z}^m\left( p, \vec{L} \right) = \left( \frac{\partial}{\partial L_x} \right)^{t_x} \left( \frac{\partial}{\partial L_y} \right)^{t_y} \left( \frac{\partial}{\partial L_z} \right)^{t_z} \left( (-2p)^m F_m\left( p\left| \vec{L} \right|^2 \right) \right)
\label{eq:R_definition}
\end{align}
The recurrence relationship for $R_{t_x,t_y,t_z}^m$ is \cite{McMurchie_Davidson_original}:
\begin{align}
R_{t_x + 1,t_y,t_z}^m\left( p, \vec{L} \right) &= t_x R_{t_x - 1,t_y,t_z}^{m + 1}\left( p, \vec{L} \right) + L_x R_{t_x,t_y,t_z}^{m + 1}\left( p, \vec{L} \right) \\
R_{t_x,t_y + 1,t_z}^m\left( p, \vec{L} \right) &= t_y R_{t_x,t_y - 1,t_z}^{m + 1}\left( p, \vec{L} \right) + L_y R_{t_x,t_y,t_z}^{m + 1}\left( p, \vec{L} \right) \\
R_{t_x,t_y,t_z + 1}^m\left( p, \vec{L} \right) &= t_z R_{t_x,t_y,t_z - 1}^{m + 1}\left( p, \vec{L} \right) + L_z R_{t_x,t_y,t_z}^{m + 1}\left( p, \vec{L} \right) \\
R_{0,0,0}^m\left( p, \vec{L} \right) &= (-2p)^m F_m\left( p\left| \vec{L} \right|^2 \right) \\
R_{t_x,t_y,t_z}^m\left( p, \vec{L} \right) &= 0 \qquad \text{ if } t_x < 0 \text{ or } t_y < 0 \text{ or } t_z < 0
\label{eq:R_recursion}
\end{align}

With the help of these auxiliary integrals, equation \ref{eq:v1e_general_step1} reduces to:
\begin{align}
V_{\mu\nu C} = q_C C_\mu C_\nu \frac{2\pi}{p} \sum^{i_x + j_x}_{t_x = 0} E^{i_x, j_x}_{t_x, x} \sum^{i_y + j_y}_{t_y = 0} E^{i_y, j_y}_{t_y, y} \sum^{i_z + j_z}_{t_z = 0} E^{i_z, j_z}_{t_z, z} R_{t_x,t_y,t_z}^0 \left( p, \vec{P} - \vec{C} \right)
\label{eq:v1e_formula}
\end{align}

\subsection{Electron repulsion integral}
The integrals considered so far only involve two atomic orbitals. Modeling electron-electron repulsion however necessitates integration over four atomic orbitals and the evaluation of the associated integrals is formally the most computationally intensive part of the SCF procedure. The general ERI is defined as:
\begin{align}
(\mu\nu|\lambda\sigma) = \iiint_{\infty} d\vec{r}_1 \iiint_{\infty} d\vec{r}_2 \ \mu(\vec{r}_1; \vec{i},a,\vec{A}) \nu(\vec{r}_1; \vec{j},b,\vec{B}) \frac{1}{\left| \vec{r}_1 - \vec{r}_2 \right|} \lambda(\vec{r}_2; \vec{k},c,\vec{C}) \sigma(\vec{r}_2; \vec{l},d,\vec{D})
\label{eq:eri_definition}
\end{align}
Here we label the four pGTOs $\mu(\vec{r}; \vec{i},a,\vec{A})$, $\nu(\vec{r}; \vec{j},b,\vec{B})$, $\lambda(\vec{r}; \vec{k},c,\vec{C})$ and $\sigma(\vec{r}; \vec{l},d,\vec{D})$; with centers $\vec{A}$, $\vec{B}$, $\vec{C}$, $\vec{D}$, exponents $a$, $b$, $c$, $d$ and angular momentum indices $\vec{i}$, $\vec{j}$, $\vec{k}$, $\vec{l}$, respectively.
The pair distribution corresponding to the first electron (``bra") is $\mu(\vec{r}_1)\nu(\vec{r}_1)$ and $\lambda(\vec{r}_2)\sigma(\vec{r}_2)$ for the second electron (``ket"). These distributions have pair centers $\vec{P} = \frac{a\vec{A} + b\vec{B}}{a+b}$ and $\vec{Q} = \frac{c\vec{C} + d\vec{D}}{c+d}$, and pair exponents $p = a+b$ and $q = c+d$, respectively.

Similar to the nuclear attraction integral, we start from the base case where all four pGTOs are $s$-orbitals:
\begin{align}
(ss|ss) = & C_\mu C_\nu C_\lambda C_\sigma e^{-\frac{ab}{a+b} \left| \vec{A} - \vec{B} \right|^2} e^{-\frac{cd}{c+d} \left| \vec{C} - \vec{D} \right|^2} \iiint_{\infty} d\vec{r}_1 \iiint_{\infty} d\vec{r}_2 \ e^{-p \left| \vec{r}_1 - \vec{P} \right|^2} e^{-q \left| \vec{r}_2 - \vec{Q} \right|^2} \frac{1}{\left| \vec{r}_1 - \vec{r}_2 \right|} \notag \\
    = & C_\mu C_\nu C_\lambda C_\sigma e^{-\frac{ab}{a+b} \left| \vec{A} - \vec{B} \right|^2} e^{-\frac{cd}{c+d} \left| \vec{C} - \vec{D} \right|^2} \frac{2\pi^{5/2}}{pq\sqrt{p+q}} F_0\left( \frac{pq}{p+q} \left| \vec{P} - \vec{Q} \right|^2 \right)
\label{eq:eri_base_case}
\end{align}
The six dimensional integral over $\vec{r}_1$ and $\vec{r}_2$ again reduces to a Boys function form. For the general case, we have:
\begin{align}
(\mu\nu|\lambda\sigma) = & C_\mu C_\nu C_\lambda C_\sigma  \notag \\
    & \left( \prod_{\tau \in \{x,y,z\}} \sum^{i_\tau + j_\tau}_{t_\tau = 0} E^{i_\tau, j_\tau}_{t_\tau, \tau}(A_\tau, B_\tau, p) \left( \frac{\partial}{\partial P_\tau} \right)^{t_\tau} \right) \notag \\
    & \left( \prod_{\tau \in \{x,y,z\}} \sum^{k_\tau + l_\tau}_{s_\tau = 0} E^{k_\tau, l_\tau}_{s_\tau, \tau}(C_\tau, D_\tau, q) \left( \frac{\partial}{\partial Q_\tau} \right)^{s_\tau} \right) \notag \\
    & \frac{2\pi^{5/2}}{pq\sqrt{p+q}} F_0\left( \frac{pq}{p+q} \left| \vec{P} - \vec{Q} \right|^2 \right) 
\label{eq:eri_step1}
\end{align}
In this equation we use parameters $\vec{A}, \vec{B}, p$ and $\vec{C}, \vec{D}, q$ to distinguish $E^{i_\tau, j_\tau}_{t_\tau, \tau}$ for the two electrons. We can further reduce this to the auxiliary integral form:
\begin{align}
(\mu\nu|\lambda\sigma) = & C_\mu C_\nu C_\lambda C_\sigma \notag \\
    & \sum^{i_x + j_x}_{t_x = 0} E^{i_x, j_x}_{t_x, x}(A_x, B_x, p) \sum^{i_y + j_y}_{t_y = 0} E^{i_y, j_y}_{t_y, y}(A_y, B_y, p) \sum^{i_z + j_z}_{t_z = 0} E^{i_z, j_z}_{t_z, z}(A_z, B_z, p) \notag \\
    & \sum^{k_x + l_x}_{s_x = 0} E^{k_x, l_x}_{s_x, x}(C_x, D_x, p) \sum^{k_y + l_y}_{s_y = 0} E^{k_y, l_y}_{s_y, y}(C_y, D_y, p) \sum^{k_z + l_z}_{s_z = 0} E^{k_z, l_z}_{s_z, z}(C_z, D_z, p) \notag \\
    & (-1)^{s_x + s_y + s_z} \frac{2\pi^{5/2}}{pq\sqrt{p+q}}  R_{t_x + s_x, t_y + s_y, t_z + s_z}^0\left( \frac{pq}{p+q}, \vec{P} - \vec{Q} \right)
\label{eq:eri_formula}
\end{align}
which is the fully-specified formula used for ERI evaluation. However, for simplicity, we subsequently abbreviate this formula as:
\begin{align}
(\mu\nu|\lambda\sigma) = \sum_P E_P^{\mu\nu} \sum_Q E_Q^{\lambda\sigma} R_{PQ}
\label{eq:eri_simplified}
\end{align}
where the summation over $\vec{t}$ and $\vec{s}$ is implied by summation over $P$ and $Q$, the $E$ terms on each side have been combined, and we use $R_{PQ}$ to represent the auxiliary integral with prefactors.

Range-separated hybrid density functionals \cite{wpbeh, lc_wpbe, cam_b3lyp, wb97} also require the following long-range ERI:
\begin{align}
\left( \mu\nu \middle| \frac{\mathrm{erf}(\omega r)}{r} \middle| \lambda\sigma \right) = \iiint_{\infty} d\vec{r}_1 \iiint_{\infty} d\vec{r}_2 \ \mu(\vec{r}_1) \nu(\vec{r}_1) \frac{\mathrm{erf}(\omega \left| \vec{r}_1 - \vec{r}_2 \right|)}{\left| \vec{r}_1 - \vec{r}_2 \right|} \lambda(\vec{r}_2) \sigma(\vec{r}_2)
\label{eq:eri_longrange_definition}
\end{align}
where the error function $\mathrm{erf}(x)$ interpolates between no repulsion at short interelectronic separation to full repulsion at long range. The full derivation is rather complicated, and we therefore only show the final expression for this long-range ERI \cite{long_range_integral_1, long_range_integral_2}:
\begin{align}
\left( \mu\nu \middle| \frac{\mathrm{erf}(\omega r)}{r} \middle| \lambda\sigma \right) = & C_\mu C_\nu C_\lambda C_\sigma \notag \\
    & \sum^{i_x + j_x}_{t_x = 0} E^{i_x, j_x}_{t_x, x}(A_x, B_x, p) \sum^{i_y + j_y}_{t_y = 0} E^{i_y, j_y}_{t_y, y}(A_y, B_y, p) \sum^{i_z + j_z}_{t_z = 0} E^{i_z, j_z}_{t_z, z}(A_z, B_z, p) \notag \\
    & \sum^{k_x + l_x}_{s_x = 0} E^{k_x, l_x}_{s_x, x}(C_x, D_x, p) \sum^{k_y + l_y}_{s_y = 0} E^{k_y, l_y}_{s_y, y}(C_y, D_y, p) \sum^{k_z + l_z}_{s_z = 0} E^{k_z, l_z}_{s_z, z}(C_z, D_z, p) \notag \\
    & (-1)^{s_x + s_y + s_z} \frac{2\pi^{5/2}\omega}{pq\sqrt{pq + p\omega^2 + q\omega^2}}  R_{t_x + s_x, t_y + s_y, t_z + s_z}^0\left( \frac{pq\omega^2}{pq + p\omega^2 + q\omega^2}, \vec{P} - \vec{Q} \right)
\label{eq:eri_longrange_formula}
\end{align}
This is quite similar to the expression for standard ERIs (only the prefactor and the input argument to auxiliary integrals $R_{t_x,t_y,t_z}^m$ are different) and will therefore be treated in a similar manner in the implementation.

In most SCF implementations, the 4-index tensor $(\mu\nu|\lambda\sigma)$ is never built explicitly, but ERIs are instead used to construct the Coulomb (classical electron-electron repulsion) and HF exchange matrices \cite{direct_scf_original}.
The Coulomb matrix $\mathbf{J}$ is defined as:
\begin{align}
J_{\mu\nu} = \sum_{\lambda\sigma} (\mu\nu|\lambda\sigma) D_{\lambda\sigma}
\label{eq:j_definition}
\end{align}
in terms of the density matrix $\mathbf{D}$. In order to accelerate the routine for high angular momentum cases, we apply equation \ref{eq:eri_simplified} and reorder the summation following the the celebrated ``J Engine'' algorithm: \cite{j_engine_hgp, j_prepostprocess_original, j_prepostprocess_family_basis_set, terachem_gpu_2, terachem_2013, terachem_regent}
\begin{align}
J_{\mu\nu} = \sum_P E_P^{\mu\nu} \sum_Q R_{PQ} \sum_{\lambda\sigma} E_Q^{\lambda\sigma} D_{\lambda\sigma}
\label{eq:j_formula}
\end{align}

Equation \ref{eq:j_formula} suggests splitting the evaluation of $J_{\mu\nu}$ into three substeps: first transform the density matrix $\mathbf{D}$ from the Cartesian Gaussian basis into the Hermite Gaussian basis (summation over $\lambda\sigma$), then compute the auxiliary integral and obtain $\mathbf{J}$ in the Hermite Gaussian basis (summation over $Q$), and finally convert the $\mathbf{J}$ matrix from the Hermite Gaussian basis back to the Cartesian Gaussian basis (summation over $P$). Implementing Equation \ref{eq:j_formula} in this way reduces both run time and compilation time, and has the added bonus of making the program more modular.

The HF exchange matrix $\mathbf{K}$ is defined as
\begin{align}
K_{\mu\lambda} = \sum_{\nu\sigma} (\mu\nu|\lambda\sigma) D_{\nu\sigma}
\label{eq:k_definition}
\end{align}
Unfortunately, the summation in $\mathbf{K}$ computation cannot be split into substeps similar to $\mathbf{J}$:
\begin{align}
K_{\mu\lambda} = \sum_P \sum_Q \sum_{\nu\sigma} E_P^{\mu\nu} R_{PQ} E_Q^{\lambda\sigma} D_{\nu\sigma}
\label{eq:k_formula}
\end{align}
As a result, $\mathbf{K}$ computation requires performing six summations (three summations each for both $P$ and $Q$ from three cartesian directions) in one step, and as a result is generally considerably more expensive than $\mathbf{J}$ computation.

\subsection{Exchange-correlation integral}
Most Kohn-Sham DFT functionals require evaluation of local exchange-correlation integrals. Within the generalized gradient approximation (GGA), the general forms of the relevant integrals are:
\begin{align}
V_{\mu\nu}^{XC,0} &= \iiint_\infty d\vec{r} \ \frac{\partial \varepsilon_{XC}}{\partial \rho}( \rho(\vec{r}), \vec{\nabla} \rho(\vec{r}) )\mu(\vec{r}) \nu(\vec{r}) \\
V_{\mu\nu}^{XC,1} &= \iiint_\infty d\vec{r} \ \frac{\partial \varepsilon_{XC}}{\partial (\vec{\nabla} \rho)}( \rho(\vec{r}), \vec{\nabla} \rho(\vec{r}) ) \cdot \vec{\nabla} (\mu(\vec{r}) \nu(\vec{r}))
\label{eq:xc_integral_definition}
\end{align}
which are computed numerically as a weighted sum over grid points:
\begin{align}
V_{\mu\nu}^{XC,0} &\approx \sum_g^{N_{grid}} w_g \frac{\partial \varepsilon_{XC}}{\partial \rho}( \rho(\vec{r}_g), \vec{\nabla} \rho(\vec{r}_g) ) \mu(\vec{r}_g) \nu(\vec{r}_g) \label{eq:xc_grid_sum_0} \\
V_{\mu\nu}^{XC,1} &\approx \sum_g^{N_{grid}} w_g \frac{\partial \varepsilon_{XC}}{\partial (\vec{\nabla} \rho)}( \rho(\vec{r}_g), \vec{\nabla} \rho(\vec{r}_g) ) \cdot \vec{\nabla} (\mu(\vec{r}_g) \nu(\vec{r}_g))\label{eq:xc_grid_sum_1}
\end{align}
where the electron density and its gradient at each point are obtained as
\begin{align}
\rho(\vec{r}) &= \sum_{\mu\nu}^{n_{AO}} D_{\mu\nu} \mu(\vec{r}) \nu(\vec{r}) \label{eq:xc_density_0} \\
\vec{\nabla} \rho(\vec{r}) &= \sum_{\mu\nu}^{n_{AO}} D_{\mu\nu} \vec{\nabla} (\mu(\vec{r}) \nu(\vec{r})) \label{eq:xc_density_1}
\end{align}
and the exchange-correlation potentials $\dfrac{\partial \varepsilon_{XC}}{\partial \rho} (\rho(\vec{r}), \vec{\nabla} \rho(\vec{r}) )$ and $\dfrac{\partial \varepsilon_{XC}}{\partial (\vec{\nabla} \rho)}( \rho(\vec{r}), \vec{\nabla} \rho(\vec{r}) )$ depend only on the density and its gradient at each grid point. \cite{dft_gradient, dft_integral, gaussian_program_dft, dft_standard_grid_1, dft_standard_grid_2_3, dft_standard_grid_0, dft_radial_grid_1, dft_radial_grid_2, dft_radial_grid_3, dft_radial_grid_4, dft_radial_grid_5, dft_radial_grid_6} The procedure for evaluating the exchange-correlation integral is: (1) a molecular grid is introduced, (2) the electron density is evaluated at each grid point, (3) the exchange-correlation term is evaluated at each grid point, and (4) for each pair of GTOs, a summation over grid points of the exchange-correlation potential is carried out.

In both equations \ref{eq:xc_grid_sum_0} and \ref{eq:xc_density_0}, the pair distribution value at a grid point $\mu(\vec{r}_g) \nu(\vec{r}_g)$ is needed, and we can use the definition in equation \ref{eq:pair_definition} to compute it. In both equation \ref{eq:xc_grid_sum_1} and \ref{eq:xc_density_1}, the derivative of pair distribution value at a grid point $\vec{\nabla} (\mu(\vec{r}_g) \nu(\vec{r}_g))$ is needed. Here we show the derivative along $x$-direction (the derivatives along $y$ and $z$ can be obtained by permutation symmetry):
\begin{align}
\frac{\partial}{\partial x} \mu(\vec{r})\nu(\vec{r}) = & C_\mu C_\nu \left( -2p (x - P_x) (x - A_x)^{i_x} (x - B_x)^{j_x} \right. \notag \\
    & \left. + i_x (x - A_x)^{i_x-1} (x - B_x)^{j_x} + j_x (x - A_x)^{i_x} (x - B_x)^{j_x-1} \right)  \notag \\
    & (y - A_y)^{i_y} (y - B_y)^{j_y} (z - A_z)^{i_z} (z - B_z)^{j_z} e^{-\frac{ab}{a+b} \left| \vec{A} - \vec{B} \right|^2} e^{-p \left| \vec{r} - \vec{P} \right|^2}
\label{eq:pair_derivative}
\end{align}

\subsection{Spatial derivatives of integrals}

In order to perform geometry optimizations or dynamics simulations, we require derivatives of the energy with respect to atomic positions \cite{pulay_hf_derivative, pople_hf_derivative, dft_gradient}. This necessitates knowledge of the derivatives of all the integrals listed above with respect to atomic center locations $\vec{A}$, $\vec{B}$, $\vec{C}$ and $\vec{D}$. The relevant formulae are provided in supporting information, as the equations are rather long. 

\section{Implementation} \label{sec:implementation}

\subsection{Equation generation}

The McMurchie-Davidson scheme splits the integral evaluation into three parts: formation of transformation coefficients $E^{i_\tau, j_\tau}_{t_\tau, \tau}$, formation of auxiliary integrals $R_{t_x,t_y,t_z}^0$, and combination of $E^{i_\tau, j_\tau}_{t_\tau, \tau}$ and $R_{t_x,t_y,t_z}^0$ terms to form the integrals.

We generate the equations for each of the relevant $E^{i_\tau, j_\tau}_{t_\tau, \tau}$ and $R_{t_x,t_y,t_z}^0$ terms by carrying out the recurrence relationship with SymPy \cite{sympy} symbolic math tools. Every term recurs down to the base case and no intermediate recursion terms are saved. We thereby avoid searching for shared intermediates, unlike many previous ERI algorithms.
\cite{head_gordon_pople_original, head_gordon_pople_flops, head_gordon_pople_recursion_net, gamess_spd, gamess_f} 
The fully expanded $E^{i_\tau, j_\tau}_{t_\tau, \tau}$ terms have only $\vec{P} - \vec{A}$, $\vec{P} - \vec{B}$, $\frac{1}{2p}$ as inputs, and $e^{ -\frac{ab}{a+b} \left|\vec{A} - \vec{B}\right|^2 }$ as a overall prefactor. The fully expanded $R_{t_x,t_y,t_z}^0\left( p,\vec{L} \right)$ terms require $p$, $\vec{L}$ and the Boys function values $F_m\left( p\left|\vec{L}\right|^2 \right)$ as inputs ($m$ ranges from 0 to the sum of angular momentum indices $l_{sum} = i_x+i_y+i_z+j_x+j_y+j_z$, so $l_{sum}+1$ Boys function values are needed).

For computing the Coulomb matrix $\mathbf{J}$, the computation of $E^{i_\tau, j_\tau}_{t_\tau, \tau}$ and $R_{t_x,t_y,t_z}^0$ terms are separated, as shown in equation \ref{eq:j_formula}, and so no additional terms need to be generated.

For overlap, electron kinetic energy and nuclear attraction integrals, we substitute the expression of $E^{i_\tau, j_\tau}_{t_\tau, \tau}$ terms into the integral equations \ref{eq:overlap_formula}, \ref{eq:kinetic_formula} and \ref{eq:v1e_formula} to obtain a set of messy ``final" equations. Without further simplification, these equations require a large number of unnecessary floating point operations (FLOP) to evaluate, because they have many subexpressions that show up more than once. We use the common subexpression elimination (CSE) routine \cite{cse_polynomial, cse_black_hole} from SymPy to extract these subexpressions and reduce the amount of repeated floating point operations. As will be mentioned in the performance section, a full CSE does not always yield the best performance, and might need to be partially ``reversed".

For computing the HF exchange matrix $\mathbf{K}$, we are not able to fully expand the six summations over three $P$ and three $Q$ indices in equation \ref{eq:k_formula} because it makes the equation too long to be handled by SymPy functions or the CUDA compiler. We instead make each term in P and Q summation a separate equation (leaving the summation over $\lambda$ and $\sigma$), and substitute the $E^{i_\tau, j_\tau}_{t_\tau, \tau}$ terms with their expanded forms. A CSE on these split equations is expensive, and preliminary testing showed no performance improvement for either compilation time or run time. Therefore, no CSE is therefore performed for $\mathbf{K}$ construction in our final implementation.

\begin{table}[htb!]
\begin{tabular}{ >{\centering\arraybackslash}p{3em} | >{\raggedleft\arraybackslash}p{5.4em} | >{\raggedleft\arraybackslash}p{5em} | >{\raggedleft\arraybackslash}p{5em} | >{\raggedleft\arraybackslash}p{5em} }
 \hline
 Kernel & FLOP count before CSE & FLOP count after CSE & FLOP count reduction & Number of temporary variables \\
 \hline
 SS & 0 & 0 & 0 & 0 \\
 SP & 11 & 9 & 2 & 0 \\
 SD & 68 & 52 & 16 & 6 \\
 SF & 262 & 146 & 116 & 31 \\
 PP & 104 & 71 & 33 & 11 \\
 PD & 515 & 267 & 248 & 64 \\
 PF & 1796 & 710 & 1086 & 179 \\
 DD & 2243 & 837 & 1406 & 187 \\
 DF & 7241 & 2122 & 5119 & 455 \\
 FF & 22064 & 4817 & 17247 & 957 \\
 \hline
\end{tabular}
\caption{FLOP counts for the nuclear attraction integral equations, where $\vec{P} - \vec{A}$, $\vec{P} - \vec{B}$, $\frac{1}{2p}$ and $R_{t_x,t_y,t_z}^0$ shows up on the righthand side of each equation. The CSE is performed simultaneously on the same type of integrals (for example $sp_x$, $sp_y$ and $sp_z$ integrals). The number of temporary variables used for CSE is also reported.}
\label{tab:E_cse_flops_count}
\end{table}

We note that there are several limitations for expression simplification using SymPy. SymPy treats power operations (such as ``x**2" in python) as a primitive operation and thus cannot find common subexpressions between two different powers of the same variable (for example ``x**2" and ``x**3"). This limits its capacity of reducing FLOP counts for the auxilliary Boys integrals $R_{t_x,t_y,t_z}^0$, which contain many power operations. For example, the FLOP counts for the fully-expanded expression for the $l_{sum} = 12$ case and the corresponding SymPy CSE result are 28816 and 6569, respectively, while the optimal tree-search algorithm reports a lower value of 2219 \cite{McMurchie_Davidson_tree_search}. We therefore placed the fully-expanded $R_{t_x,t_y,t_z}^0$ expressions into the source code and rely on the compiler to optimize simple power expressions. The SymPy CSE is better suited for the Cartesian to Hermite Gaussian transformation $E^{i_\tau, j_\tau}_{t_\tau, \tau}$ term reductions, since the expressions have more complicated forms. This is demonstrated by the FLOP count reduction by CSE for nuclear attraction integrals (reported in Table \ref{tab:E_cse_flops_count}), which shows a greater than 60\% decrease for high angular momentum kernels. It is nonetheless worth noting that all of the FLOP counts reported in the table does not take into account fused operations in CUDA optimization, which is important for performance but hard to predict at code level.

It is also important to point out that, given the limited performance of SymPy, we use python strings to represent the equations during most equation manipulation steps, and only carry out a few SymPy simplification operations and CSE at the end.

\subsection{Data structure}

Pair distributions of pGTOs are the basic data structure used in all TeraChem integral routines, which therefore all utilize a fully uncontracted scheme. Once the basis set and geometry is defined, we compute a list of all pGTO pairs and sort them by $\left| (\mu\nu|\mu\nu) \right|^{1/2}$, in order to implement screening by the ERI Cauchy-Schwarz inequality:
\begin{align}
\left| (\mu\nu|\lambda\sigma) \right| \leq \left| (\mu\nu|\mu\nu) \right|^{1/2} \left| (\lambda\sigma|\lambda\sigma) \right|^{1/2}
\label{eq:cauchy_schwarz_bound}
\end{align}
Pairs with $\left| (\mu\nu|\mu\nu) \right|^{1/2}$ smaller than a pair threshold (PQTHRE) are discarded. For a sufficiently large system and a given pair threshold, the number of significant pGTO pairs only scales as $O(N)$ with system size $N$.

For orbitals with nonzero angular momentum, the pGTOs from the same shell with the same exponent but different angular momentum indices (the $p_x$, $p_y$ and $p_z$ orbitals, for example) are usually grouped together, both because of the reduced amount of storage, and because of the shared intermediates in the integrals involved ($V_{sp_x}$, $V_{sp_y}$ and $V_{sp_z}$ for example). TeraChem stores these pGTOs as one primitive shell pair, referring to more than one pGTO. For example, a PP type primitive shell pair contains $3 \times 3 = 9$ pGTO pairs. The current implementation of TeraChem stores 10 lists of primitive shell pairs with different angular momentum combinations (SS, SP, ..., FF), exploiting the permutation symmetry of the pairs. When implementing integrals, we also prepare individual integral routines for each angular momentum combination, for example $V_{SS}$, $V_{SP}$ and $V_{PP}$, and each of these functions handles all pGTO pairs in the corresponding primitive shell pair.

In order to better support this angular pair data structure, TeraChem stores atomic orbitals with angular momentum as the leading order,\cite{terachem_gpu_1} instead of the popular choice of having the atomic center index as leading order, which is used in most other programs \cite{psi4_recent, pyscf_recent}. As a result, the Fock matrix and density matrix are easily separable into angular pair segments. For instance, given a system with only $s$ and $p$ orbitals, we can separate the AO basis density matrix into 4 submatrices, with the $SS$ and $PP$ blocks on diagonal and $SP$ and $PS$ blocks as off-diagonals. Therefore, we only need to go through the block corresponding to a particular angular pair while accessing that pair for ERI computation.

\subsection{GPU acceleration of one-electron integrals}

The overlap and electron kinetic energy integrals are computationally inexpensive and are computed only once per HF/DFT calculation. These integrals therefore are not accelerated on the GPU but are computed on the CPU.

The nuclear attraction integral is also computed only once per HF/DFT calculation. However this involves Boys function evaluations and has a formal scaling of $O(N^3)$. We therefore implement it on GPUs by assigning one primitive shell pair to one GPU thread, and having every thread loop through all point charges to form the summation in equation \ref{eq:v1e_summation_over_C}. In order to avoid random memory access problems on GPUs, we store the integral evaluation result in the same order as the sorted primitive shell pairs and accumulate them to the Fock matrix only after copying the data back to host memory.

\subsection{GPU acceleration of Coulomb and exchange matrices}

The Coulomb ($\mathbf{J}$) and HF exchange ($\mathbf{K}$) matrices are rebuilt each SCF iteration and therefore account for most of the computation cost during a SCF procedure. GPU acceleration of these routines has consequently attracted considerable attention. \cite{terachem_gpu_1, terachem_gpu_2, terachem_gpu_3, terachem_2013, terachem_dynamic_precision, brianqc, gaussian_program, gaussian_program_high_angular, gamess_rys_quadrature, gamess_sp, gamess_spd, gamess_f, fermions_linear_scaling_exchange, fermions_linear_scaling_gradient, fermions_density_fitting, quick, libintx_up_to_iiii, libintx_gpu_2, pyscf_gpu, quantum_supercharger_library, turbomole, wuhan_electronic_structure_package, brush_algorithm, gpu_chemistry_review_2010, packages_on_hardwares, gauxc} Here we briefly describe our GPU-accelerated implementation.

\subsubsection{Coulomb matrix}
Not all operations for $\mathbf{J}$ matrix construction are performed on the GPU, as we find better performance if we transform between Cartesian Gaussians and Hermite Gaussians for both density matrix $\mathbf{D}$ (``preprocessing'') and Coulomb matrix $\mathbf{J}$ (``postprocessing'') on the CPU, placing only the auxiliary integral computation on GPU. This approach reduces repeated computation. For example, the $(ss|pp)$ and $(sp|pp)$ integrals share the same list of ket shell pairs and thus the same list of Hermite density matrix elements. We compute the Hermite density for each list of primitive shell pairs once before any auxiliary integral GPU kernel function call, and copy all of them to the GPU, which will be used as an input to auxiliary integral calls. For the Hermite $\mathbf{J}$ output from the auxiliary integral calls, we allocate GPU memory for each GPU function call to avoid atomic addition operations on the GPU. But on the host side we only store one copy of Hermite $\mathbf{J}$ matrix elements, whose size is the number of bra pairs (in Hermite Gaussian basis, which requires more space than Cartesian Gaussian basis). When one of the GPU auxiliary integral calls finishes, the result is accumulated to the CPU copy of Hermite $\mathbf{J}$ under lock protection (i.e., as an atomic operation). Once all auxiliary integral calls finish, the accumulated $\mathbf{J}$ matrix in the Hermite Gaussian basis is transformed to the Cartesian Gaussian basis and summed into the Fock matrix. Similar to the density transformation (where CPU-based preprocessing saves work for integrals with shared ket pairs), this CPU-based postprocessing algorithm for $\mathbf{J}$ transformation saves work for integrals with shared bra pairs, like $(ss|pp)$ and $(ss|sp)$ integrals.

The auxiliary integrals for Coulomb matrix are grouped into angular momentum pair combinations, like $(ss|ss)$ and $(ss|sp)$, and in theory one GPU kernel function should be implemented for each 4-index combination. However, given the 8-fold symmetry relationship of ERIs with real basis functions:
\begin{align}
(\mu\nu|\lambda\sigma) &= (\nu\mu|\lambda\sigma) \notag \\
(\mu\nu|\lambda\sigma) &= (\mu\nu|\sigma\lambda) \notag \\
(\mu\nu|\lambda\sigma) &= (\lambda\sigma|\mu\nu)
\label{eq:8_fold_symmetry}
\end{align}
only the angular momentum pair combination with $L_\mu \leq L_\nu$ and $L_\lambda \leq L_\sigma$ is unique and requires a GPU kernel function (where $L_\mu$ is the angular momentum of the shell $\mu$, etc.). The bra-ket symmetry (last equation in equation \ref{eq:8_fold_symmetry}) cannot be exploited in the Coulomb matrix auxiliary integral kernel, because ket pairs share indices with $\mathbf{D}$ and bra pairs share indices with $\mathbf{J}$. As a result, we have 100 GPU kernel functions for Coulomb matrix auxiliary integrals (10 bra angular pairs $\times$ 10 ket angular pairs). In order to support calculations with mixed and dynamic precision, \cite{terachem_dynamic_precision} we generate separate kernel functions for single and double precision floating point numbers. Therefore, a total of 200 GPU kernel functions are generated for $\mathbf{J}$ construction.

\begin{figure}[htb!]
    \centering
    \includegraphics[width=0.8\linewidth]{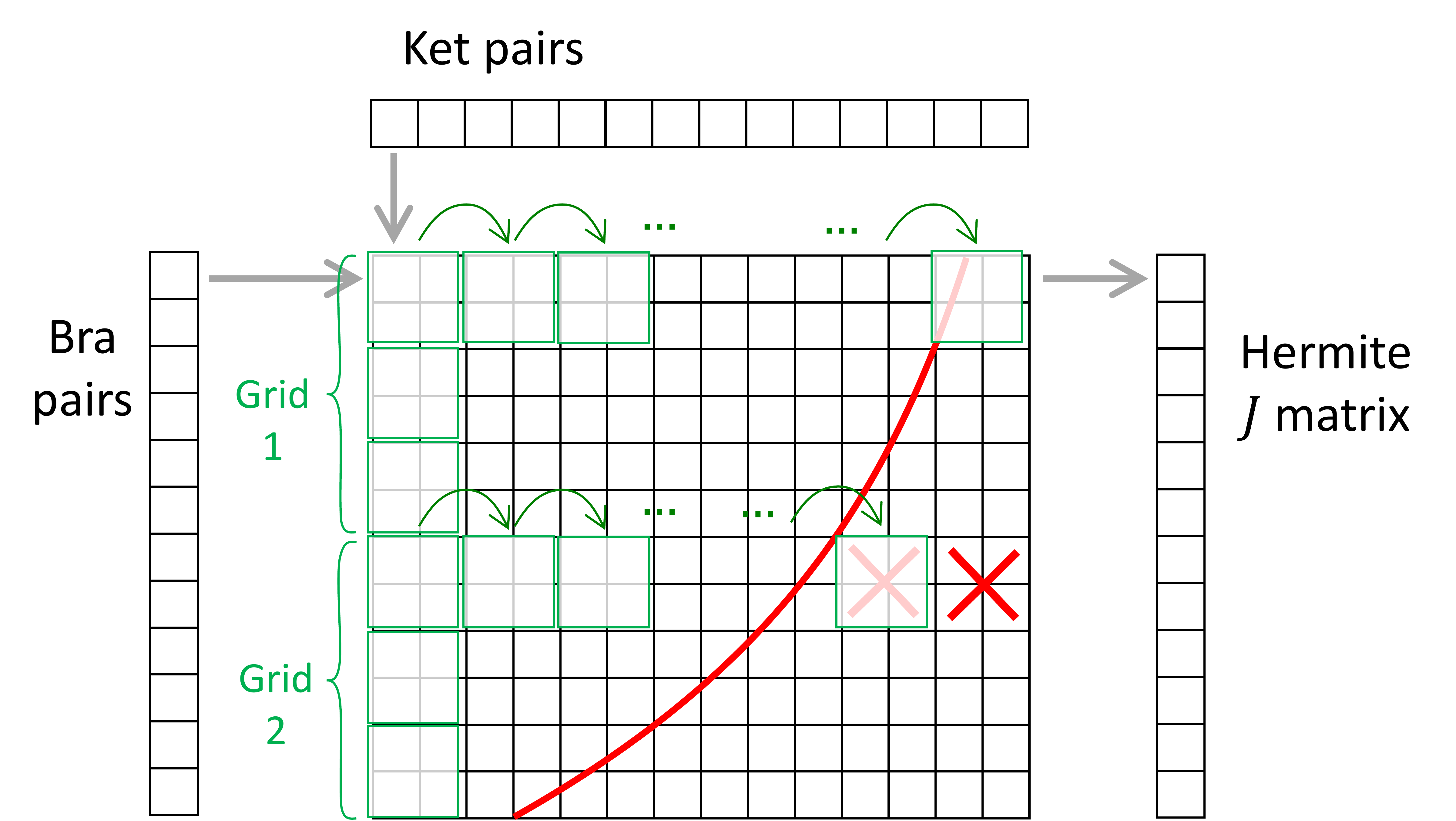}
\caption{Demonstration of $\mathbf{J}$ matrix auxiliary integral GPU kernel execution. Here we use green boxes to represent thread blocks (shown as size $2 \times 2$ for simplicity), and black boxes to represent the bra and ket pairs. The bra pairs are split into two grids. Each thread block will iterate through all ket pairs, fetching the ket pair and Hermite density inputs and computing the Hermite $\mathbf{J}$ matrix, until it reaches the end of ket pair list, or the upper-bound value of the whole thread block falls below a threshold (represented by the red line). Once done iterating through ket pairs, the thread block will perform a internal summation along ket direction, and place the result back onto GPU memory.}
\label{fig:j_kernel_demo}
\end{figure}

Fig. \ref{fig:j_kernel_demo} illustrates the GPU algorithm for auxiliary integrals: the threads in every GPU kernel are grouped into $8 \times 8$ thread blocks, where the x and y index of the thread represents the bra and ket indices of the integral. We allocate enough thread blocks ($\left\lceil \frac{n_{bra}}{8} \right\rceil$) to cover all bra primitive shell pairs, and if the number of bra pairs of an angular momentum combination is larger than the GPU thread capacity, we will split the bra pair list into smaller sub-lists, and assign one grid of thread blocks for each sub-list of bra pairs. The grids can be launched in parallel on multiple GPUs, or sequentially on one GPU. Each thread block will iterate through ket pairs, as illustrated in Fig. \ref{fig:j_kernel_demo}. At the beginning of the iterations, each thread will load its bra pair information according to its thread index, including the pair center location $\vec{P}$, pair exponent $p$, coefficient $C_\mu C_\nu$ and Cauchy-Schwarz upper-bound $\mathrm{bound}_{\mu\nu}$. In each iteration, each thread will load its ket pair information, including $\vec{Q}$, $q$, $C_\lambda C_\sigma$, $\mathrm{bound}_{\lambda\sigma}$, together with the Hermite density $D_{t_xt_yt_z}$, and the maximum absolute value of density matrix element as a density bound value $\mathrm{bound}_{D_{\lambda\sigma}}$. For nonzero angular momentum cases, one ket primitive shell pair maps to more than one density matrix element, for example, the $PD$ primitive shell pair will map to 18 density matrix elements (corresponding to the three p orbitals and six Cartesian d orbitals), and we compute the max absolute value of the set to determine the cutoff. In every iteration, once the ket bound is fetched, we check if the total upper-bound is above the threshold (THRECL):
\begin{align}
\mathrm{bound}_{\mu\nu}\mathrm{bound}_{\lambda\sigma} \mathrm{bound}_{D_{\lambda\sigma}} = \left| (\mu\nu|\mu\nu) \right|^{1/2} \left| (\lambda\sigma|\lambda\sigma) \right|^{1/2} \mathrm{max}(|D_{\lambda\sigma}|) \geq \mathrm{THRECL}
\label{eq:kernel_upperbound}
\end{align}
If the condition above is not satisfied for one thread, the thread will ignore this ket pair, neither fetching other ket data or Hermite density, nor performing any further evaluation. However, because of the strong synchronization implementation of CUDA, if other threads in the same thread block are working on significant elements, the thread with no work has to wait. In order to minimize this effect, we re-sort the list of ket pairs by the product of $\mathrm{bound}_{\lambda\sigma}$ and $\mathrm{bound}_{D_{\lambda\sigma}}$ for every SCF iteration, before the $J$ matrix construction. As a result, in all following iterations, this thread will only see ket pairs with smaller overall bounds and will drop all of them. Once the whole thread block has the total upper-bound below threshold, the iteration for this thread block will stop. Each thread will accumulate the Hermite $\mathbf{J}$ matrix element into its local register. Once the whole thread block finishes the iteration, the threads with the same $x$ index (bra index) will collectively sum their Hermite $\mathbf{J}$ value, which provides the Hermite $\mathbf{J}$ summation over all ket pairs. This value is written back to GPU memory, copied back to CPU memory, and handled by the postprocessing CPU logic mentioned above.

In terms of the ``location" of the intermediate variables for each kernel, we define all variables as register variables, and rely on the compiler to transfer data between register and cache/global memory to handle register overflow. We limit our usage of shared memory to the size of thread block, just to perform the reduction operation at the end of each kernel execution, to leave as much L1 cache as possible. This also applies to $\mathbf{K}$ construction and spatial derivative kernels.

\subsubsection{Exchange matrix}

$\mathbf{K}$ construction is more complicated because of the index mismatch between the bra ket pair and input ($\mathbf{D}$) output ($\mathbf{K}$) pair. As a consequence, we cannot easily predict which density matrix or $\mathbf{K}$ matrix element to access given the list of bra and ket pairs. This seems to require storing the whole density matrix and the whole $\mathbf{K}$ matrix of a particular angular pair in GPU memory [for example, when computing $(sp|sp)$ integrals, we need to store the $PP$ segment of the $\mathbf{D}$ matrix and $SS$ segment of the $\mathbf{K}$ matrix], and to access both of them in a random memory access pattern, given that the sorted $\mu\nu$ pairs do not preserve the order of $\mu$ or $\nu$ indices.

\begin{figure}[htb!]
    \centering
    \includegraphics[width=0.8\linewidth]{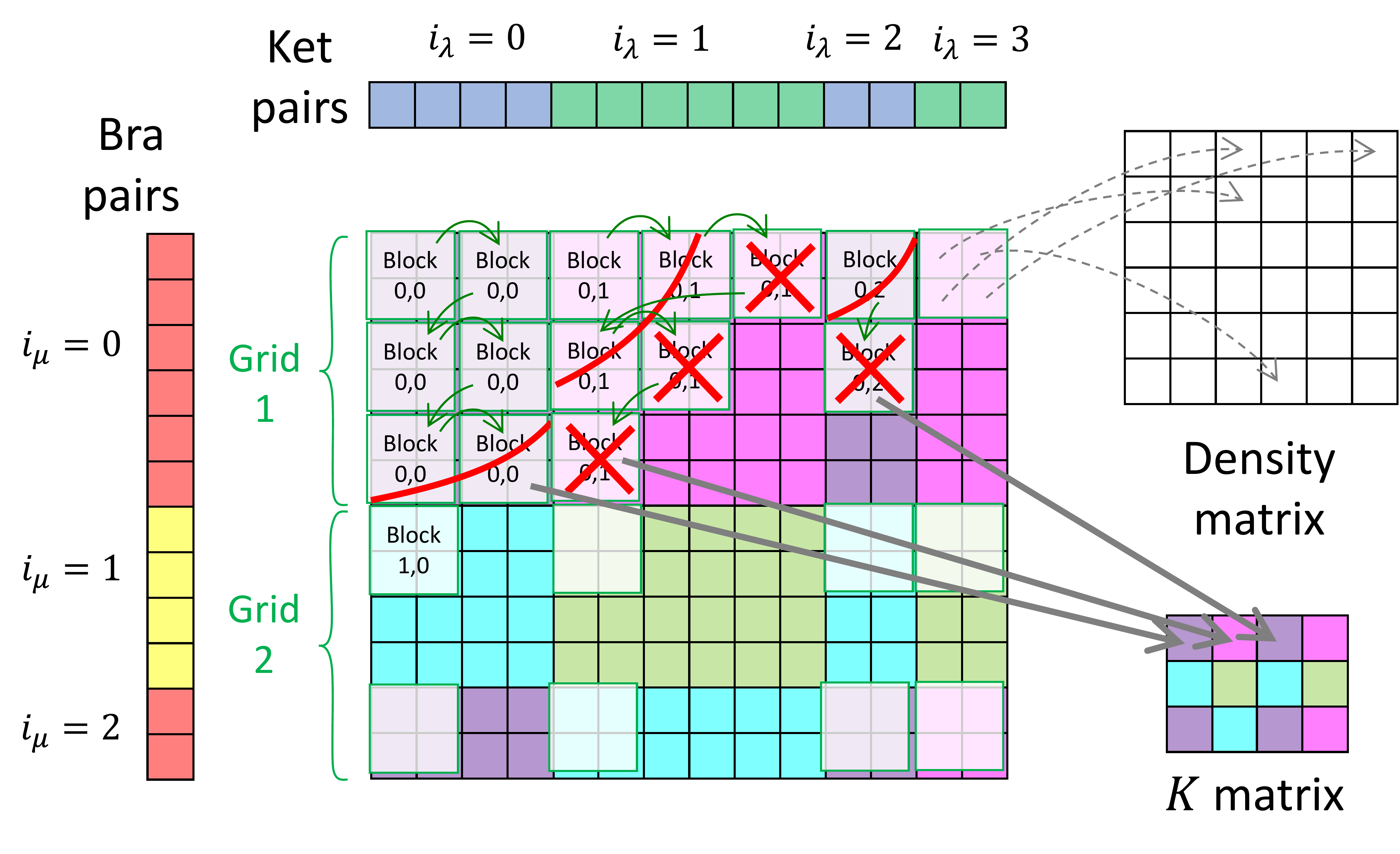}
\caption{Demonstration of exchange matrix $\mathbf{K}$ GPU kernel execution. We use green boxes to represent thread blocks (of size $2 \times 2$), and black boxes to represent the bra and ket pairs. We color the pairs according to $\mu$ and $\lambda$ indices. The bra pairs are split into two grids. Each thread block will find its $K_{\mu\lambda}$ region and iterate through this region, by first moving along the ket direction and then along the bra direction. At each iteration, each thread will fetch the bra and ket pairs, then fetch the density matrix element according to $\nu$ and $\sigma$ indices, and compute the $\mathbf{K}$ matrix element. If the whole thread block notices that the total bound of bra, ket and density is below the threshold, it will jump to the next bra pair. In that case, the iteration ends if it has never moved along the ket direction. Each $K_{\mu\lambda}$ region has a separate cutoff, represented by the red lines. Once done iterating through the $K_{\mu\lambda}$ region, the thread block will perform an internal summation along both directions, and place the result back into GPU memory.}
\label{fig:k_kernel_asymmetric_demo}
\end{figure}

In order to minimize the GPU memory usage and amount of uncoalesced memory access, we redesign the pair data structure as follows: we first sort the pair list $\mu\nu$ with the $\mu$ index as leading order, and the upper-bound value ($\mathrm{bound}_{\mu\nu}$) as secondary order. This new pair data structure groups the 4-index integrals $(\mu\nu|\lambda\sigma)$ into regions with the same $\mu$ and $\lambda$ indices, and thus regions with the same $K_{\mu\lambda}$ element, as demonstrated in Fig. \ref{fig:k_kernel_asymmetric_demo}. Then, for every $\mu$ index, we check if the number of primitive shell pairs is a multiple of 8, and if not, we pad with empty pairs (pairs with bound value of zero) until it is a multiple of 8. The padding pattern guarantees that an $8 \times 8$ thread block can iterate through the $K_{\mu\lambda}$ region without worrying about boundaries.

In Fig. \ref{fig:k_kernel_asymmetric_demo} we show the behavior of each thread block in GPU kernel: We allocate number of blocks equal to number of bra $\mu$ indices times the number of ket $\lambda$ indices. Each thread block will first find its $K_{\mu\lambda}$ region start location and size. It will then iterate through the whole region. At each iteration, each thread will fetch the pair center location $\vec{P}$, pair exponent $p$, coefficient $C$ and upper-bound for both bra and ket pairs, similar to the $\mathbf{J}$ matrix auxiliary integral implementation. In addition, each thread will fetch the $\nu$ and $\sigma$ index from bra and ket correspondingly, and use that to compute a density matrix element index. This index is unpredictable, so we require the whole density matrix of the angular pair stored on GPU memory, and each thread has to access them in an uncoalesced fashion. In order to minimize the need for this expensive density fetching, we perform the cutoff of bra and ket pair first, i.e. if the bra and ket upper-bound product is already too small, we avoid fetching the density. We also pre-compute and save the density bound ($\mathrm{bound}_{D_{\nu\sigma}}$) of each $\nu\sigma$ primitive shell pair (in the same way as described in $\mathbf{J}$ matrix build) as a single-precision floating point number, so as to minimize the amount of density fetching if its value is small. It is important to notice that cutoff will happen independently in each $K_{\mu\lambda}$ region, indicated by different background color in Fig. \ref{fig:k_kernel_asymmetric_demo}. If the total bound is greater than the threshold (equation \ref{eq:kernel_upperbound}), the ERI is computed according to equation \ref{eq:eri_formula}. Once the thread block finishes its iterations, it will sum the $K_{\mu\lambda}$ value of all 64 threads, and store them in the GPU memory. Since the output index depends only on block index (which is coupled to the $\mu$ and $\lambda$ indices), we can allocate only the necessary GPU memory, and fill the output in an ordered fashion. Although this is unfortunately not ``coalesced'' access because only one thread in a warp will write to the output, the algorithm does minimize $\mathbf{K}$ matrix output memory access. The $\mathbf{K}$ matrix elements are then copied back to the CPU, reordered and summed into the Fock matrix.

\begin{figure}[htb!]
    \centering
    \includegraphics[width=0.8\linewidth]{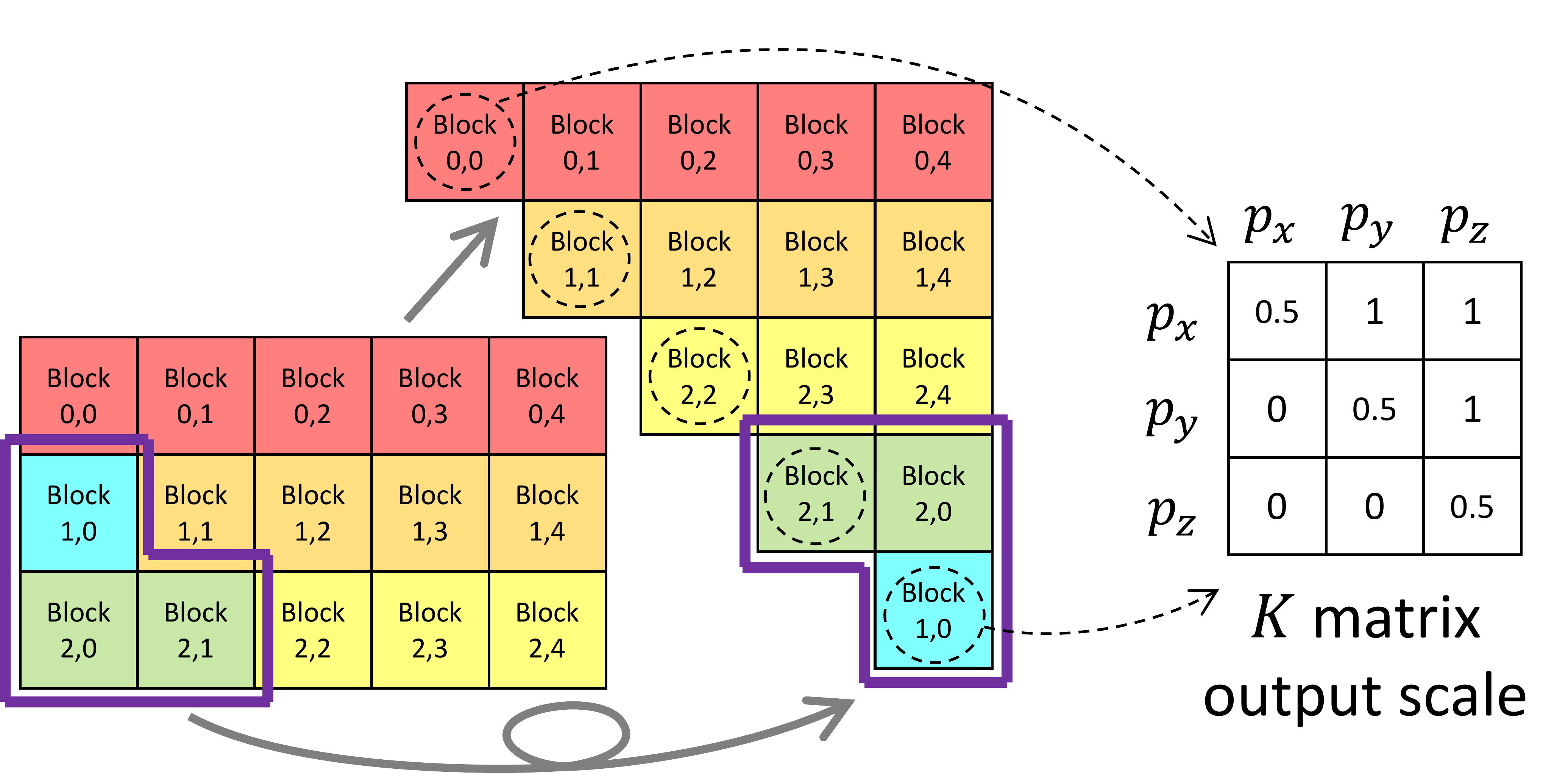}
\caption{The block index remapping pattern for $\mathbf{K}$ matrix  GPU kernel with same angular pair for bra and ket. Here we show an example with 5 $\mu$ or $\lambda$ indices. For each diagonal block after remap, we also show the $\mathbf{K}$ matrix output scaling factors for a $(p\nu|p\nu)$ kernel, where we halve the diagonal elements and zero the lower triangular elements. Other angular index combinations are similarly scaled.}
\label{fig:k_kernel_symmetric_demo}
\end{figure}

The $\mathbf{K}$  matrix kernel described above works properly if the bra angular pair is different from ket angular pair (for example $(ss|sp)$ kernel). If they are the same (for example $(sp|sp)$ kernel), we have to take care to avoid double counting. To avoid that, we remap the block indices of each thread block according to Fig. \ref{fig:k_kernel_symmetric_demo}, so that only the upper-triangular part of the ERI region is accessed. If the remapped block index is on the diagonal, we permit the double counting, but scale the $\mathbf{K}$ matrix result by half (see Fig. \ref{fig:k_kernel_symmetric_demo}) before storing it back to GPU memory.

For $\mathbf{K}$  matrix computation, we can only exploit the last symmetry in the 8-fold symmetry relationship (equation \ref{eq:8_fold_symmetry}), and that requires us to generate a far larger number of GPU kernel functions than Coulomb matrix auxiliary integral kernels. We order all 16 unsymmetrized angular pairs ($SS < SP < SD < SF < PS < ... < FF$) and generate kernels for each combination of two angular pairs, if the bra angular pair is less than or equal to the ket angular pair. This results in 136 kernels each for single and double precision.

We note that we have the choice to perform the transformation coefficient $E^{i_\tau, j_\tau}_{t_\tau, \tau}$ computation on CPU and copy it to GPU memory (as done for the $\mathbf{J}$ matrix), or just compute them on GPU. The benefit of saving repeated work by pre-computing $E^{i_\tau, j_\tau}_{t_\tau, \tau}$ terms on CPU is still valid for $\mathbf{K}$ matrix evaluation. However, it comes with an increased memory transfer cost.  Our tests indicated that, for current GPU architectures, run time is decreased by computing $E^{i_\tau, j_\tau}_{t_\tau, \tau}$ terms on the GPU. We believe this is because the $E^{i_\tau, j_\tau}_{t_\tau, \tau}$ computation can be performed while the thread is waiting to fetch the density matrix elements. However, placing the $E^{i_\tau, j_\tau}_{t_\tau, \tau}$ transformation in the GPU kernels does significantly increase the CUDA compilation time.

Another implementation detail worth mentioning is that lengthy kernel functions are expensive both in compile time and run time. For high angular momentum integrals, like $(ff|ff)$, lengthy kernels are unavoidable because of the complicated integral equations. The complexity of this kernel in fact prevented us from being able to compile the code with CUDA 12.2.1 on the 2 TB memory machines available to us. However, we can mitigate this problem by splitting the output $\mathbf{K}$ matrix elements belonging to the same primitive shell pair into more than one kernel. For example in the $(fd|fd)$ exchange matrix kernel function, the $K_{\mu\lambda}$ output refers to a $FF$ type shell pair, which maps to 100 pair components ($f_{xyz}f_{xyz}$, $f_{xyz}f_{x^2y}$, ..., $f_{z^3}f_{z^3}$) in the $\mathbf{K}$ matrix. We split each output element into a separate kernel function, and thereby utilize 100 smaller kernel functions instead of one big kernel function. This splitting decreases the register requirements and thus register spills for each of the kernels, while increasing the number of floating point operations (because many intermediates, including the Boys functions and some auxiliary integrals, are being recomputed). It is difficult to suggest well-defined criteria for kernel splitting (especially for small kernels) since the available register and cache is different for each type of GPU. A fully automated empirical testing approach could be considered here, as previously done for linear algebra \cite{atlas_linalg} (ATLAS) and fast Fourier transforms \cite{fftw} (FFTW), as well as for GPU-accelerated effective core potential integrals \cite{ace_gpu} and ERI algorithms \cite{brianqc} in quantum chemistry. For now, we content ourselves with a simple rough guideline. On an NVIDIA RTX 4090 GPU, we found that limiting the number of equations for each kernel to at most 4000 (if possible, otherwise split as much as possible), the run time is roughly optimized for a photoactive yellow protein (PYP) system (QM region 5  with 723 atoms,\cite{pyp_qm_region} HF/cc-pVDZ with 7333 basis functions). We provide profiling results for some other kernel splitting strategies in the supporting information.

\subsection{GPU acceleration of exchange-correlation integral}

It is interesting to point out that, even for a system with more than 5000 basis functions, the numerical evaluation of the exchange-correlation integrals, despite a smaller formal scaling of $O(N^3)$, has a big prefactor that makes the run time comparable to the HF exchange matrix computation. This is a consequence of the large number of grid points necessary for accurate local exchange-correlation quadrature. \cite{dft_standard_grid_1, dft_standard_grid_2_3, dft_standard_grid_0, dft_radial_grid_1, dft_radial_grid_2, dft_radial_grid_3, dft_radial_grid_4, dft_radial_grid_5, dft_radial_grid_6} We therefore also describe our GPU-accelerated algorithm of evaluating densities and density gradients on grid points and evaluating exchange-correlation integrals by summing over grid point values. The exchange-correlation term evaluation on each grid point scales linearly with molecular size and takes negligible amount of run time, so CPU parallel execution is sufficient and we do not use the GPU for this part of the computation.

When computing the density value on each grid point, we assign one grid point to each GPU thread, and each thread will iterate through the list of primitive shell pairs and sum the pair value according to equation \ref{eq:xc_density_0} and \ref{eq:xc_density_1}. In order to accelerate the operation, we assign each pair to each atomic center based on adjacency, meaning that we group together all pairs whose pair center location $\vec{P}$ is closest to a particular atomic center. Given this data structure, each thread will first loop through all atomic center, and if the center location is too far away from the grid point location, then all pairs associated with this center will be ignored. The pair groups make the cutoff logic more efficient.

When computing exchange-correlation integrals for each primitive shell pair, we assign one pair to each GPU thread, and each thread iterates through all grid points and sums the exchange-correlation term multiplied with the pair value, according to equation \ref{eq:xc_grid_sum_0} and \ref{eq:xc_grid_sum_1}. Since grid points can be very far from all atomic center locations, we instead partition the bounding volume of all grid points into boxes with fixed size, and distribute each grid point into the corresponding box. Each thread in the GPU kernel will loop through all boxes, and if any grid point in the box is too far away from the pair center location, as determined by $|\vec{r} - \vec{P}|^2 > 50 \ \mathrm{Bohr}^2$ and $p|\vec{r} - \vec{P}|^2 > 100$, then the whole box is ignored. It is important to point out that the pair group and grid point box data structure does not reduce the scaling of run time cost, and merely reduces the prefactor.

\section{Performance} \label{sec:performance}

\begin{figure}[htb!]
    \centering
    \includegraphics[width=0.6\linewidth]{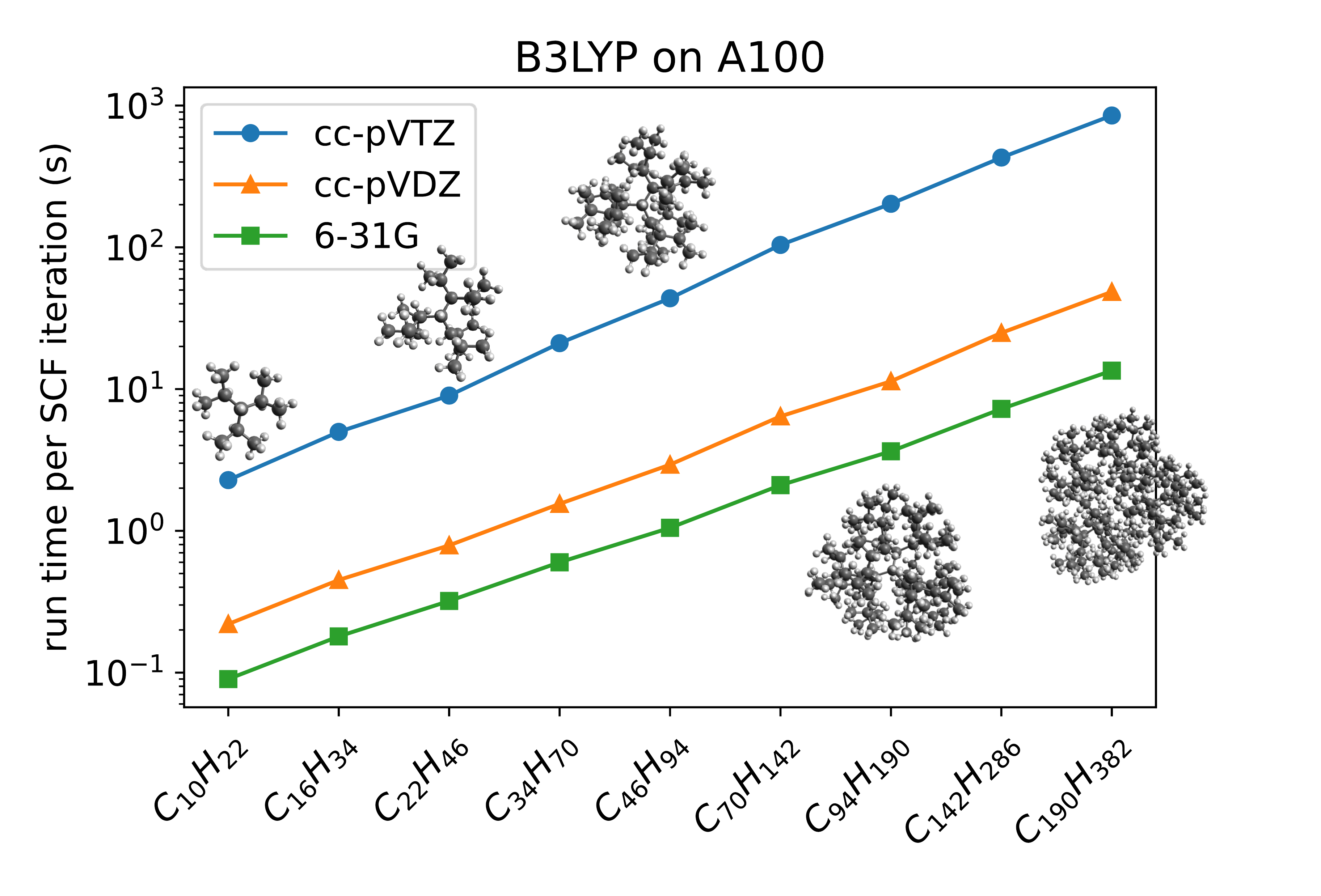}
\caption{Run time per SCF iteration (averaged over all cycles) for B3LYP \cite{b3lyp_original, b3lyp_parameter} calculations on branched alkanes, using one NVIDIA A100 GPU. We use the 6-31G \cite{6_31g_h, 6_31g} ($s$ and $p$ orbitals only), cc-pVDZ \cite{cc_pvxz} ($s,p$ and $d$ orbitals) and cc-pVTZ \cite{cc_pvxz} ($s,p,d$ and $f$ orbitals) basis sets. The scaling of run time with respect to the number of basis functions ($n_{AO}$) is $O(n_{AO}^{2.1})$ for the B3LYP/cc-pVTZ calculation, $O(n_{AO}^{1.9})$ for the B3LYP/cc-pVDZ calculation and $O(n_{AO}^{1.7})$ for the B3LYP/6-31G calculation.
}
\label{fig:alkane_overall_runtime_a100}
\end{figure}

\subsection{DFT Calculations}

We present the run time per SCF iteration for hybrid DFT (B3LYP \cite{b3lyp_original, b3lyp_parameter}) calculations on branched alkane molecules with the 6-31G\cite{6_31g_h, 6_31g}, cc-pVDZ \cite{cc_pvxz} and cc-pVTZ \cite{cc_pvxz} basis sets in Fig. \ref{fig:alkane_overall_runtime_a100}, to demonstrate the overall performance of our ERI and exchange-correlation integral implementation. All TeraChem calculations reported in this section use a mixed precision scheme for computing ERIs, unless specified otherwise. In this scheme, integrals with a density weighted Schwartz upper bound above a threshold (here, $10^{-5}$, which is the default value for TeraChem) are evaluated in double precision while the rest are evaluated in single precision\cite{terachem_dynamic_precision}. It is also possible to dynamically update the threshold during SCF\cite{terachem_dynamic_precision} but this `dynamic precision' scheme is not utilized for the results reported in the present work. The largest calculation is on the \ce{C190H382} branched alkane with the cc-pVTZ basis (12380 basis functions, in total), for which a single B3LYP SCF iteration takes 851 s on average with our TeraChem implementation on a single NVIDIA A100 GPU. 
The run time scales nearly quadratically with the number of atoms, indicating that the asymptotic cubic scaling regime (arising from matrix diagonalizations) has not been attained for the tested systems. The cc-pVTZ calculations require roughly an order of magnitude (10-18$\times$) more time per SCF iteration than cc-pVDZ, while the number of basis functions grows by a factor of 2.6. We note that unless mentioned otherwise, we utilized the ``optimized-segmented" version of the cc-pVDZ and cc-pVTZ basis sets throughout this work, which span the same single-particle space as the original formulation \cite{cc_pvxz} while being far more computationally efficient due to removal of redundant primitive Gaussians \cite{hashimoto_ccpvxz_optseg, davidson_ccpvxz_optseg}. Other software packages like Gaussian \cite{gaussian16} and Q-Chem \cite{qchem_5_or_higher} use these optimized-segmented formulations as default when Dunning basis sets are requested. We also do not include the timings for the first SCF iteration in the average SCF iteration times reported in Fig. \ref{fig:alkane_overall_runtime_a100} or subsequent analysis, as the local character of the guess density matrix obtained from the superposition of atomic densities \cite{sadguess} leads to an atypically low computation time for this initial step. Our output files and the associated basis set files are provided in a Zenodo repository,\cite{wang_2024_13328235} for further perusal by interested readers. 

\begin{figure}[htb!]
    \centering
    \includegraphics[width=0.45\linewidth]{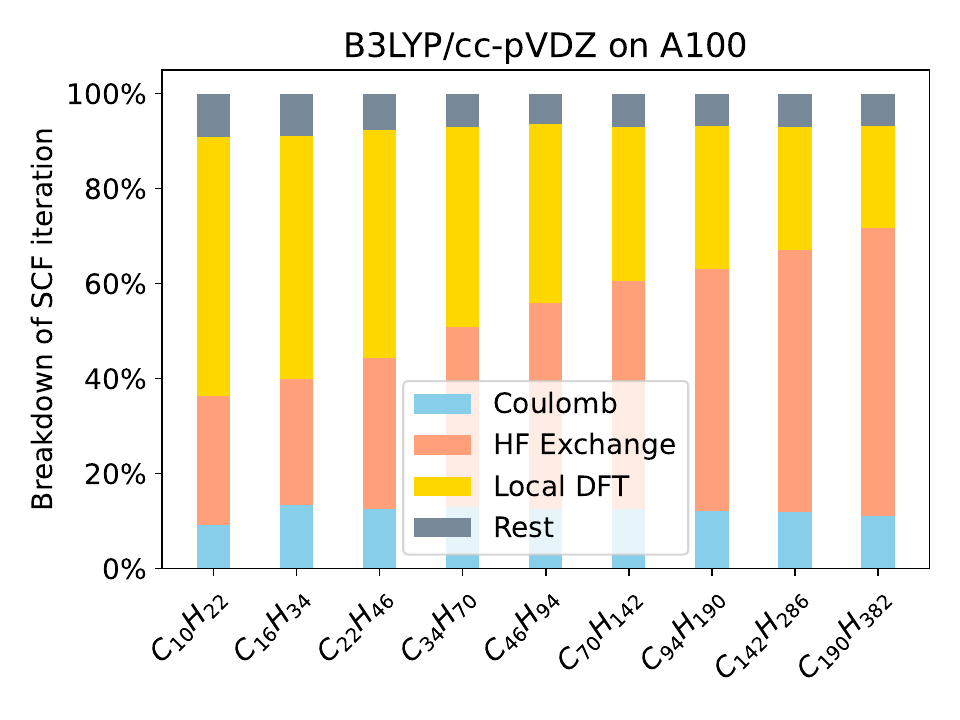}
    \includegraphics[width=0.45\linewidth]{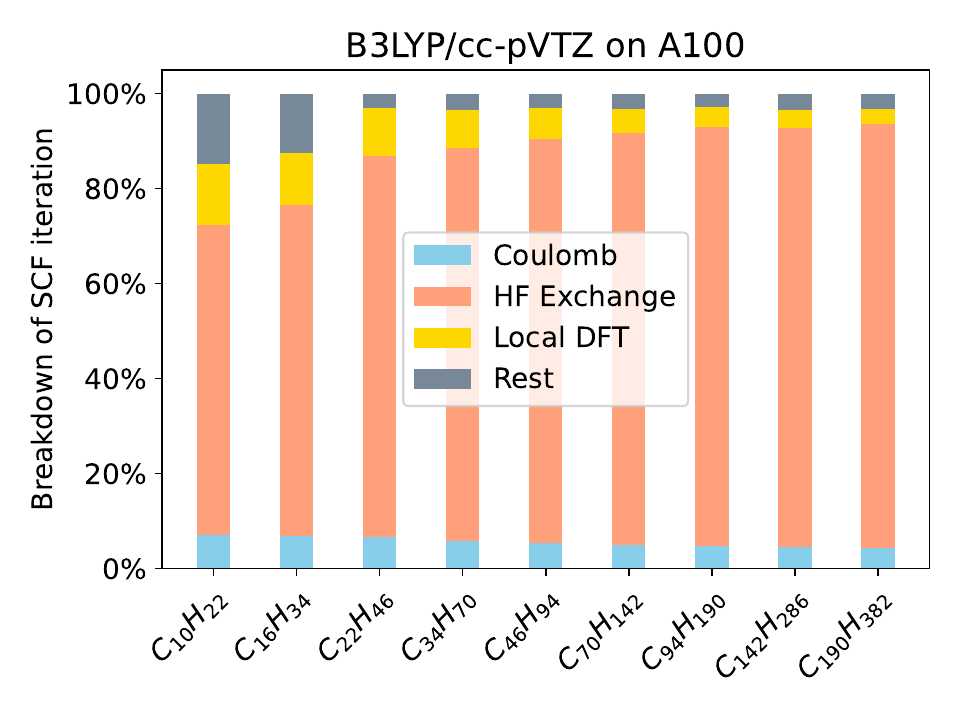}
\caption{Distribution of SCF iteration run time components (averaged over all cycles) for B3LYP/cc-pVDZ (left) and B3LYP/cc-pVTZ (right) calculations on branched alkanes, using one NVIDIA A100 GPU. The ``rest" part includes the matrix operations, such as the eigenvalue decomposition of Fock matrix.
}
\label{fig:alkane_runtime_component_a100}
\end{figure}

In Fig. \ref{fig:alkane_runtime_component_a100} we show the component distribution of every SCF iteration for the branched alkane calculations with the cc-pVDZ and cc-pVTZ basis sets. As the number of basis functions increases, the HF exchange calculation increasingly dominates the run time, while the relative contribution of local DFT exchange correlation integral evaluation decreases significantly. The proportion of the total run time required for HF exchange evaluation is much larger for the cc-pVTZ basis than cc-pVDZ, as expected. 

\begin{table}[htb!]
\begin{tabular}{c | c | r| r | r} 
 \hline
 Molecule & Basis Set & $n_{AO}$ & TeraChem (s) & BrianQC (s) \\
 \hline
 \multirow{2}{8em}{\ce{C46H94}} & cc-pVDZ & 1160 & 3 & 12 \\
 & cc-pVTZ & 3020 & 44 & 153 \\
 \hline
\multirow{2}{8em}{\ce{C70H142}} & cc-pVDZ & 1760 & 6 & 23 \\
 & cc-pVTZ & 4580 & 104 & 424 \\
 \hline
 \multirow{2}{8em}{\ce{C94H190}} & cc-pVDZ & 2360 & 11 & 41 \\
 & cc-pVTZ & 6140 & 203 & 915\\
 \hline
 \multirow{2}{8em}{fullerene (\ce{C60})} & cc-pVDZ & 900 & 2 & 4 \\
 & cc-pVTZ & 2100 & 45 & 88 \\
 \hline
 \multirow{2}{10em}{taxol (\ce{C47H51NO14})} & cc-pVDZ & 1185 & 2 & 6 \\
 & cc-pVTZ & 2935 & 35 & 98 \\
 \hline
 \multirow{2}{12em}{valinomycin (\ce{C54H90N6O18})} & cc-pVDZ & 1620 & 4 & 17 \\
 & cc-pVTZ & 4080 & 65 & 193 \\
 \hline
\end{tabular}
\caption{Run time per SCF iteration (averaged over all cycles) for B3LYP/cc-pVDZ and B3LYP/cc-pVTZ calculations on several organic molecules, compared between the present TeraChem implementation and BrianQC \cite{brianqc}. A single NVIDIA A100 GPU was used for all calculations. For BrianQC we only report the sum of the Coulomb, HF exchange, and exchange-correlation computation times, which the program reports rounded to the nearest second. The matrix linear algebra time per iteration, and other miscellaneous components is therefore not included in the BrianQC timing.}
\label{tab:comparison_to_brianqc}
\end{table}

\begin{table}[htb!]
\begin{tabular}{c | c | r| r | r} 
 \hline
 Molecule & Basis Set & $n_{AO}$ & TeraChem (s) & BrianQC (s) \\
 \hline
 \multirow{2}{18em}{Tetra-aza Co(II) complex\cite{tetraaza_complex} +\ce{CO2} \ce{[CoC16H22N4O2]+}} & def2-SVP & 476 & 1 & 2 \\
 & def2-TZVP & 976 & 6 & 12 \\
 \hline
 \multirow{2}{18em}{$\mu$-alkyl dicopper(I) complex\cite{dicopper_complex} \ce{[Cu2C35H31N6F2]+}} & def2-SVP & 872 & 2 & 6 \\
 & def2-TZVP & 1838 & 27 & 50 \\
 \hline
 \multirow{2}{18em}{MOF-5\cite{mof5_original}  Cluster model\cite{mof5_geometry}(0.5 pore) \ce{Zn8C28H34O26}} & def2-SVP & 1268 & 3 & 8 \\
 & def2-TZVP & 2588 & 24 & 68 \\
 \hline
  \multirow{2}{18em}{MOF-5\cite{mof5_original}  Cluster model\cite{mof5_geometry} (1 pore) \ce{Zn16C64H64O52}} & def2-SVP & 2636 & 9 & 35 \\
 & def2-TZVP & 5440 & 96 & 309 \\
 \hline
\end{tabular}
\caption{Run time per SCF iteration (averaged over all cycles) for B3LYP/def2-SVP\cite{def2_svp} and B3LYP/def2-TZVP\cite{def2_svp} calculations on several metal containing species, compared between the present TeraChem implementation and BrianQC \cite{brianqc}. A single NVIDIA A100 GPU was used for all calculations. For BrianQC we only report the sum of the Coulomb, HF exchange, and exchange-correlation computation times, which the program reports rounded to the nearest second. The matrix linear algebra time per iteration, and other miscellaneous components is therefore not included in the BrianQC timing.}
\label{tab:comparison_to_brianqc_metal}
\end{table}

We compare our results with the BrianQC backend (version 1.3.0) \cite{brianqc} for Q-Chem (version 6.0.1) \cite{qchem_5_or_higher}, which is another GPU-accelerated software that supports high angular momentum basis functions and evaluates ERIs with a scheme using a mixture of single and double precision. Table \ref{tab:comparison_to_brianqc} shows time per B3LYP SCF iteration on a single NVIDIA A100 GPU for some representative organic molecules\footnote{The alkanes in Table \ref{tab:comparison_to_brianqc} are a subset of the branched alkanes shown in Fig. \ref{fig:alkane_overall_runtime_a100}, while fullerene, taxol and valinomycin have been previously utilized as benchmarks for GPU accelerated quantum chemistry codes\cite{terachem_gpu_3, terachem_2013, brianqc, quick}. We take this opportunity to note that the species referred to as `taxol' in in the aforementioned works (as well as potentially elsewhere) is actually a derivative structure with the formula \ce{C45H49NO15}, as opposed to the actual taxol structure\cite{nicolaou_taxol}, which has the formula of \ce{C47H51NO14}. In this work we only report results for the proper structure of taxol, but provide output files for the derivative structure in the SI for potential comparison with previous studies.} with the cc-pVDZ and cc-pVTZ basis sets, while Table \ref{tab:comparison_to_brianqc_metal} shows the same for chemically relevant metal containing species with the def2-SVP and def2-TZVP basis sets \cite{def2_svp}. Additional results for these species on modern gaming GPUs like NVIDIA RTX 3090Ti and 4090 are included in the supporting information. 

It is apparent that for sufficiently large systems, our TeraChem implementation provides an overall faster performance (by roughly a factor of $2-4$) than BrianQC, even without utilizing TeraChem's dynamic precision and dynamic grid algorithms for acceleration. BrianQC and many other integral packages \cite{libintx_up_to_iiii, gamess_f} extensively focus on finding GPU-accelerated algorithms for computing individual ERI elements. TeraChem however also focuses heavily on the design of GPU-accelerated algorithms crafted for $\mathbf{J}$ and $\mathbf{K}$ matrix construction for SCF calculations, where each kernel function performs more operations than just elementary ERI evaluations. This permits better optimization of the SCF performance as a whole. We hypothesize that BrianQC's automated engine for selecting the best recurrence relations for each angular momentum combination of ERI and each type of GPU, in combination with our kernel splitting scheme and $\mathbf{J}$ and $\mathbf{K}$ build algorithms, may potentially lead to a more efficient electronic structure package.

\begin{figure}[htb!]
    \centering
    \includegraphics[width=0.48\linewidth]{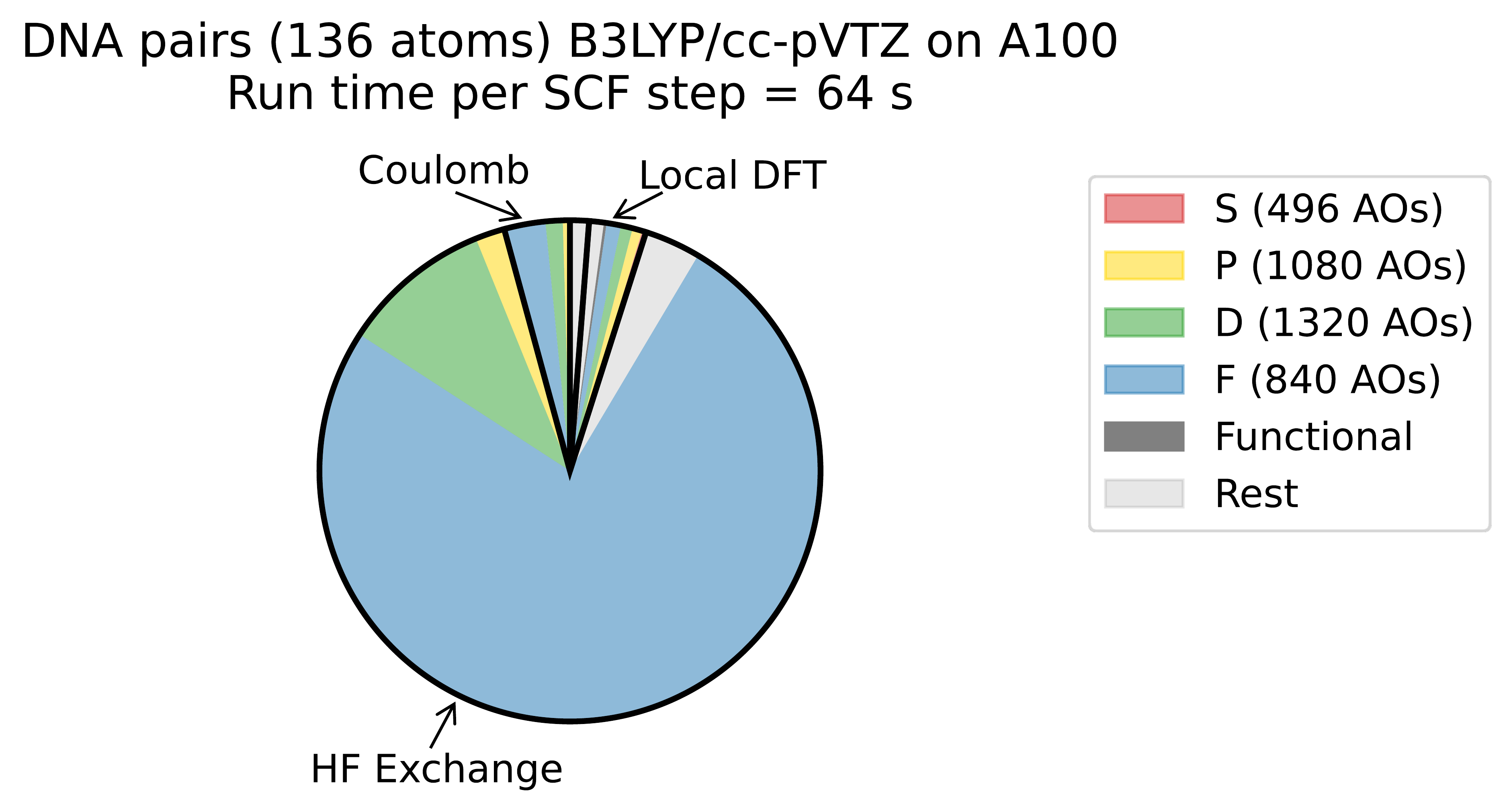}
    \includegraphics[width=0.48\linewidth]{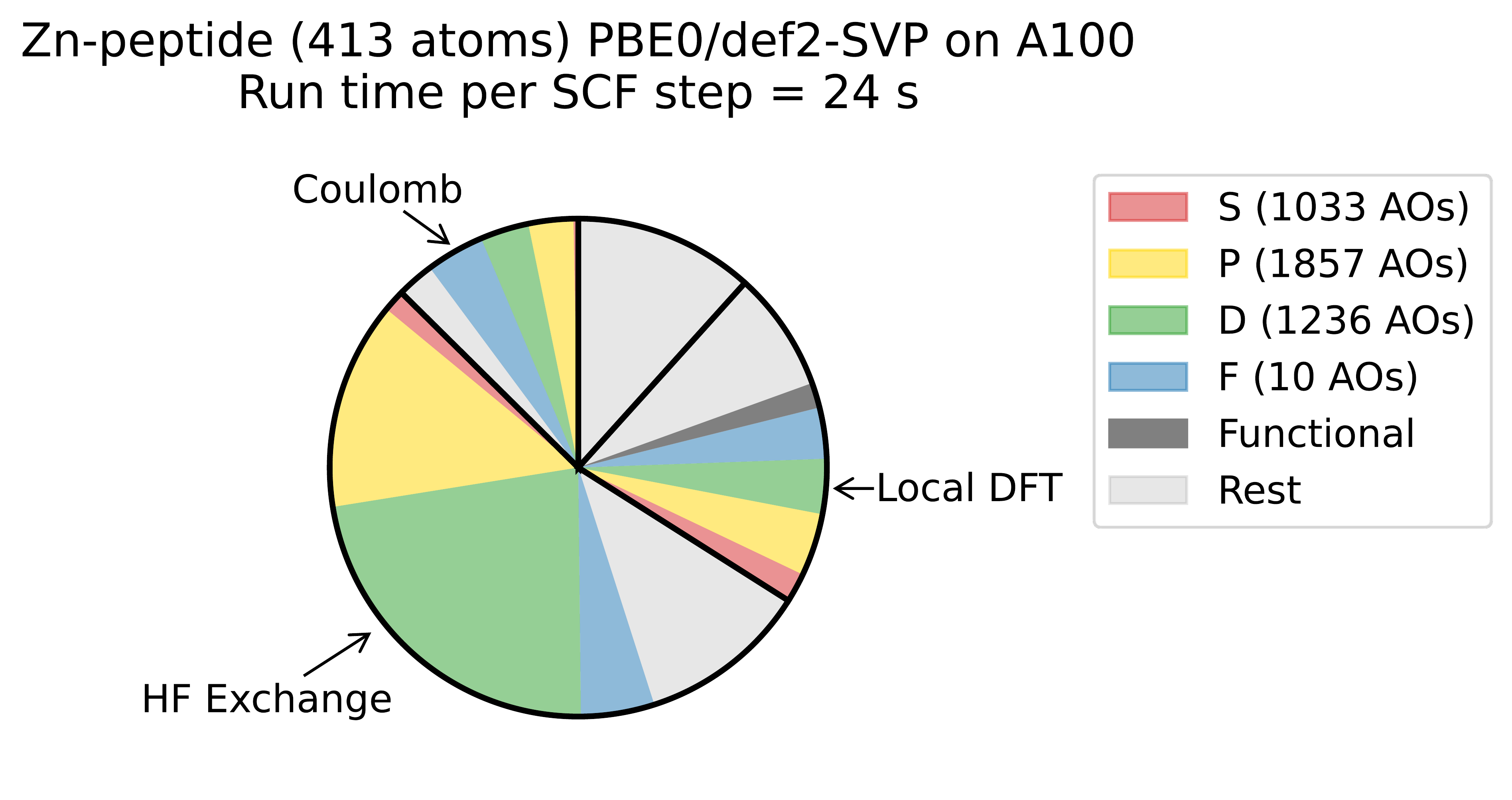}
\caption{Kernel run time distributions across SCF iterations for model organic (two DNA CG base pairs, B3LYP/cc-pVTZ) and bioinorganic systems (protein 6UFA, \cite{protein_6ufa} PBE0\cite{pbe0}/def2-SVP\cite{def2_svp}) on one NVIDIA A100 GPU. We split the run time first into Coulomb, HF exchange and local DFT (including both the density and exchange-correlation matrix element construction) parts, separated by black lines, then by the angular momentum, distinguished by color. The run time for a given angular momentum $L$ is defined as the total run time across all GPU kernels where $\mathrm{max}(L_\mu, L_\nu, L_\lambda, L_\sigma) = L$.
The ``functional" contribution (shown in dark gray) arises from the evaluation of exchange-correlation term on each grid point, which is independent of angular momentum. The remaining time (``rest", shown in light gray) arises from data pre- and post-processing on CPU and linear algebra.}
\label{fig:angular_component_a100}
\end{figure}

It is also worthwhile to consider the distribution of compute times over kernels of different angular momentum. In Fig. \ref{fig:angular_component_a100}, we show the kernel run time distributions for both an organic system (two cytosine-guanine base pairs from DNA) with B3LYP/cc-pVTZ basis and a zinc-containing peptide with PBE0/def2-SVP.
The model organic system has a large fraction of $f$ type atomic orbitals (22\%), arising from the one set of $f$ orbitals per second-period atom like C, N, and O. The model bioinorganic system, on the other hand, only has one shell of $f$ type orbitals from the single Zn atom, resulting in only 0.2\% of the total number of AOs being of the $f$ type. However, we find that $f$ orbital kernels require a considerably larger proportion of compute time in both cases, with the $f$ type $\mathbf{K}$ evaluation kernels being responsible for almost two thirds (66\%) of the SCF iteration run time for the organic system. As a consequence, it is important to optimize the performance of all integral kernels of all angular momentum combinations, regardless of the basis set composition. A more detailed runtime breakdown of individual $\mathbf{K}$ matrix kernels classified by angular momenta of all orbitals (i.e. $(ff|ff),(df|ff)$ etc.) is provided in the supporting information, as well as additional profiling results for more systems and GPU types. We note that the most computationally demanding type of kernel is system and hardware dependent, but no individual  kernel contributes more than 6\% of the $\mathbf{K}$ matrix construction time for the systems studied.

We now turn our focus to individual kernels, directing most of our attention to the nuclear attraction integrals as they are the most compute-bound kernels and there are no cutoffs inside kernels to affect compute time. Given a fixed algorithm and data structure, the only available degrees of freedom lie in the arrangement of equations, or more precisely, the number of intermediate variables we generate for each equation. As mentioned in the implementation section, we first obtained fully expanded equations for $R_{t_x t_y t_z}^0$ (as a function of $\vec{P} - \vec{C}$ and $R_{000}^m$) and $V_{\mu\nu C}$ (as a function of $\frac{1}{2p}$, $\vec{P} - \vec{A}$, $\vec{P} - \vec{B}$, $R_{t_x t_y t_z}^0$ and an accumulated coefficient $C = q_C C_\mu C_\nu e^{-\frac{ab}{a+b}|\vec{A}-\vec{B}|^2}$), and then performed CSE to obtain the maximum number of intermediate variables. We then reduce the number of intermediate variables by selecting the ones with simplest expression (as measured by the number of variables on the right hand side of the equation), and merge them into other intermediate variables or final expressions. This way we are able to systematically re-formulate an equation into a set of equations with different numbers of intermediate variables. We then measure the performance of each kernel, with exactly the same functionality but different number of intermediates and thus different compute unit/register/latency balance.

\begin{figure}[htb!]
    \centering
    \includegraphics[width=0.45\linewidth] {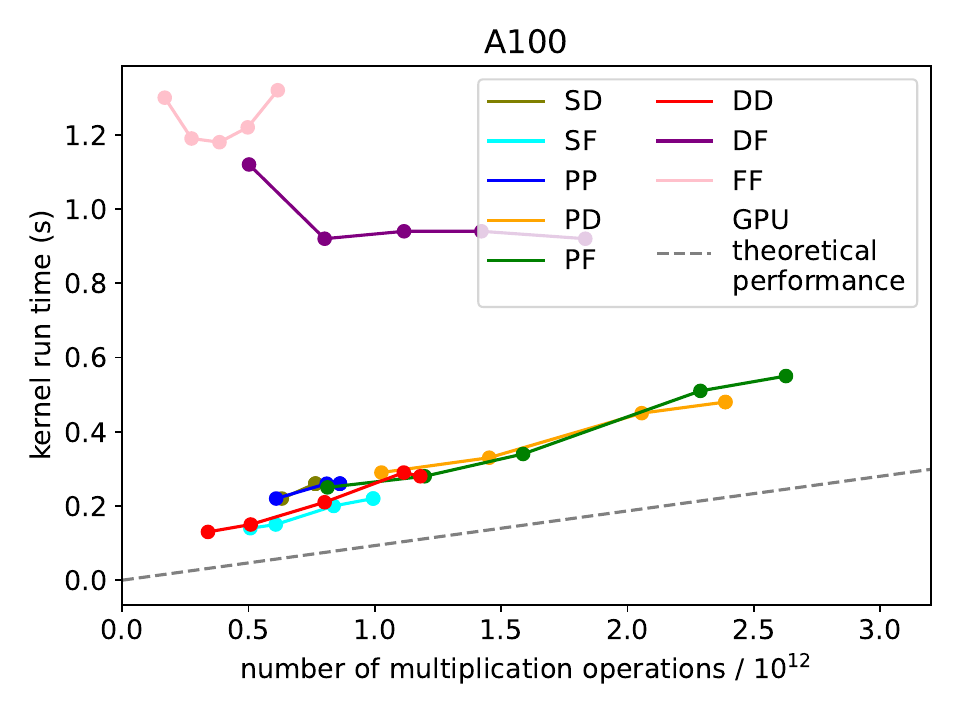}
    \includegraphics[width=0.45\linewidth] {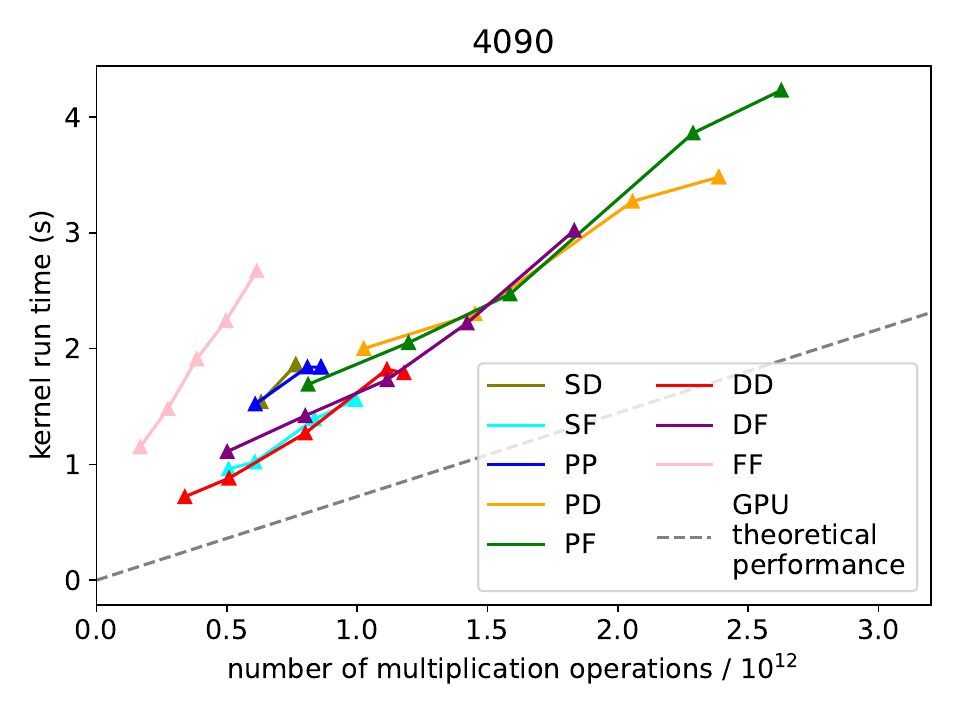}
\caption{The nuclear attraction integral evaluation time for PYP QM region 5, surrounded by a large number of point charges (the rest of the protein and solvent water molecules, which are modeled with the Amber protein force field \cite{amber_protein_force_field, amber_2023}). The original \cite{cc_pvxz} (i.e. not optimized-segemented) version of the cc-pVTZ basis was utilized, in order to maximize the number of floating point operations. Full double precision is used in nuclear attraction integral computation. The computation time is split into angular momentum pairs. Different points on the same curve represent different numbers of intermediates in the generated equations. We presented the number of multiplication operations in all intermediate and final equations on the x-axis. The number of arithmetic operations required by a set of equations decreases with increase in the number of intermediate variables. The ``GPU theoretical performance" line represents the minimal computation time possible for a given number of floating point operations for the corresponding GPU. \cite{gpu_database}
}
\label{fig:runtime_vs_nvariables}
\end{figure}

As shown in Fig. \ref{fig:runtime_vs_nvariables}, in most nuclear attraction integral kernels, the run time increases with the number of total multiplication operations, and thus decreases with the number of intermediate variables, because intermediate variables can reduce the amount of repeated computation. So, for compute-bound kernels, it is always a good idea to perform a full CSE to the integral equations. However, high angular momentum kernels like the DF and FF kernels have too many intermediates, leading to the GPU kernel using up all registers and the runtime is no longer compute-bound. \cite{nsight_compute} In such cases, fewer intermediates and more repeated computation is more time efficient, as made evident by the DF and FF curves for the NVIDIA A100 GPU in Fig. \ref{fig:runtime_vs_nvariables}. Unfortunately the crossover between the two regimes depends on the device features and compiler optimization, and is not easily predicted or rationalized. For example, we observe a shallow local minimum in the FF and DF kernel run time curves on A100, while 4090 performance is improved significantly and monotonically with reduction of floating point operations. Optimal performance therefore requires an optimization task to obtain the best number of intermediates for each type of integral kernel and each GPU model. We consequently elect to do full CSE in our present implementation for the nuclear attraction integrals, in order to minimize performance degradation on currently available GPU models.

For memory-bound kernels, such as all of the HF exchange matrix kernels, the room for optimization by manipulating the equations is small, and saving no intermediates other than the $R_{t_x t_y t_z}^0$ terms yields the best performance in most cases. The arithmetic unit utilization in these kernels is low in general (for example close to 0.1 FLOP per byte for the double precision $(sf|ff)$ kernel and 1 FLOP per byte for the single precision $(sf|ff)$ kernel, and below 0.01\% of the optimal FLOP per second for both cases, as obtained from NVIDIA Nsight Compute \cite{nsight_compute} profiling results). We thus believe repeated re-computation of terms can improve performance if it reduces global memory access. We note that the computational bottleneck for $\mathbf{K}$ matrix computation might be different for more complicated recurrence relations for auxiliary integrals, such as the Obara-Saika \cite{gamess_f, head_gordon_pople_original} or Rys quadrature \cite{pyscf_gpu, rys_quadrature} algorithms.

There are some additional points to consider about the $\mathbf{K}$ matrix GPU algorithm. As mentioned in the implementation section, we choose to sort the $\mu\nu$ pairs with $\mu$ as the primary index, which results in minimized memory access for $\mathbf{K}$ matrix output and regular amount of random memory access for density matrix input. However there exists another choice: we can sort the $\mu\nu$ pairs with $\nu$ as primary index, and the algorithm will result in a minimized memory access for density matrix input and regular amount of random memory access for $\mathbf{K}$ matrix output. The downside is that we need to perform atomic addition for $\mathbf{K}$ matrix output, which is considerably more expensive than the automatically supported atomic read operation. Not surprisingly, the overall performance of the ``inverted" algorithm is worse by about 29\% on a PYP QM region 5 system (HF/cc-pVDZ). However for some particular kernels (for example $(sp|sd)$ and $(sd|sd)$) the compute time is reduced by more than 50\%, compared to the regular algorithm. We thus believe that the ``inverted" algorithm for $\mathbf{K}$ is more efficient if the angular momenta for $\mu$ and $\lambda$ are small, due to the relatively fewer number of atomic additions being needed for such kernels. This angular momentum dependent behavior of $\mathbf{K}$ kernels  will be investigated further in future, for the purpose of developing efficient algorithms that can switch between the ``regular" and ``inverted" forms.

We also note that a scheme that takes advantage of basis function contraction may offer some advantages for HF exchange matrix computation over our fully uncontracted approach, since the former reduces the count of random memory access for density matrix fetching.  An analysis of performance with respect to basis function contraction level will also be explored in future work.

\begin{figure}[htb!]
    \centering
    \begin{minipage}{0.48\textwidth}
        \includegraphics[width=\linewidth]{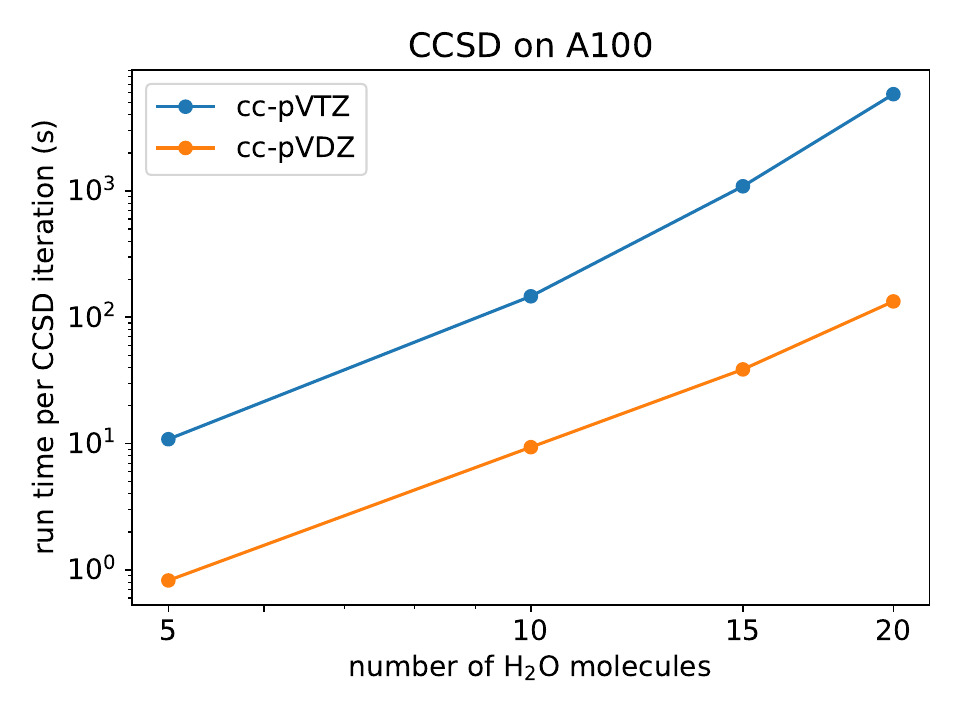}
    \end{minipage}
        \begin{minipage}{0.48\textwidth}
        \includegraphics[width=\linewidth]{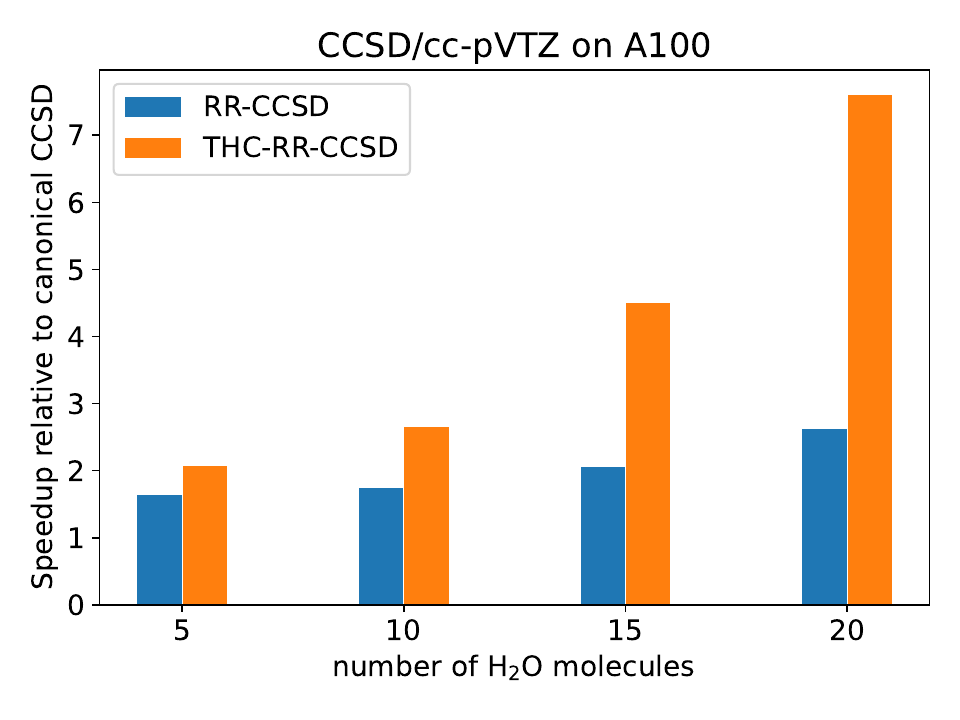}
    \end{minipage}
\caption{Average iteration timings for canonical CCSD with the cc-pVDZ and cc-pVTZ basis sets for the \ce{(H2O)5},\ce{(H2O)10},\ce{(H2O)15} and \ce{(H2O)20} water clusters (left) and speedups compared to canonical CCSD with RR-CCSD, and THC-RR-CCSD for these water clusters with the cc-pVTZ basis (right). All calculations were performed on a single A100 GPU. The CC methods in TeraChem rely on Cholesky decomposition \cite{beebe_simplifications_1977}  of the two-electron integrals, for which a threshold of $10^{-4}$ hartree was used. A rank reduction threshold of $\epsilon=10^{-5}$ was utilized for these calculations,  which refers to the threshold imposed on eigenvalues of the MP2 amplitudes used in construction of the low-rank projector in RR-CCSD and THC-RR-CCSD. Geometries were taken from Ref \citenum{terachem_ccsd_gpu}.}
\label{fig:CCSD}
\end{figure}

\subsection{Coupled Cluster Singles and Doubles Calculations}

The newly supported integrals over $f$ type orbitals also enable the use of polarized triple zeta basis sets with the GPU-accelerated CC methods already implemented in TeraChem \cite{terachem_ccsd_gpu}. This is particularly useful, as CC methods converge relatively slowly to the complete basis set limit with increasing basis size \cite{kutzelnigg1992rates, helgaker1997basis} as compared to HF/DFT \cite{jensen2005estimating}. The computational cost of the formally $O(N^6)$ scaling canonical CCSD method is dominated by the storage and manipulation of the rank-four double excitations amplitude tensor $t_{ij}^{ab}$ (where $a,b$ are unoccupied orbitals and $i,j$ are occupied). This has led to efforts to develop more computationally efficient formulations that exploit the sparsity of the $t_{ij}^{ab}$ tensor \cite{hampel1996local, kinoshita_singular_2003, subotnik2006near, neese_accurate_2009, hansen_efficient_2011, schutski_tensor-structured_2017}. GPU accelerated algorithms for rank reduced \cite{terachem_rank_reduce_cc_1, terachem_rank_reduce_cc_2} (RR-) CCSD \cite{terachem_rank_reduce_ccsd} and tensor hypercontraction \cite{hohenstein_thc_I, parrish_thc_II, hohenstein_thc_III} (THC-) based RR-CCSD \cite{terachem_rank_reduce_cc_3} have previously been implemented in TeraChem, with the latter having a formal $O(N^4)$ scaling. Their performance can now be assessed for $f$ orbital containing basis sets. 

The left panel of Fig. \ref{fig:CCSD} shows timings for a single canonical CCSD iteration for four water clusters of increasing size (\ce{(H2O)5},\ce{(H2O)10},\ce{(H2O)15} and \ce{(H2O)20}) with the cc-pVDZ ($f$ function free) and cc-pVTZ (containing one set of $f$ functions per O atom) basis sets. We find that the largest canonical CCSD calculation (\ce{(H2O)20}/cc-pVTZ) involves 1300 basis functions and requires 5830 s per CCSD iteration (on average) on a single NVIDIA A100 GPU. RR-CCSD and THC-RR-CCSD significantly reduce the run time, requiring 2220 s and 768 s on average per iteration for \ce{(H2O)20}/cc-pVTZ respectively. In general, RR-CCSD and THC-RR-CCSD lead to significant speedups relative to canonical CCSD with the cc-pVTZ basis (shown in the right panel of Fig. \ref{fig:CCSD}), while more modest acceleration is observed with the cc-pVDZ basis set ($\sim 2$ or less, as shown in the supporting information). Overall, it is quite feasible to carry out CCSD calculations with over a thousand basis functions with TeraChem on a single modern GPU at present, with low rank approximations and multi-GPU parallelization extending the domain of application even further, as will be explored in a future work.

\section{Applications}

\begin{figure}[htb!]
    \centering
    \includegraphics[width=\linewidth]{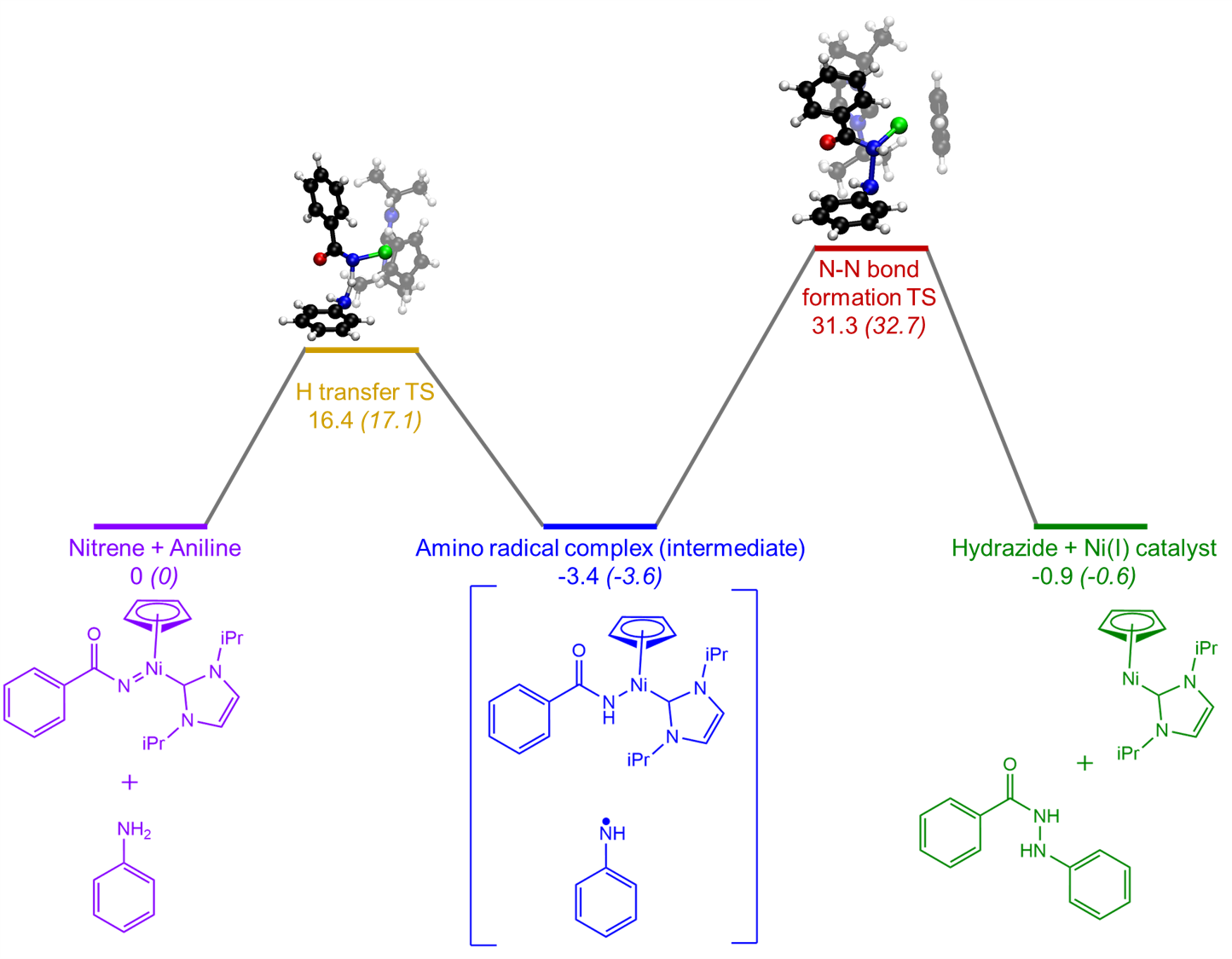}
\caption{Computed mechanism for the coupling of aniline with nitrene to form hydrazide with gas phase $\omega$B97X-D3(BJ)/pcseg-1. The reaction is found to occur via H transfer followed by N-N bond formation through radical substitution. The chemical structures for the minimum energy configurations and the computed transition state (TS) geometries connecting them are depicted (the additional ligands on Ni were made semi-transparent in the TS structures for visual clarity, full xyz files are provided in the supporting information). Free energies in kcal/mol (at 300 K temperature), relative to the reactant, are also shown \textit{(the italicized values in parentheses correspond to calculations where the f type basis functions on Ni in pcseg-1 were removed)}. The nuclear contributions to the free energy were approximated with the rigid rotor and harmonic oscillator approximations.
}
\label{fig:Stolz_step}
\end{figure}

\subsection{Nickel catalyzed N-N coupling}

We took advantage of the newly added $f$ orbital support in TeraChem to computationally explore a recently reported route for Ni(I) catalyzed hydrazide synthesis by coupling an amine to a hydroxamate \cite{nickel_catalyst_experimental}. A mechanism was proposed in the experimental work \cite{nickel_catalyst_experimental}, and here we explore the feasibility of the key step of N-N bond formation between the amine and a nitrene intermediate complexed to the active Ni(I) species (shown on the left in Fig. \ref{fig:Stolz_step}). Our calculations are in the gas phase, and utilize the D3(BJ) \cite{dispersion_d3, dispersion_d3BJ} based version of the $\omega$B97X-V functional \cite{wb97_v} (henceforth referred to as $\omega$B97X-D3(BJ) \cite{wb97x_d3bj}) and the pcseg-1 basis set \cite{pcseg_1}. Optimization of the proposed nitrene intermediate and aniline (a representative amine for this reaction) indicated the formation of a noncovalently bonded complex between them, which we henceforth refer to as the `reactant'. A similar, noncovalently bonded complex between the hydrazide product and the proposed Ni(I) catalyst was also found and will be referred to as the `product'. We note that no structure with direct coordination of the product hydrazide to the Ni center could be optimized, suggesting that the hydrazide may directly dissociate from the Ni complex upon N-N bond formation, without the formation of intermediate D in Ref \citenum{nickel_catalyst_experimental}.

The nature of the pathway between the reactant and product complexes is more interesting as both N-N bond formation and H transfer occur between the amine and the nitrene. Nudged elastic band (NEB) \cite{neb_original, dlfind, chemshell_1, chemshell_2} calculations between the two structures indicate that this is not a concerted process, but rather the amine first donates a H atom to the nitrene to become an amino radical, which subsequently undergoes a radical substitution reaction at former nitrene site to form the hydrazide and release the active form of the Ni(I) catalyst. Optimization of the transition state for the first step (H transfer) yields a free energy barrier of 16.4 kcal/mol with the $\omega$B97X-D3(BJ)/pcseg-1 model chemistry at 300 K (nuclear contributions to the free energy were modeled with the rigid rotor and harmonic oscillator approximations). The second step of N-N bond formation via radical substitution however appears to have an unfeasibly high barrier of 34.8 kcal/mol, which should not lead to appreciable product formation at room temperature on the timescale of a few hours. We notice that the transition state found by $\omega$B97X-D3(BJ)/pcseg-1 for the second step is heavily spin contaminated (having an $\langle S^2 \rangle = 1.26$ vs the ideal value of 0.75 for a doublet). This is likely a consequence of the transfer of the spin localization site from the N atom on the amino radical to the Ni atom in the final product. The large level of spin contamination suggests nontrivial multireference character for this structure (which resembles intermediate D in Ref \citenum{nickel_catalyst_experimental}) and invites use of complete active space (CAS) methods (which we did not attempt here).
We refrained from further investigation as the focus of this work is the $f$ orbital implementation rather than this specific application, but we believe that a careful study of this reaction would be interesting from a computational chemistry perspective.  We also repeated the calculations (optimizations and frequency calculations) without the $f$ type orbital on Ni in the pcseg-1 basis, and find that the relative electronic energies are largely unaffected, but the transition states are destabilized by $\sim$ 1 kcal/mol in free energy from the nuclear contributions (as shown in Fig. \ref{fig:Stolz_step} and through the individual components reported in the SI). 

\begin{figure}[htb!]
    \centering
    \begin{minipage}{0.48\textwidth}
        \includegraphics[width=\linewidth]{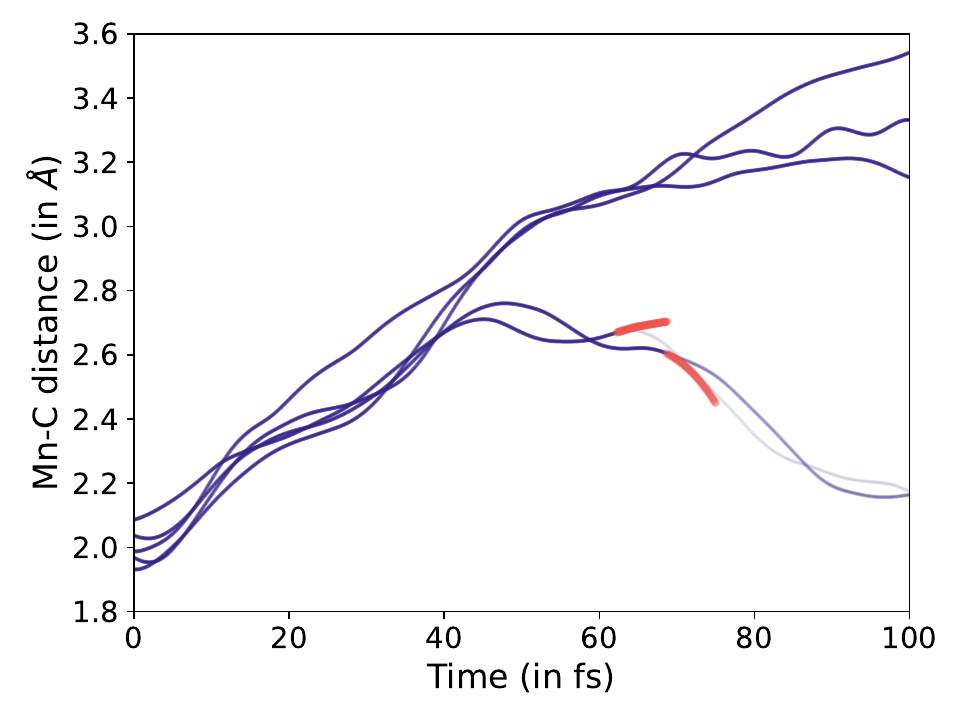}
    \end{minipage}
        \begin{minipage}{0.48\textwidth}
        \includegraphics[width=\linewidth]{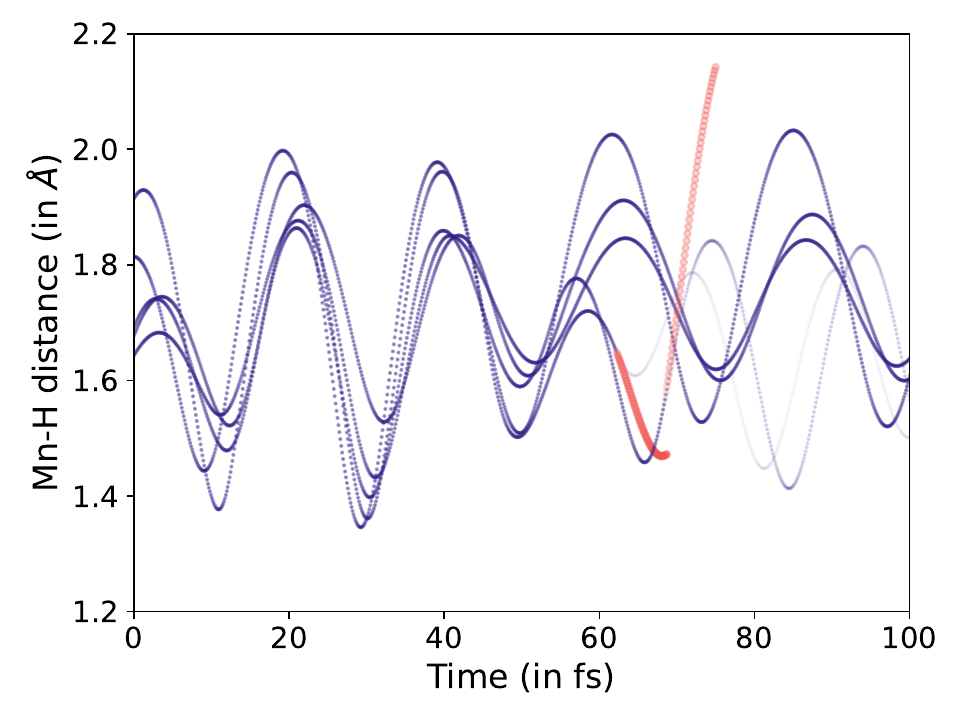}
    \end{minipage}
\caption{Evolution of Mn-C (left) and Mn-H (right) bond distances over time, from the CASSCF/AIMS simulations. Trajectories on the excited state are shown in blue and ground state trajectories are shown in red. The ground state trajectories are terminated after the energy gap with the first excited state becomes larger than 1 eV, as subsequent nonadiabatic behavior is quite unlikely.
}
\label{fig:MnHCH3}
\end{figure}

\subsection{Nonadiabatic dynamics of \ce{MnH(CH3)}}

TeraChem's ability to perform GPU-accelerated complete active space configuration interaction (CASCI) \cite{terachem_fomo_casci} and complete active space self-consistent field (CASSCF) \cite{terachem_casscf} calculations enables nonadiabatic molecular dynamics simulations. Indeed, TeraChem has been successfully employed in conjunction with the ab initio multiple spawning (AIMS) \cite{aims_1, aims_2, aims_3, aims_4, aims_5, aims_casscf, aims_fomo_casci, aims_review} method to explore the photophysics and photochemistry of many main group systems. \cite{aims_application_polariton, aims_application_gfp, aims_application_ammonia_ued, aims_application_electrocyclic} The new support for $f$ type orbitals enables the exploration of AIMS dynamics of transition metal containing species. As an example demonstration, we have simulated the ultrafast photorelaxation dynamics of the model organometallic MnH(CH$_3$) species \cite{billups1980activation, cho2013infrared} following photoexcitation from the ground (sextet) state to the first excited sextet state. We utilized CASSCF \cite{casscf_original_1, casscf_original_2, casscf_original_3, casscf_original_4} with the def2-SVP basis, utilizing an active space of 9 electrons (7 up spins and 2 down spins) in 7 orbitals (the two $\sigma$ bonding levels and the 5d orbitals), averaging over the lowest six sextet states with equal weight \cite{state_averaged_casscf}. At this level of theory, the Franck-Condon geometry excitation energy was computed to be 2.5 eV, suggesting that the first excited state can be accessed with visible light. We chose five initial conditions sampled from the harmonic Wigner distribution obtained from optimization and frequency calculations with $\omega$B97X-D3 \cite{wb97xd3} /def2-SVP. Significant nuclear dynamics were observed for all five cases, with Fig. \ref{fig:MnHCH3} showing that 3 sets of initial conditions lead to methyl group dissociation within 100 fs while remaining on the first excited state without internal conversion to the ground state. The other two sets of initial conditions do not appear to show dissociation of the methyl group on this timescale, but instead transfer 92\% and 65\% of their electronic state population back to the ground state. All five sets of initial conditions exhibit large amplitude oscillations in Mn-H bond distance, but no excited-state bond dissociation is observed within 100 fs. These significant ultrafast structural changes following the first dipole-allowed excitation indicate that this molecule could be an interesting organometallic model complex for time-resolved experiments like ultrafast electron diffraction or transient X-ray absorption, to better understand photocleavage of Mn-C bonds.

\section{Conclusion}

We have added support for $f$ type Gaussian basis functions into the GPU-accelerated TeraChem software package in order to reduce basis set incompleteness error for calculations in main group systems, and explore light transition metal chemistry with polarized bases. Our data structures and optimized GPU kernels for computing integrals with the McMurchie-Davidson approach lead to quite improved performance, being a factor of $\sim 3\times$ faster than the GPU-based BrianQC module of the Q-Chem software package for several large organic and transition metal containing systems, on a NVIDIA A100 card. A considerable proportion of the computational effort is nonetheless spent on evaluating integrals involving $f$ functions even when they make up a relatively small proportion of the overall bases. It would therefore be worthwhile to investigate further algorithmic design towards more efficient evaluation of high angular momentum integrals on GPUs. Our $f$ orbital implementation also permits efficient GPU accelerated coupled cluster calculations, which we demonstrate through timings for water clusters of increasing size.   

We used the present implementation to perform preliminary DFT explorations on the mechanism of a recently reported pathway for Ni(I) catalyzed  hydrazide formation \cite{nickel_catalyst_experimental} and found that the key N-N coupling step appears to first require a H transfer from an amine to a nitrene complexed to Ni. The subsequent N-N bond formation via radical substitution appears to involve a multireference transition state and may not be suitable for DFT investigations. We also briefly examine the photochemistry of MnH(CH$_3$) subsequent to excitation to the first excited sextet state from the ground state with CASSCF and AIMS, and find that ultrafast ($<$100 fs) Mn-C bond cleavage occurs in $\sim 60\%$ of the cases, indicating that this may be an interesting model complex to study experimentally for understanding organometallic photocatalysis. We intend to leverage TeraChem's GPU acceleration to carry out more detailed investigations of more complex metal containing systems, such as metalloproteins, in the future. Work is also underway towards the inclusion of relativistic effects, which would permit GPU-accelerated simulation of lanthanide and actinide chemistry in future.

\section*{Acknowledgment}

This research was financially supported by the Office of Naval Research (N00014-21-1-2151). D.H. is a Stanford Science Fellow. O.J.F is a U.S. Department of Energy Computational Science Graduate Fellow (Grant No. DE-SC0023112). This work used computational resources of the National Energy Research Scientific Computing Center (NERSC), a U.S. Department of Energy Office of Science User Facility located at Lawrence Berkeley National Laboratory, operated under Contract No. DE-AC02-05CH11231 using NERSC award BES-ERCAP0028744.

\section*{Data Availability}
The data that supports the findings of this study are available within the article, its supplementary material and the associated Zenodo repository.\cite{wang_2024_13328235} The Zenodo repository also contains code for computation of nuclear-electron attraction integrals. 

\section*{Supporting Information}
PDF: Implementation of nuclear gradients, additional timings.

\noindent XLXS: Breakdown of $\mathbf{K}$ construction runtime over different kernels. 

\section*{Conflicts of Interest}
T.J.M. is a co-founder of PetaChem, LLC.

\bibliography{references}

\providecommand{\latin}[1]{#1}
\makeatletter
\providecommand{\doi}
  {\begingroup\let\do\@makeother\dospecials
  \catcode`\{=1 \catcode`\}=2 \doi@aux}
\providecommand{\doi@aux}[1]{\endgroup\texttt{#1}}
\makeatother
\providecommand*\mcitethebibliography{\thebibliography}
\csname @ifundefined\endcsname{endmcitethebibliography}  {\let\endmcitethebibliography\endthebibliography}{}
\begin{mcitethebibliography}{176}
\providecommand*\natexlab[1]{#1}
\providecommand*\mciteSetBstSublistMode[1]{}
\providecommand*\mciteSetBstMaxWidthForm[2]{}
\providecommand*\mciteBstWouldAddEndPuncttrue
  {\def\EndOfBibitem{\unskip.}}
\providecommand*\mciteBstWouldAddEndPunctfalse
  {\let\EndOfBibitem\relax}
\providecommand*\mciteSetBstMidEndSepPunct[3]{}
\providecommand*\mciteSetBstSublistLabelBeginEnd[3]{}
\providecommand*\EndOfBibitem{}
\mciteSetBstSublistMode{f}
\mciteSetBstMaxWidthForm{subitem}{(\alph{mcitesubitemcount})}
\mciteSetBstSublistLabelBeginEnd
  {\mcitemaxwidthsubitemform\space}
  {\relax}
  {\relax}

\bibitem[Becke(2014)]{general_dft_review}
Becke,~A.~D. Perspective: Fifty years of density-functional theory in chemical physics. \emph{The Journal of Chemical Physics} \textbf{2014}, \emph{140}\relax
\mciteBstWouldAddEndPuncttrue
\mciteSetBstMidEndSepPunct{\mcitedefaultmidpunct}
{\mcitedefaultendpunct}{\mcitedefaultseppunct}\relax
\EndOfBibitem
\bibitem[Thiel(2014)]{computational_catalysis_review}
Thiel,~W. Computational catalysis—past, present, and future. \emph{Angewandte Chemie International Edition} \textbf{2014}, \emph{33}, 8605--8613\relax
\mciteBstWouldAddEndPuncttrue
\mciteSetBstMidEndSepPunct{\mcitedefaultmidpunct}
{\mcitedefaultendpunct}{\mcitedefaultseppunct}\relax
\EndOfBibitem
\bibitem[Houk and Liu(2017)Houk, and Liu]{computational_organic_review}
Houk,~K.; Liu,~F. Holy grails for computational organic chemistry and biochemistry. \emph{Accounts of Chemical Research} \textbf{2017}, \emph{50}, 539--543\relax
\mciteBstWouldAddEndPuncttrue
\mciteSetBstMidEndSepPunct{\mcitedefaultmidpunct}
{\mcitedefaultendpunct}{\mcitedefaultseppunct}\relax
\EndOfBibitem
\bibitem[Grimme and Schreiner(2017)Grimme, and Schreiner]{computational_chemistry_review}
Grimme,~S.; Schreiner,~P.~R. Computational Chemistry: The Fate of Current Methods and Future Challenges. \emph{Angewandte Chemie (International ed. in English)} \textbf{2017}, \emph{57}, 4170--4176\relax
\mciteBstWouldAddEndPuncttrue
\mciteSetBstMidEndSepPunct{\mcitedefaultmidpunct}
{\mcitedefaultendpunct}{\mcitedefaultseppunct}\relax
\EndOfBibitem
\bibitem[Loos \latin{et~al.}(2018)Loos, Scemama, Blondel, Garniron, Caffarel, and Jacquemin]{mountaineer_excited_state}
Loos,~P.-F.; Scemama,~A.; Blondel,~A.; Garniron,~Y.; Caffarel,~M.; Jacquemin,~D. A mountaineering strategy to excited states: Highly accurate reference energies and benchmarks. \emph{J. Chem. Theory Comput.} \textbf{2018}, \emph{14}, 4360--4379\relax
\mciteBstWouldAddEndPuncttrue
\mciteSetBstMidEndSepPunct{\mcitedefaultmidpunct}
{\mcitedefaultendpunct}{\mcitedefaultseppunct}\relax
\EndOfBibitem
\bibitem[Eriksen \latin{et~al.}(2020)Eriksen, Anderson, Deustua, Ghanem, Hait, Hoffmann, Lee, Levine, Magoulas, Shen, \latin{et~al.} others]{benzene_ground}
Eriksen,~J.~J.; Anderson,~T.~A.; Deustua,~J.~E.; Ghanem,~K.; Hait,~D.; Hoffmann,~M.~R.; Lee,~S.; Levine,~D.~S.; Magoulas,~I.; Shen,~J. \latin{et~al.}  The ground state electronic energy of benzene. \emph{Journal of Physical Chemistry letters} \textbf{2020}, \emph{11}, 8922--8929\relax
\mciteBstWouldAddEndPuncttrue
\mciteSetBstMidEndSepPunct{\mcitedefaultmidpunct}
{\mcitedefaultendpunct}{\mcitedefaultseppunct}\relax
\EndOfBibitem
\bibitem[Hohenberg and Kohn(1964)Hohenberg, and Kohn]{dft_original}
Hohenberg,~P.; Kohn,~W. Inhomogeneous electron gas. \emph{Phys. Rev.} \textbf{1964}, \emph{136}, B864\relax
\mciteBstWouldAddEndPuncttrue
\mciteSetBstMidEndSepPunct{\mcitedefaultmidpunct}
{\mcitedefaultendpunct}{\mcitedefaultseppunct}\relax
\EndOfBibitem
\bibitem[Kohn and Sham(1965)Kohn, and Sham]{kohn_sham_original}
Kohn,~W.; Sham,~L.~J. Self-consistent equations including exchange and correlation effects. \emph{Physical review} \textbf{1965}, \emph{140}, A1133\relax
\mciteBstWouldAddEndPuncttrue
\mciteSetBstMidEndSepPunct{\mcitedefaultmidpunct}
{\mcitedefaultendpunct}{\mcitedefaultseppunct}\relax
\EndOfBibitem
\bibitem[Simons(2023)]{computational_chemistry_review_complicated}
Simons,~J. Why is quantum chemistry so complicated? \emph{Journal of the American Chemical Society} \textbf{2023}, \emph{145}, 4343--4354\relax
\mciteBstWouldAddEndPuncttrue
\mciteSetBstMidEndSepPunct{\mcitedefaultmidpunct}
{\mcitedefaultendpunct}{\mcitedefaultseppunct}\relax
\EndOfBibitem
\bibitem[Austin \latin{et~al.}(2020)Austin, \latin{et~al.} others]{nersc}
Austin,~B.; others NERSC-10 Workload Analysis. 2020; \url{https://portal.nersc.gov/project/m888/nersc10/workload/N10_Workload_Analysis.latest.pdf}, Accessed: March 31, 2024\relax
\mciteBstWouldAddEndPuncttrue
\mciteSetBstMidEndSepPunct{\mcitedefaultmidpunct}
{\mcitedefaultendpunct}{\mcitedefaultseppunct}\relax
\EndOfBibitem
\bibitem[Yasuda(2008)]{gaussian_program}
Yasuda,~K. Two-electron integral evaluation on the graphics processor unit. \emph{Journal of Computational Chemistry} \textbf{2008}, \emph{29}, 334--342\relax
\mciteBstWouldAddEndPuncttrue
\mciteSetBstMidEndSepPunct{\mcitedefaultmidpunct}
{\mcitedefaultendpunct}{\mcitedefaultseppunct}\relax
\EndOfBibitem
\bibitem[Ufimtsev and Martinez(2008)Ufimtsev, and Martinez]{terachem_gpu_1}
Ufimtsev,~I.~S.; Martinez,~T.~J. Quantum chemistry on graphical processing units. 1. Strategies for two-electron integral evaluation. \emph{Journal of Chemical Theory and Computation} \textbf{2008}, \emph{4}, 222--231\relax
\mciteBstWouldAddEndPuncttrue
\mciteSetBstMidEndSepPunct{\mcitedefaultmidpunct}
{\mcitedefaultendpunct}{\mcitedefaultseppunct}\relax
\EndOfBibitem
\bibitem[Ufimtsev and Martinez(2009)Ufimtsev, and Martinez]{terachem_gpu_2}
Ufimtsev,~I.~S.; Martinez,~T.~J. Quantum chemistry on graphical processing units. 2. Direct self-consistent-field implementation. \emph{Journal of Chemical Theory and Computation} \textbf{2009}, \emph{5}, 1004--1015\relax
\mciteBstWouldAddEndPuncttrue
\mciteSetBstMidEndSepPunct{\mcitedefaultmidpunct}
{\mcitedefaultendpunct}{\mcitedefaultseppunct}\relax
\EndOfBibitem
\bibitem[Ufimtsev and Martinez(2009)Ufimtsev, and Martinez]{terachem_gpu_3}
Ufimtsev,~I.~S.; Martinez,~T.~J. Quantum chemistry on graphical processing units. 3. Analytical energy gradients, geometry optimization, and first principles molecular dynamics. \emph{Journal of Chemical Theory and Computation} \textbf{2009}, \emph{5}, 2619--2628\relax
\mciteBstWouldAddEndPuncttrue
\mciteSetBstMidEndSepPunct{\mcitedefaultmidpunct}
{\mcitedefaultendpunct}{\mcitedefaultseppunct}\relax
\EndOfBibitem
\bibitem[Titov \latin{et~al.}(2013)Titov, Ufimtsev, Luehr, and Martinez]{terachem_2013}
Titov,~A.~V.; Ufimtsev,~I.~S.; Luehr,~N.; Martinez,~T.~J. Generating efficient quantum chemistry codes for novel architectures. \emph{Journal of Chemical Theory and Computation} \textbf{2013}, \emph{9}, 213--221\relax
\mciteBstWouldAddEndPuncttrue
\mciteSetBstMidEndSepPunct{\mcitedefaultmidpunct}
{\mcitedefaultendpunct}{\mcitedefaultseppunct}\relax
\EndOfBibitem
\bibitem[Seritan \latin{et~al.}(2021)Seritan, Bannwarth, Fales, Hohenstein, Isborn, Kokkila-Schumacher, Li, Liu, Luehr, Snyder~Jr, \latin{et~al.} others]{terachem_2021}
Seritan,~S.; Bannwarth,~C.; Fales,~B.~S.; Hohenstein,~E.~G.; Isborn,~C.~M.; Kokkila-Schumacher,~S.~I.; Li,~X.; Liu,~F.; Luehr,~N.; Snyder~Jr,~J.~W. \latin{et~al.}  TeraChem: A graphical processing unit-accelerated electronic structure package for large-scale ab initio molecular dynamics. \emph{Wiley Interdisciplinary Reviews: Computational Molecular Science} \textbf{2021}, \emph{11}, e1494\relax
\mciteBstWouldAddEndPuncttrue
\mciteSetBstMidEndSepPunct{\mcitedefaultmidpunct}
{\mcitedefaultendpunct}{\mcitedefaultseppunct}\relax
\EndOfBibitem
\bibitem[Fales \latin{et~al.}(2020)Fales, Curtis, Johnson, Lahana, Seritan, Wang, Weir, Mart{\'\i}nez, and Hohenstein]{terachem_ccsd_gpu}
Fales,~B.~S.; Curtis,~E.~R.; Johnson,~K.~G.; Lahana,~D.; Seritan,~S.; Wang,~Y.; Weir,~H.; Mart{\'\i}nez,~T.~J.; Hohenstein,~E.~G. Performance of coupled-cluster singles and doubles on modern stream processing architectures. \emph{Journal of Chemical Theory and Computation} \textbf{2020}, \emph{16}, 4021--4028\relax
\mciteBstWouldAddEndPuncttrue
\mciteSetBstMidEndSepPunct{\mcitedefaultmidpunct}
{\mcitedefaultendpunct}{\mcitedefaultseppunct}\relax
\EndOfBibitem
\bibitem[Parrish \latin{et~al.}(2019)Parrish, Zhao, Hohenstein, and Mart{\'\i}nez]{terachem_rank_reduce_cc_1}
Parrish,~R.~M.; Zhao,~Y.; Hohenstein,~E.~G.; Mart{\'\i}nez,~T.~J. Rank reduced coupled cluster theory. I. Ground state energies and wavefunctions. \emph{The Journal of Chemical Physics} \textbf{2019}, \emph{150}\relax
\mciteBstWouldAddEndPuncttrue
\mciteSetBstMidEndSepPunct{\mcitedefaultmidpunct}
{\mcitedefaultendpunct}{\mcitedefaultseppunct}\relax
\EndOfBibitem
\bibitem[Hohenstein \latin{et~al.}(2019)Hohenstein, Zhao, Parrish, and Mart{\'\i}nez]{terachem_rank_reduce_cc_2}
Hohenstein,~E.~G.; Zhao,~Y.; Parrish,~R.~M.; Mart{\'\i}nez,~T.~J. Rank reduced coupled cluster theory. II. Equation-of-motion coupled-cluster singles and doubles. \emph{The Journal of Chemical Physics} \textbf{2019}, \emph{151}\relax
\mciteBstWouldAddEndPuncttrue
\mciteSetBstMidEndSepPunct{\mcitedefaultmidpunct}
{\mcitedefaultendpunct}{\mcitedefaultseppunct}\relax
\EndOfBibitem
\bibitem[Hohenstein \latin{et~al.}(2022)Hohenstein, Fales, Parrish, and Mart{\'\i}nez]{terachem_rank_reduce_cc_3}
Hohenstein,~E.~G.; Fales,~B.~S.; Parrish,~R.~M.; Mart{\'\i}nez,~T.~J. Rank-reduced coupled-cluster. III. Tensor hypercontraction of the doubles amplitudes. \emph{The Journal of Chemical Physics} \textbf{2022}, \emph{156}\relax
\mciteBstWouldAddEndPuncttrue
\mciteSetBstMidEndSepPunct{\mcitedefaultmidpunct}
{\mcitedefaultendpunct}{\mcitedefaultseppunct}\relax
\EndOfBibitem
\bibitem[Hohenstein and Mart{\'\i}nez(2021)Hohenstein, and Mart{\'\i}nez]{terachem_rank_reduce_ccsd}
Hohenstein,~E.~G.; Mart{\'\i}nez,~T.~J. GPU acceleration of rank-reduced coupled-cluster singles and doubles. \emph{The Journal of Chemical Physics} \textbf{2021}, \emph{155}\relax
\mciteBstWouldAddEndPuncttrue
\mciteSetBstMidEndSepPunct{\mcitedefaultmidpunct}
{\mcitedefaultendpunct}{\mcitedefaultseppunct}\relax
\EndOfBibitem
\bibitem[Song and Martinez(2016)Song, and Martinez]{song_sosmp2_part1}
Song,~C.; Martinez,~T.~J. Atomic orbital-based SOS-MP2 with tensor hypercontraction. I. GPU-based tensor construction and exploiting sparsity. \emph{The Journal of Chemical Physics} \textbf{2016}, \emph{144}, 174111\relax
\mciteBstWouldAddEndPuncttrue
\mciteSetBstMidEndSepPunct{\mcitedefaultmidpunct}
{\mcitedefaultendpunct}{\mcitedefaultseppunct}\relax
\EndOfBibitem
\bibitem[Song and Martinez(2017)Song, and Martinez]{song_sosmp2_part2}
Song,~C.; Martinez,~T.~J. Atomic orbital-based SOS-MP2 with tensor hypercontraction. II. Local tensor hypercontraction. \emph{The Journal of Chemical Physics} \textbf{2017}, \emph{146}, 034104\relax
\mciteBstWouldAddEndPuncttrue
\mciteSetBstMidEndSepPunct{\mcitedefaultmidpunct}
{\mcitedefaultendpunct}{\mcitedefaultseppunct}\relax
\EndOfBibitem
\bibitem[Song and Martinez(2017)Song, and Martinez]{song_mp2_gradient}
Song,~C.; Martinez,~T.~J. Analytical gradients for tensor hyper-contracted MP2 and SOS-MP2 on graphical processing units. \emph{Journal of Chemical Physics} \textbf{2017}, \emph{147}, 161723\relax
\mciteBstWouldAddEndPuncttrue
\mciteSetBstMidEndSepPunct{\mcitedefaultmidpunct}
{\mcitedefaultendpunct}{\mcitedefaultseppunct}\relax
\EndOfBibitem
\bibitem[Snyder \latin{et~al.}(2017)Snyder, Fales, Hohenstein, Levine, and Mart{\'\i}nez]{terachem_casscf}
Snyder,~J.~W.; Fales,~B.~S.; Hohenstein,~E.~G.; Levine,~B.~G.; Mart{\'\i}nez,~T.~J. A direct-compatible formulation of the coupled perturbed complete active space self-consistent field equations on graphical processing units. \emph{The Journal of Chemical Physics} \textbf{2017}, \emph{146}\relax
\mciteBstWouldAddEndPuncttrue
\mciteSetBstMidEndSepPunct{\mcitedefaultmidpunct}
{\mcitedefaultendpunct}{\mcitedefaultseppunct}\relax
\EndOfBibitem
\bibitem[Slav{\'\i}{\v{c}}ek and Mart{\'\i}nez(2010)Slav{\'\i}{\v{c}}ek, and Mart{\'\i}nez]{terachem_fomo_casci}
Slav{\'\i}{\v{c}}ek,~P.; Mart{\'\i}nez,~T.~J. Ab initio floating occupation molecular orbital-complete active space configuration interaction: An efficient approximation to CASSCF. \emph{The Journal of Chemical Physics} \textbf{2010}, \emph{132}\relax
\mciteBstWouldAddEndPuncttrue
\mciteSetBstMidEndSepPunct{\mcitedefaultmidpunct}
{\mcitedefaultendpunct}{\mcitedefaultseppunct}\relax
\EndOfBibitem
\bibitem[Fales and Levine(2015)Fales, and Levine]{terachem_direct_ci_1}
Fales,~B.~S.; Levine,~B.~G. Nanoscale multireference quantum chemistry: Full configuration interaction on graphical processing units. \emph{Journal of Chemical Theory and Computation} \textbf{2015}, \emph{11}, 4708--4716\relax
\mciteBstWouldAddEndPuncttrue
\mciteSetBstMidEndSepPunct{\mcitedefaultmidpunct}
{\mcitedefaultendpunct}{\mcitedefaultseppunct}\relax
\EndOfBibitem
\bibitem[Fales and Mart{\'\i}nez(2020)Fales, and Mart{\'\i}nez]{terachem_direct_ci_2}
Fales,~B.~S.; Mart{\'\i}nez,~T.~J. Fast transformations between configuration state function and Slater determinant bases for direct configuration interaction. \emph{The Journal of Chemical Physics} \textbf{2020}, \emph{152}\relax
\mciteBstWouldAddEndPuncttrue
\mciteSetBstMidEndSepPunct{\mcitedefaultmidpunct}
{\mcitedefaultendpunct}{\mcitedefaultseppunct}\relax
\EndOfBibitem
\bibitem[Fales and Mart{\'\i}nez(2020)Fales, and Mart{\'\i}nez]{terachem_direct_ci_3}
Fales,~B.~S.; Mart{\'\i}nez,~T.~J. Efficient treatment of large active spaces through multi-GPU parallel implementation of direct configuration interaction. \emph{Journal of Chemical Theory and Computation} \textbf{2020}, \emph{16}, 1586--1596\relax
\mciteBstWouldAddEndPuncttrue
\mciteSetBstMidEndSepPunct{\mcitedefaultmidpunct}
{\mcitedefaultendpunct}{\mcitedefaultseppunct}\relax
\EndOfBibitem
\bibitem[Fales \latin{et~al.}(2018)Fales, Seritan, Settje, Levine, Koch, and Mart{\'\i}nez]{terachem_rank_reduced_ci}
Fales,~B.~S.; Seritan,~S.; Settje,~N.~F.; Levine,~B.~G.; Koch,~H.; Mart{\'\i}nez,~T.~J. Large-scale electron correlation calculations: Rank-reduced full configuration interaction. \emph{Journal of Chemical Theory and Computation} \textbf{2018}, \emph{14}, 4139--4150\relax
\mciteBstWouldAddEndPuncttrue
\mciteSetBstMidEndSepPunct{\mcitedefaultmidpunct}
{\mcitedefaultendpunct}{\mcitedefaultseppunct}\relax
\EndOfBibitem
\bibitem[Song and Martinez(2018)Song, and Martinez]{song_caspt2}
Song,~C.; Martinez,~T.~J. Reduced scaling CASPT2 using supporting subspaces and tensor hyper-contraction. \emph{The Journal of Chemical Physics} \textbf{2018}, \emph{149}, 044108\relax
\mciteBstWouldAddEndPuncttrue
\mciteSetBstMidEndSepPunct{\mcitedefaultmidpunct}
{\mcitedefaultendpunct}{\mcitedefaultseppunct}\relax
\EndOfBibitem
\bibitem[Song \latin{et~al.}(2021)Song, Neaton, and Martinez]{song_caspt2_gradient}
Song,~C.; Neaton,~J.~B.; Martinez,~T.~J. Reduced scaling formulation of CASPT2 analytical gradients using the supporting subspace method. \emph{The Journal of Chemical Physics} \textbf{2021}, \emph{154}, 014103\relax
\mciteBstWouldAddEndPuncttrue
\mciteSetBstMidEndSepPunct{\mcitedefaultmidpunct}
{\mcitedefaultendpunct}{\mcitedefaultseppunct}\relax
\EndOfBibitem
\bibitem[Song and Martinez(2020)Song, and Martinez]{song_xmspt2}
Song,~C.; Martinez,~T.~J. Reduced scaling extended multi-state CASPT2 (XMS-CASPT2) using supporting subspaces and tensor hyper-contraction. \emph{The Journal of Chemical Physics} \textbf{2020}, \emph{152}, 234113\relax
\mciteBstWouldAddEndPuncttrue
\mciteSetBstMidEndSepPunct{\mcitedefaultmidpunct}
{\mcitedefaultendpunct}{\mcitedefaultseppunct}\relax
\EndOfBibitem
\bibitem[Ufimtsev \latin{et~al.}(2011)Ufimtsev, Luehr, and Martinez]{protein_gpu_1}
Ufimtsev,~I.~S.; Luehr,~N.; Martinez,~T.~J. Charge transfer and polarization in solvated proteins from ab initio molecular dynamics. \emph{Journal of Physical Chemistry Letters} \textbf{2011}, \emph{2}, 1789--1793\relax
\mciteBstWouldAddEndPuncttrue
\mciteSetBstMidEndSepPunct{\mcitedefaultmidpunct}
{\mcitedefaultendpunct}{\mcitedefaultseppunct}\relax
\EndOfBibitem
\bibitem[Kulik \latin{et~al.}(2012)Kulik, Luehr, Ufimtsev, and Martinez]{protein_gpu_2}
Kulik,~H.~J.; Luehr,~N.; Ufimtsev,~I.~S.; Martinez,~T.~J. Ab initio quantum chemistry for protein structures. \emph{Journal of Physical Chemistry B} \textbf{2012}, \emph{116}, 12501--12509\relax
\mciteBstWouldAddEndPuncttrue
\mciteSetBstMidEndSepPunct{\mcitedefaultmidpunct}
{\mcitedefaultendpunct}{\mcitedefaultseppunct}\relax
\EndOfBibitem
\bibitem[Jones \latin{et~al.}(2022)Jones, List, and Mart{\'\i}nez]{protein_gpu_3}
Jones,~C.~M.; List,~N.~H.; Mart{\'\i}nez,~T.~J. Steric and Electronic Origins of Fluorescence in GFP and GFP-like Proteins. \emph{Journal of the American Chemical Society} \textbf{2022}, \emph{144}, 12732--12746\relax
\mciteBstWouldAddEndPuncttrue
\mciteSetBstMidEndSepPunct{\mcitedefaultmidpunct}
{\mcitedefaultendpunct}{\mcitedefaultseppunct}\relax
\EndOfBibitem
\bibitem[Yasuda(2008)]{gaussian_program_dft}
Yasuda,~K. Accelerating density functional calculations with graphics processing unit. \emph{Journal of Chemical Theory and Computation} \textbf{2008}, \emph{4}, 1230--1236\relax
\mciteBstWouldAddEndPuncttrue
\mciteSetBstMidEndSepPunct{\mcitedefaultmidpunct}
{\mcitedefaultendpunct}{\mcitedefaultseppunct}\relax
\EndOfBibitem
\bibitem[Yasuda and Maruoka(2014)Yasuda, and Maruoka]{gaussian_program_high_angular}
Yasuda,~K.; Maruoka,~H. Efficient calculation of two-electron integrals for high angular basis functions. \emph{International Journal of Quantum Chemistry} \textbf{2014}, \emph{114}, 543--552\relax
\mciteBstWouldAddEndPuncttrue
\mciteSetBstMidEndSepPunct{\mcitedefaultmidpunct}
{\mcitedefaultendpunct}{\mcitedefaultseppunct}\relax
\EndOfBibitem
\bibitem[Asadchev \latin{et~al.}(2010)Asadchev, Allada, Felder, Bode, Gordon, and Windus]{gamess_rys_quadrature}
Asadchev,~A.; Allada,~V.; Felder,~J.; Bode,~B.~M.; Gordon,~M.~S.; Windus,~T.~L. Uncontracted Rys quadrature implementation of up to G functions on graphical processing units. \emph{Journal of Chemical Theory and Computation} \textbf{2010}, \emph{6}, 696--704\relax
\mciteBstWouldAddEndPuncttrue
\mciteSetBstMidEndSepPunct{\mcitedefaultmidpunct}
{\mcitedefaultendpunct}{\mcitedefaultseppunct}\relax
\EndOfBibitem
\bibitem[Barca \latin{et~al.}(2020)Barca, Galvez-Vallejo, Poole, Rendell, and Gordon]{gamess_sp}
Barca,~G.~M.; Galvez-Vallejo,~J.~L.; Poole,~D.~L.; Rendell,~A.~P.; Gordon,~M.~S. High-performance, graphics processing unit-accelerated fock build algorithm. \emph{Journal of Chemical Theory and Computation} \textbf{2020}, \emph{16}, 7232--7238\relax
\mciteBstWouldAddEndPuncttrue
\mciteSetBstMidEndSepPunct{\mcitedefaultmidpunct}
{\mcitedefaultendpunct}{\mcitedefaultseppunct}\relax
\EndOfBibitem
\bibitem[Barca \latin{et~al.}(2021)Barca, Alkan, Galvez-Vallejo, Poole, Rendell, and Gordon]{gamess_spd}
Barca,~G.~M.; Alkan,~M.; Galvez-Vallejo,~J.~L.; Poole,~D.~L.; Rendell,~A.~P.; Gordon,~M.~S. Faster self-consistent field (SCF) calculations on GPU clusters. \emph{Journal of Chemical Theory and Computation} \textbf{2021}, \emph{17}, 7486--7503\relax
\mciteBstWouldAddEndPuncttrue
\mciteSetBstMidEndSepPunct{\mcitedefaultmidpunct}
{\mcitedefaultendpunct}{\mcitedefaultseppunct}\relax
\EndOfBibitem
\bibitem[Galvez~Vallejo \latin{et~al.}(2023)Galvez~Vallejo, Barca, and Gordon]{gamess_f}
Galvez~Vallejo,~J.~L.; Barca,~G.~M.; Gordon,~M.~S. High-performance GPU-accelerated evaluation of electron repulsion integrals. \emph{Molecular Physics} \textbf{2023}, \emph{121}, e2112987\relax
\mciteBstWouldAddEndPuncttrue
\mciteSetBstMidEndSepPunct{\mcitedefaultmidpunct}
{\mcitedefaultendpunct}{\mcitedefaultseppunct}\relax
\EndOfBibitem
\bibitem[Epifanovsky \latin{et~al.}(2021)Epifanovsky, Gilbert, Feng, Lee, Mao, Mardirossian, Pokhilko, White, Coons, Dempwolff, \latin{et~al.} others]{qchem_5_or_higher}
Epifanovsky,~E.; Gilbert,~A.~T.; Feng,~X.; Lee,~J.; Mao,~Y.; Mardirossian,~N.; Pokhilko,~P.; White,~A.~F.; Coons,~M.~P.; Dempwolff,~A.~L. \latin{et~al.}  Software for the frontiers of quantum chemistry: An overview of developments in the Q-Chem 5 package. \emph{The Journal of Chemical Physics} \textbf{2021}, \emph{155}\relax
\mciteBstWouldAddEndPuncttrue
\mciteSetBstMidEndSepPunct{\mcitedefaultmidpunct}
{\mcitedefaultendpunct}{\mcitedefaultseppunct}\relax
\EndOfBibitem
\bibitem[Tornai \latin{et~al.}(2019)Tornai, Ladj{\'a}nszki, R{\'a}k, Kis, and Cserey]{brianqc}
Tornai,~G.~J.; Ladj{\'a}nszki,~I.; R{\'a}k,~{\'A}.; Kis,~G.; Cserey,~G. Calculation of quantum chemical two-electron integrals by applying compiler technology on GPU. \emph{Journal of Chemical Theory and Computation} \textbf{2019}, \emph{15}, 5319--5331\relax
\mciteBstWouldAddEndPuncttrue
\mciteSetBstMidEndSepPunct{\mcitedefaultmidpunct}
{\mcitedefaultendpunct}{\mcitedefaultseppunct}\relax
\EndOfBibitem
\bibitem[Wu \latin{et~al.}(2024)Wu, Sun, Pu, Zheng, Ma, Yan, Yu, Wu, Huo, Li, \latin{et~al.} others]{pyscf_gpu}
Wu,~X.; Sun,~Q.; Pu,~Z.; Zheng,~T.; Ma,~W.; Yan,~W.; Yu,~X.; Wu,~Z.; Huo,~M.; Li,~X. \latin{et~al.}  Python-Based Quantum Chemistry Calculations with GPU Acceleration. \emph{arXiv preprint arXiv:2404.09452} \textbf{2024}, \relax
\mciteBstWouldAddEndPunctfalse
\mciteSetBstMidEndSepPunct{\mcitedefaultmidpunct}
{}{\mcitedefaultseppunct}\relax
\EndOfBibitem
\bibitem[Holzer(2020)]{turbomole}
Holzer,~C. An improved seminumerical Coulomb and exchange algorithm for properties and excited states in modern density functional theory. \emph{The Journal of Chemical Physics} \textbf{2020}, \emph{153}\relax
\mciteBstWouldAddEndPuncttrue
\mciteSetBstMidEndSepPunct{\mcitedefaultmidpunct}
{\mcitedefaultendpunct}{\mcitedefaultseppunct}\relax
\EndOfBibitem
\bibitem[Asadchev and Valeev(2023)Asadchev, and Valeev]{libintx_up_to_iiii}
Asadchev,~A.; Valeev,~E.~F. High-performance evaluation of high angular momentum 4-center Gaussian integrals on modern accelerated processors. \emph{The Journal of Physical Chemistry A} \textbf{2023}, \emph{127}, 10889--10895\relax
\mciteBstWouldAddEndPuncttrue
\mciteSetBstMidEndSepPunct{\mcitedefaultmidpunct}
{\mcitedefaultendpunct}{\mcitedefaultseppunct}\relax
\EndOfBibitem
\bibitem[Asadchev and Valeev(2024)Asadchev, and Valeev]{libintx_gpu_2}
Asadchev,~A.; Valeev,~E.~F. 3-center and 4-center 2-particle Gaussian AO integrals on modern accelerated processors. \emph{The Journal of Chemical Physics} \textbf{2024}, \emph{160}\relax
\mciteBstWouldAddEndPuncttrue
\mciteSetBstMidEndSepPunct{\mcitedefaultmidpunct}
{\mcitedefaultendpunct}{\mcitedefaultseppunct}\relax
\EndOfBibitem
\bibitem[Dunning~Jr(1989)]{cc_pvxz}
Dunning~Jr,~T.~H. Gaussian basis sets for use in correlated molecular calculations. I. The atoms boron through neon and hydrogen. \emph{The Journal of Chemical Physics} \textbf{1989}, \emph{90}, 1007--1023\relax
\mciteBstWouldAddEndPuncttrue
\mciteSetBstMidEndSepPunct{\mcitedefaultmidpunct}
{\mcitedefaultendpunct}{\mcitedefaultseppunct}\relax
\EndOfBibitem
\bibitem[Jensen(2017)]{red_book}
Jensen,~F. \emph{Introduction to Computational Chemistry}; John Wiley \& Sons, 2017\relax
\mciteBstWouldAddEndPuncttrue
\mciteSetBstMidEndSepPunct{\mcitedefaultmidpunct}
{\mcitedefaultendpunct}{\mcitedefaultseppunct}\relax
\EndOfBibitem
\bibitem[Bursch \latin{et~al.}(2022)Bursch, Mewes, Hansen, and Grimme]{dft_best_practice}
Bursch,~M.; Mewes,~J.-M.; Hansen,~A.; Grimme,~S. Best-practice DFT protocols for basic molecular computational chemistry. \emph{Angewandte Chemie International Edition} \textbf{2022}, \emph{61}, e202205735\relax
\mciteBstWouldAddEndPuncttrue
\mciteSetBstMidEndSepPunct{\mcitedefaultmidpunct}
{\mcitedefaultendpunct}{\mcitedefaultseppunct}\relax
\EndOfBibitem
\bibitem[Kutzelnigg and Morgan~III(1992)Kutzelnigg, and Morgan~III]{kutzelnigg1992rates}
Kutzelnigg,~W.; Morgan~III,~J.~D. Rates of convergence of the partial-wave expansions of atomic correlation energies. \emph{The Journal of Chemical Physics} \textbf{1992}, \emph{96}, 4484--4508\relax
\mciteBstWouldAddEndPuncttrue
\mciteSetBstMidEndSepPunct{\mcitedefaultmidpunct}
{\mcitedefaultendpunct}{\mcitedefaultseppunct}\relax
\EndOfBibitem
\bibitem[Helgaker \latin{et~al.}(1997)Helgaker, Klopper, Koch, and Noga]{helgaker1997basis}
Helgaker,~T.; Klopper,~W.; Koch,~H.; Noga,~J. Basis-set convergence of correlated calculations on water. \emph{The Journal of Chemical Physics} \textbf{1997}, \emph{106}, 9639--9646\relax
\mciteBstWouldAddEndPuncttrue
\mciteSetBstMidEndSepPunct{\mcitedefaultmidpunct}
{\mcitedefaultendpunct}{\mcitedefaultseppunct}\relax
\EndOfBibitem
\bibitem[Szabo and Ostlund(1996)Szabo, and Ostlund]{blue_book}
Szabo,~A.; Ostlund,~N.~S. \emph{Modern Quantum Chemistry: Introduction to Advanced Electronic Structure Theory}; Dover Publication, 1996\relax
\mciteBstWouldAddEndPuncttrue
\mciteSetBstMidEndSepPunct{\mcitedefaultmidpunct}
{\mcitedefaultendpunct}{\mcitedefaultseppunct}\relax
\EndOfBibitem
\bibitem[Purvis and Bartlett(1982)Purvis, and Bartlett]{purvis_full_1982}
Purvis,~G.~D.; Bartlett,~R.~J. A full coupled-cluster singles and doubles model: The inclusion of disconnected triples. \emph{The Journal of Chemical Physics} \textbf{1982}, \emph{76}, 1910--1918\relax
\mciteBstWouldAddEndPuncttrue
\mciteSetBstMidEndSepPunct{\mcitedefaultmidpunct}
{\mcitedefaultendpunct}{\mcitedefaultseppunct}\relax
\EndOfBibitem
\bibitem[Bartlett and Musia{\l}(2007)Bartlett, and Musia{\l}]{cc_review}
Bartlett,~R.~J.; Musia{\l},~M. Coupled-cluster theory in quantum chemistry. \emph{Reviews of Modern Physics} \textbf{2007}, \emph{79}, 291\relax
\mciteBstWouldAddEndPuncttrue
\mciteSetBstMidEndSepPunct{\mcitedefaultmidpunct}
{\mcitedefaultendpunct}{\mcitedefaultseppunct}\relax
\EndOfBibitem
\bibitem[Roos \latin{et~al.}(1980)Roos, Taylor, and Sigbahn]{casscf_original_1}
Roos,~B.~O.; Taylor,~P.~R.; Sigbahn,~P.~E. A complete active space SCF method (CASSCF) using a density matrix formulated super-CI approach. \emph{Chemical Physics} \textbf{1980}, \emph{48}, 157--173\relax
\mciteBstWouldAddEndPuncttrue
\mciteSetBstMidEndSepPunct{\mcitedefaultmidpunct}
{\mcitedefaultendpunct}{\mcitedefaultseppunct}\relax
\EndOfBibitem
\bibitem[Roos(1980)]{casscf_original_2}
Roos,~B.~O. The complete active space SCF method in a fock-matrix-based super-CI formulation. \emph{International Journal of Quantum Chemistry} \textbf{1980}, \emph{18}, 175--189\relax
\mciteBstWouldAddEndPuncttrue
\mciteSetBstMidEndSepPunct{\mcitedefaultmidpunct}
{\mcitedefaultendpunct}{\mcitedefaultseppunct}\relax
\EndOfBibitem
\bibitem[Siegbahn \latin{et~al.}(1980)Siegbahn, Heiberg, Roos, and Levy]{casscf_original_3}
Siegbahn,~P.; Heiberg,~A.; Roos,~B.; Levy,~B. A comparison of the super-CI and the Newton-Raphson scheme in the complete active space SCF method. \emph{Physica Scripta} \textbf{1980}, \emph{21}, 323\relax
\mciteBstWouldAddEndPuncttrue
\mciteSetBstMidEndSepPunct{\mcitedefaultmidpunct}
{\mcitedefaultendpunct}{\mcitedefaultseppunct}\relax
\EndOfBibitem
\bibitem[Siegbahn \latin{et~al.}(1981)Siegbahn, Alml{\"o}f, Heiberg, and Roos]{casscf_original_4}
Siegbahn,~P.~E.; Alml{\"o}f,~J.; Heiberg,~A.; Roos,~B.~O. The complete active space SCF (CASSCF) method in a Newton--Raphson formulation with application to the HNO molecule. \emph{The Journal of Chemical Physics} \textbf{1981}, \emph{74}, 2384--2396\relax
\mciteBstWouldAddEndPuncttrue
\mciteSetBstMidEndSepPunct{\mcitedefaultmidpunct}
{\mcitedefaultendpunct}{\mcitedefaultseppunct}\relax
\EndOfBibitem
\bibitem[Gill(1994)]{head_gordon_pople_flops}
Gill,~P.~M. Molecular integrals over gaussian basis functions. \emph{Advances in Quantum Chemistry} \textbf{1994}, \emph{25}, 141--205\relax
\mciteBstWouldAddEndPuncttrue
\mciteSetBstMidEndSepPunct{\mcitedefaultmidpunct}
{\mcitedefaultendpunct}{\mcitedefaultseppunct}\relax
\EndOfBibitem
\bibitem[Helgaker and Taylor(1995)Helgaker, and Taylor]{helgaker_taylor_book}
Helgaker,~T.; Taylor,~P.~R. Gaussian Basis Sets and Molecular Integrals. In \emph{Modern Electronic Structure Theory, Part II}; Yarkony,~D.~R., Ed.; World Scientific, 1995; pp 725--856\relax
\mciteBstWouldAddEndPuncttrue
\mciteSetBstMidEndSepPunct{\mcitedefaultmidpunct}
{\mcitedefaultendpunct}{\mcitedefaultseppunct}\relax
\EndOfBibitem
\bibitem[Helgaker \latin{et~al.}(2013)Helgaker, Jorgensen, and Olsen]{purple_book}
Helgaker,~T.; Jorgensen,~P.; Olsen,~J. \emph{Molecular electronic-structure theory}; John Wiley \& Sons, 2013\relax
\mciteBstWouldAddEndPuncttrue
\mciteSetBstMidEndSepPunct{\mcitedefaultmidpunct}
{\mcitedefaultendpunct}{\mcitedefaultseppunct}\relax
\EndOfBibitem
\bibitem[Fermann and Valeev(2020)Fermann, and Valeev]{integral_fundamental}
Fermann,~J.~T.; Valeev,~E.~F. Fundamentals of molecular integrals evaluation. \emph{arXiv preprint arXiv:2007.12057} \textbf{2020}, \relax
\mciteBstWouldAddEndPunctfalse
\mciteSetBstMidEndSepPunct{\mcitedefaultmidpunct}
{}{\mcitedefaultseppunct}\relax
\EndOfBibitem
\bibitem[Komornicki and King(2011)Komornicki, and King]{n_electron_integral}
Komornicki,~A.; King,~H.~F. A general formulation for the efficient evaluation of n-electron integrals over products of Gaussian charge distributions with Gaussian geminal functions. \emph{The Journal of Chemical Physics} \textbf{2011}, \emph{134}\relax
\mciteBstWouldAddEndPuncttrue
\mciteSetBstMidEndSepPunct{\mcitedefaultmidpunct}
{\mcitedefaultendpunct}{\mcitedefaultseppunct}\relax
\EndOfBibitem
\bibitem[McMurchie and Davidson(1978)McMurchie, and Davidson]{McMurchie_Davidson_original}
McMurchie,~L.~E.; Davidson,~E.~R. One-and two-electron integrals over Cartesian Gaussian functions. \emph{Journal of Computational Physics} \textbf{1978}, \emph{26}, 218--231\relax
\mciteBstWouldAddEndPuncttrue
\mciteSetBstMidEndSepPunct{\mcitedefaultmidpunct}
{\mcitedefaultendpunct}{\mcitedefaultseppunct}\relax
\EndOfBibitem
\bibitem[Boys(1950)]{boys_original}
Boys,~S.~F. Electronic wave functions-I. A general method of calculation for the stationary states of any molecular system. \emph{Proceedings of the Royal Society of London. Series A. Mathematical and Physical Sciences} \textbf{1950}, \emph{200}, 542--554\relax
\mciteBstWouldAddEndPuncttrue
\mciteSetBstMidEndSepPunct{\mcitedefaultmidpunct}
{\mcitedefaultendpunct}{\mcitedefaultseppunct}\relax
\EndOfBibitem
\bibitem[Fr{\"u}chtl and Otto(1994)Fr{\"u}chtl, and Otto]{boys_interpolation}
Fr{\"u}chtl,~H.; Otto,~P. A new algorithm for the evaluation of the incomplete gamma function on vector computers. \emph{ACM Transactions on Mathematical Software (TOMS)} \textbf{1994}, \emph{20}, 436--446\relax
\mciteBstWouldAddEndPuncttrue
\mciteSetBstMidEndSepPunct{\mcitedefaultmidpunct}
{\mcitedefaultendpunct}{\mcitedefaultseppunct}\relax
\EndOfBibitem
\bibitem[Mazur \latin{et~al.}(2016)Mazur, Makowski, and {\L}azarski]{boys_gpu}
Mazur,~G.; Makowski,~M.; {\L}azarski,~R. Boys function evaluation on graphical processing units. \emph{Journal of Mathematical Chemistry} \textbf{2016}, \emph{54}, 2022--2047\relax
\mciteBstWouldAddEndPuncttrue
\mciteSetBstMidEndSepPunct{\mcitedefaultmidpunct}
{\mcitedefaultendpunct}{\mcitedefaultseppunct}\relax
\EndOfBibitem
\bibitem[Heyd \latin{et~al.}(2003)Heyd, Scuseria, and Ernzerhof]{wpbeh}
Heyd,~J.; Scuseria,~G.~E.; Ernzerhof,~M. Hybrid functionals based on a screened Coulomb potential. \emph{The Journal of Chemical Physics} \textbf{2003}, \emph{118}, 8207--8215\relax
\mciteBstWouldAddEndPuncttrue
\mciteSetBstMidEndSepPunct{\mcitedefaultmidpunct}
{\mcitedefaultendpunct}{\mcitedefaultseppunct}\relax
\EndOfBibitem
\bibitem[Vydrov and Scuseria(2006)Vydrov, and Scuseria]{lc_wpbe}
Vydrov,~O.~A.; Scuseria,~G.~E. Assessment of a long-range corrected hybrid functional. \emph{The Journal of Chemical Physics} \textbf{2006}, \emph{125}, 234109\relax
\mciteBstWouldAddEndPuncttrue
\mciteSetBstMidEndSepPunct{\mcitedefaultmidpunct}
{\mcitedefaultendpunct}{\mcitedefaultseppunct}\relax
\EndOfBibitem
\bibitem[Yanai \latin{et~al.}(2004)Yanai, Tew, and Handy]{cam_b3lyp}
Yanai,~T.; Tew,~D.~P.; Handy,~N.~C. A new hybrid exchange--correlation functional using the Coulomb-attenuating method (CAM-B3LYP). \emph{Chemical Physics Letters} \textbf{2004}, \emph{393}, 51--57\relax
\mciteBstWouldAddEndPuncttrue
\mciteSetBstMidEndSepPunct{\mcitedefaultmidpunct}
{\mcitedefaultendpunct}{\mcitedefaultseppunct}\relax
\EndOfBibitem
\bibitem[Chai and Head-Gordon(2008)Chai, and Head-Gordon]{wb97}
Chai,~J.-D.; Head-Gordon,~M. Systematic optimization of long-range corrected hybrid density functionals. \emph{The Journal of Chemical Physics} \textbf{2008}, \emph{128}, 084106\relax
\mciteBstWouldAddEndPuncttrue
\mciteSetBstMidEndSepPunct{\mcitedefaultmidpunct}
{\mcitedefaultendpunct}{\mcitedefaultseppunct}\relax
\EndOfBibitem
\bibitem[Gill and Adamson(1996)Gill, and Adamson]{long_range_integral_1}
Gill,~P.~M.; Adamson,~R.~D. A family of attenuated Coulomb operators. \emph{Chemical Physics Letters} \textbf{1996}, \emph{261}, 105--110\relax
\mciteBstWouldAddEndPuncttrue
\mciteSetBstMidEndSepPunct{\mcitedefaultmidpunct}
{\mcitedefaultendpunct}{\mcitedefaultseppunct}\relax
\EndOfBibitem
\bibitem[Adamson \latin{et~al.}(1999)Adamson, Dombroski, and Gill]{long_range_integral_2}
Adamson,~R.~D.; Dombroski,~J.~P.; Gill,~P.~M. Efficient calculation of short-range Coulomb energies. \emph{Journal of Computational Chemistry} \textbf{1999}, \emph{20}, 921--927\relax
\mciteBstWouldAddEndPuncttrue
\mciteSetBstMidEndSepPunct{\mcitedefaultmidpunct}
{\mcitedefaultendpunct}{\mcitedefaultseppunct}\relax
\EndOfBibitem
\bibitem[Alml{\"o}f \latin{et~al.}(1982)Alml{\"o}f, F{\ae}gri~Jr, and Korsell]{direct_scf_original}
Alml{\"o}f,~J.; F{\ae}gri~Jr,~K.; Korsell,~K. Principles for a direct SCF approach to LICAO--MOab-initio calculations. \emph{Journal of Computational Chemistry} \textbf{1982}, \emph{3}, 385--399\relax
\mciteBstWouldAddEndPuncttrue
\mciteSetBstMidEndSepPunct{\mcitedefaultmidpunct}
{\mcitedefaultendpunct}{\mcitedefaultseppunct}\relax
\EndOfBibitem
\bibitem[White and Head-Gordon(1996)White, and Head-Gordon]{j_engine_hgp}
White,~C.~A.; Head-Gordon,~M. A J matrix engine for density functional theory calculations. \emph{The Journal of Chemical Physics} \textbf{1996}, \emph{104}, 2620--2629\relax
\mciteBstWouldAddEndPuncttrue
\mciteSetBstMidEndSepPunct{\mcitedefaultmidpunct}
{\mcitedefaultendpunct}{\mcitedefaultseppunct}\relax
\EndOfBibitem
\bibitem[Shao and Head-Gordon(2000)Shao, and Head-Gordon]{j_prepostprocess_original}
Shao,~Y.; Head-Gordon,~M. An improved J matrix engine for density functional theory calculations. \emph{Chemical Physics Letters} \textbf{2000}, \emph{323}, 425--433\relax
\mciteBstWouldAddEndPuncttrue
\mciteSetBstMidEndSepPunct{\mcitedefaultmidpunct}
{\mcitedefaultendpunct}{\mcitedefaultseppunct}\relax
\EndOfBibitem
\bibitem[Ahmadi and Alml{\"o}f(1995)Ahmadi, and Alml{\"o}f]{j_prepostprocess_family_basis_set}
Ahmadi,~G.~R.; Alml{\"o}f,~J. The Coulomb operator in a Gaussian product basis. \emph{Chemical Physics Letters} \textbf{1995}, \emph{246}, 364--370\relax
\mciteBstWouldAddEndPuncttrue
\mciteSetBstMidEndSepPunct{\mcitedefaultmidpunct}
{\mcitedefaultendpunct}{\mcitedefaultseppunct}\relax
\EndOfBibitem
\bibitem[Johnson \latin{et~al.}(2022)Johnson, Mirchandaney, Hoag, Heirich, Aiken, and Mart{\'\i}nez]{terachem_regent}
Johnson,~K.~G.; Mirchandaney,~S.; Hoag,~E.; Heirich,~A.; Aiken,~A.; Mart{\'\i}nez,~T.~J. Multinode multi-GPU two-electron integrals: Code generation using the regent language. \emph{Journal of Chemical Theory and Computation} \textbf{2022}, \emph{18}, 6522--6536\relax
\mciteBstWouldAddEndPuncttrue
\mciteSetBstMidEndSepPunct{\mcitedefaultmidpunct}
{\mcitedefaultendpunct}{\mcitedefaultseppunct}\relax
\EndOfBibitem
\bibitem[Johnson \latin{et~al.}(1993)Johnson, Gill, and Pople]{dft_gradient}
Johnson,~B.~G.; Gill,~P.~M.; Pople,~J.~A. The performance of a family of density functional methods. \emph{The Journal of Chemical Physics} \textbf{1993}, \emph{98}, 5612--5626\relax
\mciteBstWouldAddEndPuncttrue
\mciteSetBstMidEndSepPunct{\mcitedefaultmidpunct}
{\mcitedefaultendpunct}{\mcitedefaultseppunct}\relax
\EndOfBibitem
\bibitem[Murray \latin{et~al.}(1993)Murray, Handy, and Laming]{dft_integral}
Murray,~C.~W.; Handy,~N.~C.; Laming,~G.~J. Quadrature schemes for integrals of density functional theory. \emph{Molecular Physics} \textbf{1993}, \emph{78}, 997--1014\relax
\mciteBstWouldAddEndPuncttrue
\mciteSetBstMidEndSepPunct{\mcitedefaultmidpunct}
{\mcitedefaultendpunct}{\mcitedefaultseppunct}\relax
\EndOfBibitem
\bibitem[Gill \latin{et~al.}(1993)Gill, Johnson, and Pople]{dft_standard_grid_1}
Gill,~P.~M.; Johnson,~B.~G.; Pople,~J.~A. A standard grid for density functional calculations. \emph{Chemical Physics Letters} \textbf{1993}, \emph{209}, 506--512\relax
\mciteBstWouldAddEndPuncttrue
\mciteSetBstMidEndSepPunct{\mcitedefaultmidpunct}
{\mcitedefaultendpunct}{\mcitedefaultseppunct}\relax
\EndOfBibitem
\bibitem[Dasgupta and Herbert(2017)Dasgupta, and Herbert]{dft_standard_grid_2_3}
Dasgupta,~S.; Herbert,~J.~M. Standard grids for high-precision integration of modern density functionals: SG-2 and SG-3. \emph{Journal of Computational Chemistry} \textbf{2017}, \emph{38}, 869--882\relax
\mciteBstWouldAddEndPuncttrue
\mciteSetBstMidEndSepPunct{\mcitedefaultmidpunct}
{\mcitedefaultendpunct}{\mcitedefaultseppunct}\relax
\EndOfBibitem
\bibitem[Chien and Gill(2006)Chien, and Gill]{dft_standard_grid_0}
Chien,~S.-H.; Gill,~P.~M. SG-0: a small standard grid for DFT quadrature on large systems. \emph{Journal of Computational Chemistry} \textbf{2006}, \emph{27}, 730--739\relax
\mciteBstWouldAddEndPuncttrue
\mciteSetBstMidEndSepPunct{\mcitedefaultmidpunct}
{\mcitedefaultendpunct}{\mcitedefaultseppunct}\relax
\EndOfBibitem
\bibitem[Treutler and Ahlrichs(1995)Treutler, and Ahlrichs]{dft_radial_grid_1}
Treutler,~O.; Ahlrichs,~R. Efficient molecular numerical integration schemes. \emph{The Journal of Chemical Physics} \textbf{1995}, \emph{102}, 346--354\relax
\mciteBstWouldAddEndPuncttrue
\mciteSetBstMidEndSepPunct{\mcitedefaultmidpunct}
{\mcitedefaultendpunct}{\mcitedefaultseppunct}\relax
\EndOfBibitem
\bibitem[Mura and Knowles(1996)Mura, and Knowles]{dft_radial_grid_2}
Mura,~M.~E.; Knowles,~P.~J. Improved radial grids for quadrature in molecular density-functional calculations. \emph{The Journal of Chemical Physics} \textbf{1996}, \emph{104}, 9848--9858\relax
\mciteBstWouldAddEndPuncttrue
\mciteSetBstMidEndSepPunct{\mcitedefaultmidpunct}
{\mcitedefaultendpunct}{\mcitedefaultseppunct}\relax
\EndOfBibitem
\bibitem[Lindh \latin{et~al.}(2001)Lindh, Malmqvist, and Gagliardi]{dft_radial_grid_3}
Lindh,~R.; Malmqvist,~P.-{\AA}.; Gagliardi,~L. Molecular integrals by numerical quadrature. I. Radial integration. \emph{Theoretical Chemistry Accounts} \textbf{2001}, \emph{106}, 178--187\relax
\mciteBstWouldAddEndPuncttrue
\mciteSetBstMidEndSepPunct{\mcitedefaultmidpunct}
{\mcitedefaultendpunct}{\mcitedefaultseppunct}\relax
\EndOfBibitem
\bibitem[Gill and Chien(2003)Gill, and Chien]{dft_radial_grid_4}
Gill,~P.~M.; Chien,~S.-H. Radial quadrature for multiexponential integrands. \emph{Journal of Computational Chemistry} \textbf{2003}, \emph{24}, 732--740\relax
\mciteBstWouldAddEndPuncttrue
\mciteSetBstMidEndSepPunct{\mcitedefaultmidpunct}
{\mcitedefaultendpunct}{\mcitedefaultseppunct}\relax
\EndOfBibitem
\bibitem[Mitani(2011)]{dft_radial_grid_5}
Mitani,~M. An application of double exponential formula to radial quadrature grid in density functional calculation. \emph{Theoretical Chemistry Accounts} \textbf{2011}, \emph{130}, 645--669\relax
\mciteBstWouldAddEndPuncttrue
\mciteSetBstMidEndSepPunct{\mcitedefaultmidpunct}
{\mcitedefaultendpunct}{\mcitedefaultseppunct}\relax
\EndOfBibitem
\bibitem[Mitani and Yoshioka(2012)Mitani, and Yoshioka]{dft_radial_grid_6}
Mitani,~M.; Yoshioka,~Y. Numerical integration of atomic electron density with double exponential formula for density functional calculation. \emph{Theoretical Chemistry Accounts} \textbf{2012}, \emph{131}, 1--15\relax
\mciteBstWouldAddEndPuncttrue
\mciteSetBstMidEndSepPunct{\mcitedefaultmidpunct}
{\mcitedefaultendpunct}{\mcitedefaultseppunct}\relax
\EndOfBibitem
\bibitem[Pulay(1969)]{pulay_hf_derivative}
Pulay,~P. Ab initio calculation of force constants and equilibrium geometries in polyatomic molecules: I. Theory. \emph{Molecular Physics} \textbf{1969}, \emph{17}, 197--204\relax
\mciteBstWouldAddEndPuncttrue
\mciteSetBstMidEndSepPunct{\mcitedefaultmidpunct}
{\mcitedefaultendpunct}{\mcitedefaultseppunct}\relax
\EndOfBibitem
\bibitem[Pople \latin{et~al.}(1979)Pople, Krishnan, Schlegel, and Binkley]{pople_hf_derivative}
Pople,~J.; Krishnan,~R.; Schlegel,~H.; Binkley,~J.~S. Derivative studies in hartree-fock and m{\o}ller-plesset theories. \emph{International Journal of Quantum Chemistry} \textbf{1979}, \emph{16}, 225--241\relax
\mciteBstWouldAddEndPuncttrue
\mciteSetBstMidEndSepPunct{\mcitedefaultmidpunct}
{\mcitedefaultendpunct}{\mcitedefaultseppunct}\relax
\EndOfBibitem
\bibitem[Meurer \latin{et~al.}(2017)Meurer, Smith, Paprocki, {\v{C}}ert{\'\i}k, Kirpichev, Rocklin, Kumar, Ivanov, Moore, Singh, \latin{et~al.} others]{sympy}
Meurer,~A.; Smith,~C.~P.; Paprocki,~M.; {\v{C}}ert{\'\i}k,~O.; Kirpichev,~S.~B.; Rocklin,~M.; Kumar,~A.; Ivanov,~S.; Moore,~J.~K.; Singh,~S. \latin{et~al.}  SymPy: symbolic computing in Python. \emph{PeerJ Computer Science} \textbf{2017}, \emph{3}, e103\relax
\mciteBstWouldAddEndPuncttrue
\mciteSetBstMidEndSepPunct{\mcitedefaultmidpunct}
{\mcitedefaultendpunct}{\mcitedefaultseppunct}\relax
\EndOfBibitem
\bibitem[Head-Gordon and Pople(1988)Head-Gordon, and Pople]{head_gordon_pople_original}
Head-Gordon,~M.; Pople,~J.~A. A method for two-electron Gaussian integral and integral derivative evaluation using recurrence relations. \emph{The Journal of Chemical Physics} \textbf{1988}, \emph{89}, 5777--5786\relax
\mciteBstWouldAddEndPuncttrue
\mciteSetBstMidEndSepPunct{\mcitedefaultmidpunct}
{\mcitedefaultendpunct}{\mcitedefaultseppunct}\relax
\EndOfBibitem
\bibitem[Barca and Gill(2016)Barca, and Gill]{head_gordon_pople_recursion_net}
Barca,~G.~M.; Gill,~P.~M. Two-electron integrals over gaussian geminals. \emph{Journal of Chemical Theory and Computation} \textbf{2016}, \emph{12}, 4915--4924\relax
\mciteBstWouldAddEndPuncttrue
\mciteSetBstMidEndSepPunct{\mcitedefaultmidpunct}
{\mcitedefaultendpunct}{\mcitedefaultseppunct}\relax
\EndOfBibitem
\bibitem[Hosangadi \latin{et~al.}(2006)Hosangadi, Fallah, and Kastner]{cse_polynomial}
Hosangadi,~A.; Fallah,~F.; Kastner,~R. Optimizing polynomial expressions by algebraic factorization and common subexpression elimination. \emph{IEEE Transactions on Computer-Aided Design of Integrated Circuits and Systems} \textbf{2006}, \emph{25}, 2012--2022\relax
\mciteBstWouldAddEndPuncttrue
\mciteSetBstMidEndSepPunct{\mcitedefaultmidpunct}
{\mcitedefaultendpunct}{\mcitedefaultseppunct}\relax
\EndOfBibitem
\bibitem[King(2019)]{cse_black_hole}
King,~R. Common Sub-Expression Elimination Using Subtree Isomorphisms. Bachelor's Thesis in Computer Science, University of Utah, 2019\relax
\mciteBstWouldAddEndPuncttrue
\mciteSetBstMidEndSepPunct{\mcitedefaultmidpunct}
{\mcitedefaultendpunct}{\mcitedefaultseppunct}\relax
\EndOfBibitem
\bibitem[Johnson \latin{et~al.}(1991)Johnson, Gill, and Pople]{McMurchie_Davidson_tree_search}
Johnson,~B.~G.; Gill,~P.~M.; Pople,~J.~A. Exact and approximate solutions to the one-center McMurchie--Davidson tree-search problem. \emph{International Journal of Quantum Chemistry} \textbf{1991}, \emph{40}, 809--827\relax
\mciteBstWouldAddEndPuncttrue
\mciteSetBstMidEndSepPunct{\mcitedefaultmidpunct}
{\mcitedefaultendpunct}{\mcitedefaultseppunct}\relax
\EndOfBibitem
\bibitem[Smith \latin{et~al.}(2020)Smith, Burns, Simmonett, Parrish, Schieber, Galvelis, Kraus, Kruse, Di~Remigio, Alenaizan, \latin{et~al.} others]{psi4_recent}
Smith,~D.~G.; Burns,~L.~A.; Simmonett,~A.~C.; Parrish,~R.~M.; Schieber,~M.~C.; Galvelis,~R.; Kraus,~P.; Kruse,~H.; Di~Remigio,~R.; Alenaizan,~A. \latin{et~al.}  PSI4 1.4: Open-source software for high-throughput quantum chemistry. \emph{The Journal of Chemical Physics} \textbf{2020}, \emph{152}, 184108\relax
\mciteBstWouldAddEndPuncttrue
\mciteSetBstMidEndSepPunct{\mcitedefaultmidpunct}
{\mcitedefaultendpunct}{\mcitedefaultseppunct}\relax
\EndOfBibitem
\bibitem[Sun \latin{et~al.}(2020)Sun, Zhang, Banerjee, Bao, Barbry, Blunt, Bogdanov, Booth, Chen, Cui, \latin{et~al.} others]{pyscf_recent}
Sun,~Q.; Zhang,~X.; Banerjee,~S.; Bao,~P.; Barbry,~M.; Blunt,~N.~S.; Bogdanov,~N.~A.; Booth,~G.~H.; Chen,~J.; Cui,~Z.-H. \latin{et~al.}  Recent developments in the PySCF program package. \emph{The Journal of Chemical Physics} \textbf{2020}, \emph{153}, 024109\relax
\mciteBstWouldAddEndPuncttrue
\mciteSetBstMidEndSepPunct{\mcitedefaultmidpunct}
{\mcitedefaultendpunct}{\mcitedefaultseppunct}\relax
\EndOfBibitem
\bibitem[Luehr \latin{et~al.}(2011)Luehr, Ufimtsev, and Martinez]{terachem_dynamic_precision}
Luehr,~N.; Ufimtsev,~I.~S.; Martinez,~T.~J. Dynamic precision for electron repulsion integral evaluation on graphical processing units (GPUs). \emph{Journal of Chemical Theory and Computation} \textbf{2011}, \emph{7}, 949--954\relax
\mciteBstWouldAddEndPuncttrue
\mciteSetBstMidEndSepPunct{\mcitedefaultmidpunct}
{\mcitedefaultendpunct}{\mcitedefaultseppunct}\relax
\EndOfBibitem
\bibitem[Kussmann and Ochsenfeld(2013)Kussmann, and Ochsenfeld]{fermions_linear_scaling_exchange}
Kussmann,~J.; Ochsenfeld,~C. Pre-selective screening for matrix elements in linear-scaling exact exchange calculations. \emph{The Journal of Chemical Physics} \textbf{2013}, \emph{138}, 134114\relax
\mciteBstWouldAddEndPuncttrue
\mciteSetBstMidEndSepPunct{\mcitedefaultmidpunct}
{\mcitedefaultendpunct}{\mcitedefaultseppunct}\relax
\EndOfBibitem
\bibitem[Kussmann and Ochsenfeld(2015)Kussmann, and Ochsenfeld]{fermions_linear_scaling_gradient}
Kussmann,~J.; Ochsenfeld,~C. Preselective screening for linear-scaling exact exchange-gradient calculations for graphics processing units and general strong-scaling massively parallel calculations. \emph{Journal of Chemical Theory and Computation} \textbf{2015}, \emph{11}, 918--922\relax
\mciteBstWouldAddEndPuncttrue
\mciteSetBstMidEndSepPunct{\mcitedefaultmidpunct}
{\mcitedefaultendpunct}{\mcitedefaultseppunct}\relax
\EndOfBibitem
\bibitem[Kussmann \latin{et~al.}(2021)Kussmann, Laqua, and Ochsenfeld]{fermions_density_fitting}
Kussmann,~J.; Laqua,~H.; Ochsenfeld,~C. Highly efficient resolution-of-identity density functional theory calculations on central and graphics processing units. \emph{Journal of Chemical Theory and Computation} \textbf{2021}, \emph{17}, 1512--1521\relax
\mciteBstWouldAddEndPuncttrue
\mciteSetBstMidEndSepPunct{\mcitedefaultmidpunct}
{\mcitedefaultendpunct}{\mcitedefaultseppunct}\relax
\EndOfBibitem
\bibitem[Miao and Merz~Jr(2013)Miao, and Merz~Jr]{quick}
Miao,~Y.; Merz~Jr,~K.~M. Acceleration of electron repulsion integral evaluation on graphics processing units via use of recurrence relations. \emph{Journal of Chemical Theory and Computation} \textbf{2013}, \emph{9}, 965--976\relax
\mciteBstWouldAddEndPuncttrue
\mciteSetBstMidEndSepPunct{\mcitedefaultmidpunct}
{\mcitedefaultendpunct}{\mcitedefaultseppunct}\relax
\EndOfBibitem
\bibitem[Fernandes \latin{et~al.}(2015)Fernandes, Renison, and Naidoo]{quantum_supercharger_library}
Fernandes,~K.~D.; Renison,~C.~A.; Naidoo,~K.~J. Quantum supercharger library: Hyper-parallelism of the Hartree--Fock method. \emph{Journal of Computational Chemistry} \textbf{2015}, \emph{36}, 1399--1409\relax
\mciteBstWouldAddEndPuncttrue
\mciteSetBstMidEndSepPunct{\mcitedefaultmidpunct}
{\mcitedefaultendpunct}{\mcitedefaultseppunct}\relax
\EndOfBibitem
\bibitem[Qi \latin{et~al.}(2023)Qi, Zhang, and Yang]{wuhan_electronic_structure_package}
Qi,~J.; Zhang,~Y.; Yang,~M. A hybrid CPU/GPU method for Hartree--Fock self-consistent-field calculation. \emph{The Journal of Chemical Physics} \textbf{2023}, \emph{159}, 104101\relax
\mciteBstWouldAddEndPuncttrue
\mciteSetBstMidEndSepPunct{\mcitedefaultmidpunct}
{\mcitedefaultendpunct}{\mcitedefaultseppunct}\relax
\EndOfBibitem
\bibitem[R{\'a}k and Cserey(2015)R{\'a}k, and Cserey]{brush_algorithm}
R{\'a}k,~{\'A}.; Cserey,~G. The BRUSH algorithm for two-electron integrals on GPU. \emph{Chemical Physics Letters} \textbf{2015}, \emph{622}, 92--98\relax
\mciteBstWouldAddEndPuncttrue
\mciteSetBstMidEndSepPunct{\mcitedefaultmidpunct}
{\mcitedefaultendpunct}{\mcitedefaultseppunct}\relax
\EndOfBibitem
\bibitem[G{\"o}tz \latin{et~al.}(2010)G{\"o}tz, W{\"o}lfle, and Walker]{gpu_chemistry_review_2010}
G{\"o}tz,~A.~W.; W{\"o}lfle,~T.; Walker,~R.~C. Quantum chemistry on graphics processing units. In \emph{Annual Reports in Computational Chemistry}; Elsevier, 2010; Vol.~6; pp 21--35\relax
\mciteBstWouldAddEndPuncttrue
\mciteSetBstMidEndSepPunct{\mcitedefaultmidpunct}
{\mcitedefaultendpunct}{\mcitedefaultseppunct}\relax
\EndOfBibitem
\bibitem[Gordon \latin{et~al.}(2020)Gordon, Barca, Leang, Poole, Rendell, Galvez~Vallejo, and Westheimer]{packages_on_hardwares}
Gordon,~M.~S.; Barca,~G.; Leang,~S.~S.; Poole,~D.; Rendell,~A.~P.; Galvez~Vallejo,~J.~L.; Westheimer,~B. Novel computer architectures and quantum chemistry. \emph{Journal of Physical Chemistry A} \textbf{2020}, \emph{124}, 4557--4582\relax
\mciteBstWouldAddEndPuncttrue
\mciteSetBstMidEndSepPunct{\mcitedefaultmidpunct}
{\mcitedefaultendpunct}{\mcitedefaultseppunct}\relax
\EndOfBibitem
\bibitem[Williams-Young \latin{et~al.}(2023)Williams-Young, Asadchev, Popovici, Clark, Waldrop, Windus, Valeev, and de~Jong]{gauxc}
Williams-Young,~D.~B.; Asadchev,~A.; Popovici,~D.~T.; Clark,~D.; Waldrop,~J.; Windus,~T.~L.; Valeev,~E.~F.; de~Jong,~W.~A. Distributed memory, GPU accelerated Fock construction for hybrid, Gaussian basis density functional theory. \emph{The Journal of Chemical Physics} \textbf{2023}, \emph{158}\relax
\mciteBstWouldAddEndPuncttrue
\mciteSetBstMidEndSepPunct{\mcitedefaultmidpunct}
{\mcitedefaultendpunct}{\mcitedefaultseppunct}\relax
\EndOfBibitem
\bibitem[Whaley and Dongarra(1998)Whaley, and Dongarra]{atlas_linalg}
Whaley,~R.; Dongarra,~J. Automatically Tuned Linear Algebra Software. SC '98: Proceedings of the 1998 ACM/IEEE Conference on Supercomputing. 1998; p~38\relax
\mciteBstWouldAddEndPuncttrue
\mciteSetBstMidEndSepPunct{\mcitedefaultmidpunct}
{\mcitedefaultendpunct}{\mcitedefaultseppunct}\relax
\EndOfBibitem
\bibitem[Frigo and Johnson(2005)Frigo, and Johnson]{fftw}
Frigo,~M.; Johnson,~S.~G. The design and implementation of FFTW3. \emph{Proceedings of the IEEE} \textbf{2005}, \emph{93}, 216\relax
\mciteBstWouldAddEndPuncttrue
\mciteSetBstMidEndSepPunct{\mcitedefaultmidpunct}
{\mcitedefaultendpunct}{\mcitedefaultseppunct}\relax
\EndOfBibitem
\bibitem[Song \latin{et~al.}(2016)Song, Wang, and Martinez]{ace_gpu}
Song,~C.; Wang,~L.-P.; Martinez,~T. Automated code engine for graphical processing units: Application to the effective core potential integrals and gradients. \emph{Journal of Chemical Theory and Computation} \textbf{2016}, \emph{12}, 92--106\relax
\mciteBstWouldAddEndPuncttrue
\mciteSetBstMidEndSepPunct{\mcitedefaultmidpunct}
{\mcitedefaultendpunct}{\mcitedefaultseppunct}\relax
\EndOfBibitem
\bibitem[Isborn \latin{et~al.}(2012)Isborn, Gotz, Clark, Walker, and Mart{\'\i}nez]{pyp_qm_region}
Isborn,~C.~M.; Gotz,~A.~W.; Clark,~M.~A.; Walker,~R.~C.; Mart{\'\i}nez,~T.~J. Electronic absorption spectra from MM and ab initio QM/MM molecular dynamics: Environmental effects on the absorption spectrum of photoactive yellow protein. \emph{Journal of Chemical Theory and Computation} \textbf{2012}, \emph{8}, 5092--5106\relax
\mciteBstWouldAddEndPuncttrue
\mciteSetBstMidEndSepPunct{\mcitedefaultmidpunct}
{\mcitedefaultendpunct}{\mcitedefaultseppunct}\relax
\EndOfBibitem
\bibitem[Becke(1993)]{b3lyp_original}
Becke,~A.~D. Density-functional thermochemistry. III. The role of exact exchange. \emph{The Journal of Chemical Physics} \textbf{1993}, \emph{98}, 5648--6\relax
\mciteBstWouldAddEndPuncttrue
\mciteSetBstMidEndSepPunct{\mcitedefaultmidpunct}
{\mcitedefaultendpunct}{\mcitedefaultseppunct}\relax
\EndOfBibitem
\bibitem[Stephens \latin{et~al.}(1994)Stephens, Devlin, Chabalowski, and Frisch]{b3lyp_parameter}
Stephens,~P.~J.; Devlin,~F.~J.; Chabalowski,~C.~F.; Frisch,~M.~J. Ab initio calculation of vibrational absorption and circular dichroism spectra using density functional force fields. \emph{Journal of Physical Chemistry} \textbf{1994}, \emph{98}, 11623--11627\relax
\mciteBstWouldAddEndPuncttrue
\mciteSetBstMidEndSepPunct{\mcitedefaultmidpunct}
{\mcitedefaultendpunct}{\mcitedefaultseppunct}\relax
\EndOfBibitem
\bibitem[Ditchfield \latin{et~al.}(1971)Ditchfield, Hehre, and Pople]{6_31g_h}
Ditchfield,~R.; Hehre,~W.~J.; Pople,~J.~A. Self-consistent molecular-orbital methods. IX. An extended Gaussian-type basis for molecular-orbital studies of organic molecules. \emph{The Journal of Chemical Physics} \textbf{1971}, \emph{54}, 724--728\relax
\mciteBstWouldAddEndPuncttrue
\mciteSetBstMidEndSepPunct{\mcitedefaultmidpunct}
{\mcitedefaultendpunct}{\mcitedefaultseppunct}\relax
\EndOfBibitem
\bibitem[Hehre \latin{et~al.}(1972)Hehre, Ditchfield, and Pople]{6_31g}
Hehre,~W.~J.; Ditchfield,~R.; Pople,~J.~A. Self—consistent molecular orbital methods. XII. Further extensions of Gaussian—type basis sets for use in molecular orbital studies of organic molecules. \emph{The Journal of Chemical Physics} \textbf{1972}, \emph{56}, 2257--2261\relax
\mciteBstWouldAddEndPuncttrue
\mciteSetBstMidEndSepPunct{\mcitedefaultmidpunct}
{\mcitedefaultendpunct}{\mcitedefaultseppunct}\relax
\EndOfBibitem
\bibitem[Hashimoto \latin{et~al.}(1995)Hashimoto, Hirao, and Tatewaki]{hashimoto_ccpvxz_optseg}
Hashimoto,~T.; Hirao,~K.; Tatewaki,~H. Comment on Dunning's correlation-consistent basis sets. \emph{Chemical Physics Letters} \textbf{1995}, \emph{243}, 190--192\relax
\mciteBstWouldAddEndPuncttrue
\mciteSetBstMidEndSepPunct{\mcitedefaultmidpunct}
{\mcitedefaultendpunct}{\mcitedefaultseppunct}\relax
\EndOfBibitem
\bibitem[Davidson(1996)]{davidson_ccpvxz_optseg}
Davidson,~E.~R. Comment on “Comment on Dunning's correlation-consistent basis sets”. \emph{Chemical Physics Letters} \textbf{1996}, \emph{260}, 514--518\relax
\mciteBstWouldAddEndPuncttrue
\mciteSetBstMidEndSepPunct{\mcitedefaultmidpunct}
{\mcitedefaultendpunct}{\mcitedefaultseppunct}\relax
\EndOfBibitem
\bibitem[Frisch \latin{et~al.}(2016)Frisch, Trucks, Schlegel, Scuseria, Robb, Cheeseman, Scalmani, Barone, Petersson, Nakatsuji, Li, Caricato, Marenich, Bloino, Janesko, Gomperts, Mennucci, Hratchian, Ortiz, Izmaylov, Sonnenberg, Williams-Young, Ding, Lipparini, Egidi, Goings, Peng, Petrone, Henderson, Ranasinghe, Zakrzewski, Gao, Rega, Zheng, Liang, Hada, Ehara, Toyota, Fukuda, Hasegawa, Ishida, Nakajima, Honda, Kitao, Nakai, Vreven, Throssell, Montgomery, Peralta, Ogliaro, Bearpark, Heyd, Brothers, Kudin, Staroverov, Keith, Kobayashi, Normand, Raghavachari, Rendell, Burant, Iyengar, Tomasi, Cossi, Millam, Klene, Adamo, Cammi, Ochterski, Martin, Morokuma, Farkas, Foresman, and Fox]{gaussian16}
Frisch,~M.~J.; Trucks,~G.~W.; Schlegel,~H.~B.; Scuseria,~G.~E.; Robb,~M.~A.; Cheeseman,~J.~R.; Scalmani,~G.; Barone,~V.; Petersson,~G.~A.; Nakatsuji,~H. \latin{et~al.}  Gaussian˜16 {R}evision {C}.01. 2016; Gaussian Inc. Wallingford CT\relax
\mciteBstWouldAddEndPuncttrue
\mciteSetBstMidEndSepPunct{\mcitedefaultmidpunct}
{\mcitedefaultendpunct}{\mcitedefaultseppunct}\relax
\EndOfBibitem
\bibitem[Van~Lenthe \latin{et~al.}(2006)Van~Lenthe, Zwaans, Van~Dam, and Guest]{sadguess}
Van~Lenthe,~J.; Zwaans,~R.; Van~Dam,~H.~J.; Guest,~M. Starting SCF calculations by superposition of atomic densities. \emph{Journal of Computational Chemistry} \textbf{2006}, \emph{27}, 926--932\relax
\mciteBstWouldAddEndPuncttrue
\mciteSetBstMidEndSepPunct{\mcitedefaultmidpunct}
{\mcitedefaultendpunct}{\mcitedefaultseppunct}\relax
\EndOfBibitem
\bibitem[Wang \latin{et~al.}(2024)Wang, Hait, Johnson, Fajen, Guerrero, and Martinez]{wang_2024_13328235}
Wang,~Y.; Hait,~D.; Johnson,~K.~G.; Fajen,~O.~J.; Guerrero,~R.~D.; Martinez,~T.~J. {Supporting Data for "Extending GPU-Accelerated Gaussian Integrals in the TeraChem Software Package to f Type Orbitals: Implementation and Applications."}. 2024; \url{https://doi.org/10.5281/zenodo.13328235}\relax
\mciteBstWouldAddEndPuncttrue
\mciteSetBstMidEndSepPunct{\mcitedefaultmidpunct}
{\mcitedefaultendpunct}{\mcitedefaultseppunct}\relax
\EndOfBibitem
\bibitem[Zhang \latin{et~al.}(2015)Zhang, El-Roz, Frei, Mendoza-Cortes, Head-Gordon, Lacy, and Peters]{tetraaza_complex}
Zhang,~M.; El-Roz,~M.; Frei,~H.; Mendoza-Cortes,~J.~L.; Head-Gordon,~M.; Lacy,~D.~C.; Peters,~J.~C. Visible light sensitized CO2 activation by the tetraaza [CoIIN4H (MeCN)] 2+ complex investigated by FT-IR spectroscopy and DFT calculations. \emph{The Journal of Physical Chemistry C} \textbf{2015}, \emph{119}, 4645--4654\relax
\mciteBstWouldAddEndPuncttrue
\mciteSetBstMidEndSepPunct{\mcitedefaultmidpunct}
{\mcitedefaultendpunct}{\mcitedefaultseppunct}\relax
\EndOfBibitem
\bibitem[Ziegler \latin{et~al.}(2018)Ziegler, Torquato, Levine, Nicolay, Celik, and Tilley]{dicopper_complex}
Ziegler,~M.~S.; Torquato,~N.~A.; Levine,~D.~S.; Nicolay,~A.; Celik,~H.; Tilley,~T.~D. Dicopper alkyl complexes: synthesis, structure, and unexpected persistence. \emph{Organometallics} \textbf{2018}, \emph{37}, 2807--2823\relax
\mciteBstWouldAddEndPuncttrue
\mciteSetBstMidEndSepPunct{\mcitedefaultmidpunct}
{\mcitedefaultendpunct}{\mcitedefaultseppunct}\relax
\EndOfBibitem
\bibitem[Li \latin{et~al.}(1999)Li, Eddaoudi, O'Keeffe, and Yaghi]{mof5_original}
Li,~H.; Eddaoudi,~M.; O'Keeffe,~M.; Yaghi,~O.~M. Design and synthesis of an exceptionally stable and highly porous metal-organic framework. \emph{Nature} \textbf{1999}, \emph{402}, 276--279\relax
\mciteBstWouldAddEndPuncttrue
\mciteSetBstMidEndSepPunct{\mcitedefaultmidpunct}
{\mcitedefaultendpunct}{\mcitedefaultseppunct}\relax
\EndOfBibitem
\bibitem[Spicher \latin{et~al.}(2020)Spicher, Bursch, and Grimme]{mof5_geometry}
Spicher,~S.; Bursch,~M.; Grimme,~S. Efficient calculation of small molecule binding in metal--organic frameworks and porous organic cages. \emph{The Journal of Physical Chemistry C} \textbf{2020}, \emph{124}, 27529--27541\relax
\mciteBstWouldAddEndPuncttrue
\mciteSetBstMidEndSepPunct{\mcitedefaultmidpunct}
{\mcitedefaultendpunct}{\mcitedefaultseppunct}\relax
\EndOfBibitem
\bibitem[Weigend and Ahlrichs(2005)Weigend, and Ahlrichs]{def2_svp}
Weigend,~F.; Ahlrichs,~R. Balanced basis sets of split valence, triple zeta valence and quadruple zeta valence quality for H to Rn: Design and assessment of accuracy. \emph{Physical Chemistry Chemical Physics} \textbf{2005}, \emph{7}, 3297--3305\relax
\mciteBstWouldAddEndPuncttrue
\mciteSetBstMidEndSepPunct{\mcitedefaultmidpunct}
{\mcitedefaultendpunct}{\mcitedefaultseppunct}\relax
\EndOfBibitem
\bibitem[Nicolaou \latin{et~al.}(1994)Nicolaou, Yang, Liu, Ueno, Nantermet, Guy, Claiborne, Renaud, Couladouros, Paulvannan, \latin{et~al.} others]{nicolaou_taxol}
Nicolaou,~K.; Yang,~Z.; Liu,~J.; Ueno,~H.; Nantermet,~P.; Guy,~R.; Claiborne,~C.; Renaud,~J.; Couladouros,~E.; Paulvannan,~K. \latin{et~al.}  Total synthesis of taxol. \emph{Nature} \textbf{1994}, \emph{367}, 630--634\relax
\mciteBstWouldAddEndPuncttrue
\mciteSetBstMidEndSepPunct{\mcitedefaultmidpunct}
{\mcitedefaultendpunct}{\mcitedefaultseppunct}\relax
\EndOfBibitem
\bibitem[Mulligan \latin{et~al.}(2020)Mulligan, Kang, Sawaya, Rettie, Li, Antselovich, Craven, Watkins, Labonte, DiMaio, \latin{et~al.} others]{protein_6ufa}
Mulligan,~V.~K.; Kang,~C.~S.; Sawaya,~M.~R.; Rettie,~S.; Li,~X.; Antselovich,~I.; Craven,~T.~W.; Watkins,~A.~M.; Labonte,~J.~W.; DiMaio,~F. \latin{et~al.}  Computational design of mixed chirality peptide macrocycles with internal symmetry. \emph{Protein Science} \textbf{2020}, \emph{29}, 2433--2445\relax
\mciteBstWouldAddEndPuncttrue
\mciteSetBstMidEndSepPunct{\mcitedefaultmidpunct}
{\mcitedefaultendpunct}{\mcitedefaultseppunct}\relax
\EndOfBibitem
\bibitem[Adamo and Barone(1999)Adamo, and Barone]{pbe0}
Adamo,~C.; Barone,~V. Toward reliable density functional methods without adjustable parameters: The PBE0 model. \emph{The Journal of Chemical Physics} \textbf{1999}, \emph{110}, 6158--6170\relax
\mciteBstWouldAddEndPuncttrue
\mciteSetBstMidEndSepPunct{\mcitedefaultmidpunct}
{\mcitedefaultendpunct}{\mcitedefaultseppunct}\relax
\EndOfBibitem
\bibitem[Tian \latin{et~al.}(2019)Tian, Kasavajhala, Belfon, Raguette, Huang, Migues, Bickel, Wang, Pincay, Wu, \latin{et~al.} others]{amber_protein_force_field}
Tian,~C.; Kasavajhala,~K.; Belfon,~K.~A.; Raguette,~L.; Huang,~H.; Migues,~A.~N.; Bickel,~J.; Wang,~Y.; Pincay,~J.; Wu,~Q. \latin{et~al.}  ff19SB: Amino-acid-specific protein backbone parameters trained against quantum mechanics energy surfaces in solution. \emph{Journal of Chemical Theory and Computation} \textbf{2019}, \emph{16}, 528--552\relax
\mciteBstWouldAddEndPuncttrue
\mciteSetBstMidEndSepPunct{\mcitedefaultmidpunct}
{\mcitedefaultendpunct}{\mcitedefaultseppunct}\relax
\EndOfBibitem
\bibitem[Case \latin{et~al.}(2023)Case, Aktulga, Belfon, Cerutti, Cisneros, Cruzeiro, Forouzesh, Giese, Gotz, Gohlke, \latin{et~al.} others]{amber_2023}
Case,~D.~A.; Aktulga,~H.~M.; Belfon,~K.; Cerutti,~D.~S.; Cisneros,~G.~A.; Cruzeiro,~V. W.~D.; Forouzesh,~N.; Giese,~T.~J.; Gotz,~A.~W.; Gohlke,~H. \latin{et~al.}  AmberTools. \emph{Journal of chemical information and modeling} \textbf{2023}, \emph{63}, 6183--6191\relax
\mciteBstWouldAddEndPuncttrue
\mciteSetBstMidEndSepPunct{\mcitedefaultmidpunct}
{\mcitedefaultendpunct}{\mcitedefaultseppunct}\relax
\EndOfBibitem
\bibitem[TechPowerUp(2024)]{gpu_database}
TechPowerUp GPU Specs Database. 2024; \url{https://www.techpowerup.com/gpu-specs/}, Accessed: March 31, 2024\relax
\mciteBstWouldAddEndPuncttrue
\mciteSetBstMidEndSepPunct{\mcitedefaultmidpunct}
{\mcitedefaultendpunct}{\mcitedefaultseppunct}\relax
\EndOfBibitem
\bibitem[nsi(2024)]{nsight_compute}
NVIDIA Nsight Compute. 2024; \url{https://developer.nvidia.com/nsight-compute}, Accessed: March 31, 2024\relax
\mciteBstWouldAddEndPuncttrue
\mciteSetBstMidEndSepPunct{\mcitedefaultmidpunct}
{\mcitedefaultendpunct}{\mcitedefaultseppunct}\relax
\EndOfBibitem
\bibitem[Flocke and Lotrich(2008)Flocke, and Lotrich]{rys_quadrature}
Flocke,~N.; Lotrich,~V. Efficient electronic integrals and their generalized derivatives for object oriented implementations of electronic structure calculations. \emph{Journal of Computational Chemistry} \textbf{2008}, \emph{29}, 2722--2736\relax
\mciteBstWouldAddEndPuncttrue
\mciteSetBstMidEndSepPunct{\mcitedefaultmidpunct}
{\mcitedefaultendpunct}{\mcitedefaultseppunct}\relax
\EndOfBibitem
\bibitem[Beebe and Linderberg(1977)Beebe, and Linderberg]{beebe_simplifications_1977}
Beebe,~N.~H.; Linderberg,~J. Simplifications in the generation and transformation of two-electron integrals in molecular calculations. \emph{International Journal of Quantum Chemistry} \textbf{1977}, \emph{12}, 683--705\relax
\mciteBstWouldAddEndPuncttrue
\mciteSetBstMidEndSepPunct{\mcitedefaultmidpunct}
{\mcitedefaultendpunct}{\mcitedefaultseppunct}\relax
\EndOfBibitem
\bibitem[Jensen(2005)]{jensen2005estimating}
Jensen,~F. Estimating the Hartree—Fock limit from finite basis set calculations. \emph{Theoretical Chemistry Accounts} \textbf{2005}, \emph{113}, 267--273\relax
\mciteBstWouldAddEndPuncttrue
\mciteSetBstMidEndSepPunct{\mcitedefaultmidpunct}
{\mcitedefaultendpunct}{\mcitedefaultseppunct}\relax
\EndOfBibitem
\bibitem[Hampel and Werner(1996)Hampel, and Werner]{hampel1996local}
Hampel,~C.; Werner,~H.-J. Local treatment of electron correlation in coupled cluster theory. \emph{The Journal of Chemical Physics} \textbf{1996}, \emph{104}, 6286--6297\relax
\mciteBstWouldAddEndPuncttrue
\mciteSetBstMidEndSepPunct{\mcitedefaultmidpunct}
{\mcitedefaultendpunct}{\mcitedefaultseppunct}\relax
\EndOfBibitem
\bibitem[Kinoshita \latin{et~al.}(2003)Kinoshita, Hino, and Bartlett]{kinoshita_singular_2003}
Kinoshita,~T.; Hino,~O.; Bartlett,~R.~J. Singular value decomposition approach for the approximate coupled-cluster method. \emph{The Journal of Chemical Physics} \textbf{2003}, \emph{119}, 7756--7762\relax
\mciteBstWouldAddEndPuncttrue
\mciteSetBstMidEndSepPunct{\mcitedefaultmidpunct}
{\mcitedefaultendpunct}{\mcitedefaultseppunct}\relax
\EndOfBibitem
\bibitem[Subotnik \latin{et~al.}(2006)Subotnik, Sodt, and Head-Gordon]{subotnik2006near}
Subotnik,~J.~E.; Sodt,~A.; Head-Gordon,~M. A near linear-scaling smooth local coupled cluster algorithm for electronic structure. \emph{The Journal of Chemical Physics} \textbf{2006}, \emph{125}\relax
\mciteBstWouldAddEndPuncttrue
\mciteSetBstMidEndSepPunct{\mcitedefaultmidpunct}
{\mcitedefaultendpunct}{\mcitedefaultseppunct}\relax
\EndOfBibitem
\bibitem[Neese \latin{et~al.}(2009)Neese, Hansen, Wennmohs, and Grimme]{neese_accurate_2009}
Neese,~F.; Hansen,~A.; Wennmohs,~F.; Grimme,~S. Accurate theoretical chemistry with coupled pair models. \emph{Accounts of Chemical Research} \textbf{2009}, \emph{42}, 641--648\relax
\mciteBstWouldAddEndPuncttrue
\mciteSetBstMidEndSepPunct{\mcitedefaultmidpunct}
{\mcitedefaultendpunct}{\mcitedefaultseppunct}\relax
\EndOfBibitem
\bibitem[Hansen \latin{et~al.}(2011)Hansen, Liakos, and Neese]{hansen_efficient_2011}
Hansen,~A.; Liakos,~D.~G.; Neese,~F. Efficient and accurate local single reference correlation methods for high-spin open-shell molecules using pair natural orbitals. \emph{The Journal of Chemical Physics} \textbf{2011}, \emph{135}\relax
\mciteBstWouldAddEndPuncttrue
\mciteSetBstMidEndSepPunct{\mcitedefaultmidpunct}
{\mcitedefaultendpunct}{\mcitedefaultseppunct}\relax
\EndOfBibitem
\bibitem[Schutski \latin{et~al.}(2017)Schutski, Zhao, Henderson, and Scuseria]{schutski_tensor-structured_2017}
Schutski,~R.; Zhao,~J.; Henderson,~T.~M.; Scuseria,~G.~E. Tensor-structured coupled cluster theory. \emph{The Journal of Chemical Physics} \textbf{2017}, \emph{147}\relax
\mciteBstWouldAddEndPuncttrue
\mciteSetBstMidEndSepPunct{\mcitedefaultmidpunct}
{\mcitedefaultendpunct}{\mcitedefaultseppunct}\relax
\EndOfBibitem
\bibitem[Hohenstein \latin{et~al.}(2012)Hohenstein, Parrish, and Mart{\'\i}nez]{hohenstein_thc_I}
Hohenstein,~E.~G.; Parrish,~R.~M.; Mart{\'\i}nez,~T.~J. Tensor hypercontraction density fitting. I. Quartic scaling second-and third-order M{\o}ller-Plesset perturbation theory. \emph{The Journal of Chemical Physics} \textbf{2012}, \emph{137}\relax
\mciteBstWouldAddEndPuncttrue
\mciteSetBstMidEndSepPunct{\mcitedefaultmidpunct}
{\mcitedefaultendpunct}{\mcitedefaultseppunct}\relax
\EndOfBibitem
\bibitem[Parrish \latin{et~al.}(2012)Parrish, Hohenstein, Mart{\'\i}nez, and Sherrill]{parrish_thc_II}
Parrish,~R.~M.; Hohenstein,~E.~G.; Mart{\'\i}nez,~T.~J.; Sherrill,~C.~D. Tensor hypercontraction. II. Least-squares renormalization. \emph{The Journal of Chemical Physics} \textbf{2012}, \emph{137}\relax
\mciteBstWouldAddEndPuncttrue
\mciteSetBstMidEndSepPunct{\mcitedefaultmidpunct}
{\mcitedefaultendpunct}{\mcitedefaultseppunct}\relax
\EndOfBibitem
\bibitem[Hohenstein \latin{et~al.}(2012)Hohenstein, Parrish, Sherrill, and Mart{\'\i}nez]{hohenstein_thc_III}
Hohenstein,~E.~G.; Parrish,~R.~M.; Sherrill,~C.~D.; Mart{\'\i}nez,~T.~J. Communication: Tensor hypercontraction. III. Least-squares tensor hypercontraction for the determination of correlated wavefunctions. \emph{The Journal of Chemical Physics} \textbf{2012}, \emph{137}\relax
\mciteBstWouldAddEndPuncttrue
\mciteSetBstMidEndSepPunct{\mcitedefaultmidpunct}
{\mcitedefaultendpunct}{\mcitedefaultseppunct}\relax
\EndOfBibitem
\bibitem[Barbor \latin{et~al.}(2023)Barbor, Nair, Sharp, Lohrey, Dibrell, Shah, Walsh, Reisman, and Stoltz]{nickel_catalyst_experimental}
Barbor,~J.~P.; Nair,~V.~N.; Sharp,~K.~R.; Lohrey,~T.~D.; Dibrell,~S.~E.; Shah,~T.~K.; Walsh,~M.~J.; Reisman,~S.~E.; Stoltz,~B.~M. Development of a Nickel-Catalyzed N--N Coupling for the Synthesis of Hydrazides. \emph{Journal of the American Chemical Society} \textbf{2023}, \emph{145}, 15071--15077\relax
\mciteBstWouldAddEndPuncttrue
\mciteSetBstMidEndSepPunct{\mcitedefaultmidpunct}
{\mcitedefaultendpunct}{\mcitedefaultseppunct}\relax
\EndOfBibitem
\bibitem[Grimme \latin{et~al.}(2010)Grimme, Antony, Ehrlich, and Krieg]{dispersion_d3}
Grimme,~S.; Antony,~J.; Ehrlich,~S.; Krieg,~H. A consistent and accurate ab initio parametrization of density functional dispersion correction (DFT-D) for the 94 elements H-Pu. \emph{The Journal of Chemical Physics} \textbf{2010}, \emph{132}, 154104\relax
\mciteBstWouldAddEndPuncttrue
\mciteSetBstMidEndSepPunct{\mcitedefaultmidpunct}
{\mcitedefaultendpunct}{\mcitedefaultseppunct}\relax
\EndOfBibitem
\bibitem[Grimme \latin{et~al.}(2011)Grimme, Ehrlich, and Goerigk]{dispersion_d3BJ}
Grimme,~S.; Ehrlich,~S.; Goerigk,~L. Effect of the damping function in dispersion corrected density functional theory. \emph{Journal of Computational Chemistry} \textbf{2011}, \emph{32}, 1456--1465\relax
\mciteBstWouldAddEndPuncttrue
\mciteSetBstMidEndSepPunct{\mcitedefaultmidpunct}
{\mcitedefaultendpunct}{\mcitedefaultseppunct}\relax
\EndOfBibitem
\bibitem[Mardirossian and Head-Gordon(2014)Mardirossian, and Head-Gordon]{wb97_v}
Mardirossian,~N.; Head-Gordon,~M. $\omega$B97X-V: A 10-parameter, range-separated hybrid, generalized gradient approximation density functional with nonlocal correlation, designed by a survival-of-the-fittest strategy. \emph{Physical Chemistry Chemical Physics} \textbf{2014}, \emph{16}, 9904--9924\relax
\mciteBstWouldAddEndPuncttrue
\mciteSetBstMidEndSepPunct{\mcitedefaultmidpunct}
{\mcitedefaultendpunct}{\mcitedefaultseppunct}\relax
\EndOfBibitem
\bibitem[Najibi and Goerigk(2018)Najibi, and Goerigk]{wb97x_d3bj}
Najibi,~A.; Goerigk,~L. The nonlocal kernel in van der Waals density functionals as an additive correction: An extensive analysis with special emphasis on the B97M-V and $\omega$B97M-V approaches. \emph{Journal of Chemical Theory and Computation} \textbf{2018}, \emph{14}, 5725--5738\relax
\mciteBstWouldAddEndPuncttrue
\mciteSetBstMidEndSepPunct{\mcitedefaultmidpunct}
{\mcitedefaultendpunct}{\mcitedefaultseppunct}\relax
\EndOfBibitem
\bibitem[Jensen(2014)]{pcseg_1}
Jensen,~F. Unifying general and segmented contracted basis sets. Segmented polarization consistent basis sets. \emph{Journal of Chemical Theory and Computation} \textbf{2014}, \emph{10}, 1074--1085\relax
\mciteBstWouldAddEndPuncttrue
\mciteSetBstMidEndSepPunct{\mcitedefaultmidpunct}
{\mcitedefaultendpunct}{\mcitedefaultseppunct}\relax
\EndOfBibitem
\bibitem[Henkelman and J{\'o}nsson(2000)Henkelman, and J{\'o}nsson]{neb_original}
Henkelman,~G.; J{\'o}nsson,~H. Improved tangent estimate in the nudged elastic band method for finding minimum energy paths and saddle points. \emph{The Journal of Chemical Physics} \textbf{2000}, \emph{113}, 9978--9985\relax
\mciteBstWouldAddEndPuncttrue
\mciteSetBstMidEndSepPunct{\mcitedefaultmidpunct}
{\mcitedefaultendpunct}{\mcitedefaultseppunct}\relax
\EndOfBibitem
\bibitem[Kastner \latin{et~al.}(2009)Kastner, Carr, Keal, Thiel, Wander, and Sherwood]{dlfind}
Kastner,~J.; Carr,~J.~M.; Keal,~T.~W.; Thiel,~W.; Wander,~A.; Sherwood,~P. DL-FIND: an open-source geometry optimizer for atomistic simulations. \emph{Journal of Physical Chemistry A} \textbf{2009}, \emph{113}, 11856--11865\relax
\mciteBstWouldAddEndPuncttrue
\mciteSetBstMidEndSepPunct{\mcitedefaultmidpunct}
{\mcitedefaultendpunct}{\mcitedefaultseppunct}\relax
\EndOfBibitem
\bibitem[Sherwood \latin{et~al.}(2003)Sherwood, de~Vries, Guest, Schreckenbach, Catlow, French, Sokol, Bromley, Thiel, Turner, \latin{et~al.} others]{chemshell_1}
Sherwood,~P.; de~Vries,~A.~H.; Guest,~M.~F.; Schreckenbach,~G.; Catlow,~C. R.~A.; French,~S.~A.; Sokol,~A.~A.; Bromley,~S.~T.; Thiel,~W.; Turner,~A.~J. \latin{et~al.}  QUASI: A general purpose implementation of the QM/MM approach and its application to problems in catalysis. \emph{Journal of Molecular Structure: THEOCHEM} \textbf{2003}, \emph{632}, 1--28\relax
\mciteBstWouldAddEndPuncttrue
\mciteSetBstMidEndSepPunct{\mcitedefaultmidpunct}
{\mcitedefaultendpunct}{\mcitedefaultseppunct}\relax
\EndOfBibitem
\bibitem[Metz \latin{et~al.}(2014)Metz, K{\"a}stner, Sokol, Keal, and Sherwood]{chemshell_2}
Metz,~S.; K{\"a}stner,~J.; Sokol,~A.~A.; Keal,~T.~W.; Sherwood,~P. ChemShell—a modular software package for QM/MM simulations. \emph{Wiley Interdisciplinary Reviews: Computational Molecular Science} \textbf{2014}, \emph{4}, 101--110\relax
\mciteBstWouldAddEndPuncttrue
\mciteSetBstMidEndSepPunct{\mcitedefaultmidpunct}
{\mcitedefaultendpunct}{\mcitedefaultseppunct}\relax
\EndOfBibitem
\bibitem[Ben-Nun \latin{et~al.}(2000)Ben-Nun, Quenneville, and Mart{\'\i}nez]{aims_1}
Ben-Nun,~M.; Quenneville,~J.; Mart{\'\i}nez,~T.~J. Ab initio multiple spawning: Photochemistry from first principles quantum molecular dynamics. \emph{Journal of Physical Chemistry A} \textbf{2000}, \emph{104}, 5161--5175\relax
\mciteBstWouldAddEndPuncttrue
\mciteSetBstMidEndSepPunct{\mcitedefaultmidpunct}
{\mcitedefaultendpunct}{\mcitedefaultseppunct}\relax
\EndOfBibitem
\bibitem[Ben-Nun and Mart{\'\i}nez(2002)Ben-Nun, and Mart{\'\i}nez]{aims_2}
Ben-Nun,~M.; Mart{\'\i}nez,~T.~J. Ab initio quantum molecular dynamics. \emph{Advances in Chemical Physics} \textbf{2002}, \emph{121}, 439--512\relax
\mciteBstWouldAddEndPuncttrue
\mciteSetBstMidEndSepPunct{\mcitedefaultmidpunct}
{\mcitedefaultendpunct}{\mcitedefaultseppunct}\relax
\EndOfBibitem
\bibitem[Virshup \latin{et~al.}(2012)Virshup, Chen, and Mart{\'\i}nez]{aims_3}
Virshup,~A.~M.; Chen,~J.; Mart{\'\i}nez,~T.~J. Nonlinear dimensionality reduction for nonadiabatic dynamics: The influence of conical intersection topography on population transfer rates. \emph{The Journal of Chemical Physics} \textbf{2012}, \emph{137}, 22A519\relax
\mciteBstWouldAddEndPuncttrue
\mciteSetBstMidEndSepPunct{\mcitedefaultmidpunct}
{\mcitedefaultendpunct}{\mcitedefaultseppunct}\relax
\EndOfBibitem
\bibitem[Makhov \latin{et~al.}(2014)Makhov, Glover, Martinez, and Shalashilin]{aims_4}
Makhov,~D.~V.; Glover,~W.~J.; Martinez,~T.~J.; Shalashilin,~D.~V. Ab initio multiple cloning algorithm for quantum nonadiabatic molecular dynamics. \emph{The Journal of Chemical Physics} \textbf{2014}, \emph{141}, 054110\relax
\mciteBstWouldAddEndPuncttrue
\mciteSetBstMidEndSepPunct{\mcitedefaultmidpunct}
{\mcitedefaultendpunct}{\mcitedefaultseppunct}\relax
\EndOfBibitem
\bibitem[Curchod and Mart{\'\i}nez(2018)Curchod, and Mart{\'\i}nez]{aims_5}
Curchod,~B.~F.; Mart{\'\i}nez,~T.~J. Ab initio nonadiabatic quantum molecular dynamics. \emph{Chemical Reviews} \textbf{2018}, \emph{118}, 3305--3336\relax
\mciteBstWouldAddEndPuncttrue
\mciteSetBstMidEndSepPunct{\mcitedefaultmidpunct}
{\mcitedefaultendpunct}{\mcitedefaultseppunct}\relax
\EndOfBibitem
\bibitem[Snyder \latin{et~al.}(2015)Snyder, Hohenstein, Luehr, and Mart{\'\i}nez]{aims_casscf}
Snyder,~J.~W.; Hohenstein,~E.~G.; Luehr,~N.; Mart{\'\i}nez,~T.~J. An atomic orbital-based formulation of analytical gradients and nonadiabatic coupling vector elements for the state-averaged complete active space self-consistent field method on graphical processing units. \emph{The Journal of Chemical Physics} \textbf{2015}, \emph{143}, 154107\relax
\mciteBstWouldAddEndPuncttrue
\mciteSetBstMidEndSepPunct{\mcitedefaultmidpunct}
{\mcitedefaultendpunct}{\mcitedefaultseppunct}\relax
\EndOfBibitem
\bibitem[Hollas \latin{et~al.}(2018)Hollas, Sistik, Hohenstein, Martinez, and Slavicek]{aims_fomo_casci}
Hollas,~D.; Sistik,~L.; Hohenstein,~E.~G.; Martinez,~T.~J.; Slavicek,~P. Nonadiabatic ab initio molecular dynamics with the floating occupation molecular orbital-complete active space configuration interaction method. \emph{Journal of Chemical Theory and Computation} \textbf{2018}, \emph{14}, 339--350\relax
\mciteBstWouldAddEndPuncttrue
\mciteSetBstMidEndSepPunct{\mcitedefaultmidpunct}
{\mcitedefaultendpunct}{\mcitedefaultseppunct}\relax
\EndOfBibitem
\bibitem[Curchod and Mart{\'\i}nez(2018)Curchod, and Mart{\'\i}nez]{aims_review}
Curchod,~B.~F.; Mart{\'\i}nez,~T.~J. Ab initio nonadiabatic quantum molecular dynamics. \emph{Chemical Reviews} \textbf{2018}, \emph{118}, 3305--3336\relax
\mciteBstWouldAddEndPuncttrue
\mciteSetBstMidEndSepPunct{\mcitedefaultmidpunct}
{\mcitedefaultendpunct}{\mcitedefaultseppunct}\relax
\EndOfBibitem
\bibitem[Rana \latin{et~al.}(2023)Rana, Hohenstein, and Mart{\'\i}nez]{aims_application_polariton}
Rana,~B.; Hohenstein,~E.~G.; Mart{\'\i}nez,~T.~J. Simulating the Excited-State Dynamics of Polaritons with Ab Initio Multiple Spawning. \emph{Journal of Physical Chemistry A} \textbf{2023}, \emph{128}, 139--151\relax
\mciteBstWouldAddEndPuncttrue
\mciteSetBstMidEndSepPunct{\mcitedefaultmidpunct}
{\mcitedefaultendpunct}{\mcitedefaultseppunct}\relax
\EndOfBibitem
\bibitem[List \latin{et~al.}(2024)List, Jones, and Mart{\'\i}nez]{aims_application_gfp}
List,~N.~H.; Jones,~C.~M.; Mart{\'\i}nez,~T.~J. Chemical control of excited-state reactivity of the anionic green fluorescent protein chromophore. \emph{Communications Chemistry} \textbf{2024}, \emph{7}, 25\relax
\mciteBstWouldAddEndPuncttrue
\mciteSetBstMidEndSepPunct{\mcitedefaultmidpunct}
{\mcitedefaultendpunct}{\mcitedefaultseppunct}\relax
\EndOfBibitem
\bibitem[Champenois \latin{et~al.}(2023)Champenois, List, Ware, Britton, Bucksbaum, Cheng, Centurion, Cryan, Forbes, Gabalski, \latin{et~al.} others]{aims_application_ammonia_ued}
Champenois,~E.~G.; List,~N.~H.; Ware,~M.; Britton,~M.; Bucksbaum,~P.~H.; Cheng,~X.; Centurion,~M.; Cryan,~J.~P.; Forbes,~R.; Gabalski,~I. \latin{et~al.}  Femtosecond electronic and hydrogen structural dynamics in ammonia imaged with ultrafast electron diffraction. \emph{Physical Review Letters} \textbf{2023}, \emph{131}, 143001\relax
\mciteBstWouldAddEndPuncttrue
\mciteSetBstMidEndSepPunct{\mcitedefaultmidpunct}
{\mcitedefaultendpunct}{\mcitedefaultseppunct}\relax
\EndOfBibitem
\bibitem[Liu \latin{et~al.}(2023)Liu, Sanchez, Ware, Champenois, Yang, Nunes, Attar, Centurion, Cryan, Forbes, \latin{et~al.} others]{aims_application_electrocyclic}
Liu,~Y.; Sanchez,~D.~M.; Ware,~M.~R.; Champenois,~E.~G.; Yang,~J.; Nunes,~J. P.~F.; Attar,~A.; Centurion,~M.; Cryan,~J.~P.; Forbes,~R. \latin{et~al.}  Rehybridization dynamics into the pericyclic minimum of an electrocyclic reaction imaged in real-time. \emph{Nature Communications} \textbf{2023}, \emph{14}, 2795\relax
\mciteBstWouldAddEndPuncttrue
\mciteSetBstMidEndSepPunct{\mcitedefaultmidpunct}
{\mcitedefaultendpunct}{\mcitedefaultseppunct}\relax
\EndOfBibitem
\bibitem[Billups \latin{et~al.}(1980)Billups, Konarski, Hauge, and Margrave]{billups1980activation}
Billups,~W.; Konarski,~M.~M.; Hauge,~R.~H.; Margrave,~J.~L. Activation of methane with photoexcited metal atoms. \emph{Journal of the American Chemical Society} \textbf{1980}, \emph{102}, 7393--7394\relax
\mciteBstWouldAddEndPuncttrue
\mciteSetBstMidEndSepPunct{\mcitedefaultmidpunct}
{\mcitedefaultendpunct}{\mcitedefaultseppunct}\relax
\EndOfBibitem
\bibitem[Cho and Andrews(2013)Cho, and Andrews]{cho2013infrared}
Cho,~H.-G.; Andrews,~L. Infrared Spectra of Manganese Insertion, Vinyl, and Cyclic Complexes Prepared in Reactions of Laser-Ablated Mn Atoms with Methane, Ethane, Ethyl Chloride, and 1, 2-Dichloroethane. \emph{Organometallics} \textbf{2013}, \emph{32}, 3458--3468\relax
\mciteBstWouldAddEndPuncttrue
\mciteSetBstMidEndSepPunct{\mcitedefaultmidpunct}
{\mcitedefaultendpunct}{\mcitedefaultseppunct}\relax
\EndOfBibitem
\bibitem[Diffenderfer and Yarkony(1982)Diffenderfer, and Yarkony]{state_averaged_casscf}
Diffenderfer,~R.~N.; Yarkony,~D.~R. Use of the state-averaged MCSCF procedure: application to radiative transitions in magnesium oxide. \emph{Journal of Physical Chemistry} \textbf{1982}, \emph{86}, 5098--5105\relax
\mciteBstWouldAddEndPuncttrue
\mciteSetBstMidEndSepPunct{\mcitedefaultmidpunct}
{\mcitedefaultendpunct}{\mcitedefaultseppunct}\relax
\EndOfBibitem
\bibitem[Lin \latin{et~al.}(2013)Lin, Li, Mao, and Chai]{wb97xd3}
Lin,~Y.-S.; Li,~G.-D.; Mao,~S.-P.; Chai,~J.-D. Long-range corrected hybrid density functionals with improved dispersion corrections. \emph{Journal of Chemical Theory and Computation} \textbf{2013}, \emph{9}, 263--272\relax
\mciteBstWouldAddEndPuncttrue
\mciteSetBstMidEndSepPunct{\mcitedefaultmidpunct}
{\mcitedefaultendpunct}{\mcitedefaultseppunct}\relax
\EndOfBibitem
\end{mcitethebibliography}


\providecommand{\latin}[1]{#1}
\makeatletter
\providecommand{\doi}
  {\begingroup\let\do\@makeother\dospecials
  \catcode`\{=1 \catcode`\}=2 \doi@aux}
\providecommand{\doi@aux}[1]{\endgroup\texttt{#1}}
\makeatother
\providecommand*\mcitethebibliography{\thebibliography}
\csname @ifundefined\endcsname{endmcitethebibliography}  {\let\endmcitethebibliography\endthebibliography}{}
\begin{mcitethebibliography}{12}
\providecommand*\natexlab[1]{#1}
\providecommand*\mciteSetBstSublistMode[1]{}
\providecommand*\mciteSetBstMaxWidthForm[2]{}
\providecommand*\mciteBstWouldAddEndPuncttrue
  {\def\EndOfBibitem{\unskip.}}
\providecommand*\mciteBstWouldAddEndPunctfalse
  {\let\EndOfBibitem\relax}
\providecommand*\mciteSetBstMidEndSepPunct[3]{}
\providecommand*\mciteSetBstSublistLabelBeginEnd[3]{}
\providecommand*\EndOfBibitem{}
\mciteSetBstSublistMode{f}
\mciteSetBstMaxWidthForm{subitem}{(\alph{mcitesubitemcount})}
\mciteSetBstSublistLabelBeginEnd
  {\mcitemaxwidthsubitemform\space}
  {\relax}
  {\relax}

\bibitem[Helgaker and Taylor(1992)Helgaker, and Taylor]{pair_gradient}
Helgaker,~T.; Taylor,~P.~R. On the evaluation of derivatives of Gaussian integrals. \emph{Theoretica chimica acta} \textbf{1992}, \emph{83}, 177--183\relax
\mciteBstWouldAddEndPuncttrue
\mciteSetBstMidEndSepPunct{\mcitedefaultmidpunct}
{\mcitedefaultendpunct}{\mcitedefaultseppunct}\relax
\EndOfBibitem
\bibitem[Ufimtsev and Martinez(2009)Ufimtsev, and Martinez]{terachem_gpu_3}
Ufimtsev,~I.~S.; Martinez,~T.~J. Quantum chemistry on graphical processing units. 3. Analytical energy gradients, geometry optimization, and first principles molecular dynamics. \emph{Journal of Chemical Theory and Computation} \textbf{2009}, \emph{5}, 2619--2628\relax
\mciteBstWouldAddEndPuncttrue
\mciteSetBstMidEndSepPunct{\mcitedefaultmidpunct}
{\mcitedefaultendpunct}{\mcitedefaultseppunct}\relax
\EndOfBibitem
\bibitem[Pulay(1969)]{pulay_hf_derivative}
Pulay,~P. Ab initio calculation of force constants and equilibrium geometries in polyatomic molecules: I. Theory. \emph{Molecular Physics} \textbf{1969}, \emph{17}, 197--204\relax
\mciteBstWouldAddEndPuncttrue
\mciteSetBstMidEndSepPunct{\mcitedefaultmidpunct}
{\mcitedefaultendpunct}{\mcitedefaultseppunct}\relax
\EndOfBibitem
\bibitem[Pople \latin{et~al.}(1979)Pople, Krishnan, Schlegel, and Binkley]{pople_hf_derivative}
Pople,~J.; Krishnan,~R.; Schlegel,~H.; Binkley,~J.~S. Derivative studies in hartree-fock and m{\o}ller-plesset theories. \emph{International Journal of Quantum Chemistry} \textbf{1979}, \emph{16}, 225--241\relax
\mciteBstWouldAddEndPuncttrue
\mciteSetBstMidEndSepPunct{\mcitedefaultmidpunct}
{\mcitedefaultendpunct}{\mcitedefaultseppunct}\relax
\EndOfBibitem
\bibitem[Johnson \latin{et~al.}(1993)Johnson, Gill, and Pople]{dft_gradient}
Johnson,~B.~G.; Gill,~P.~M.; Pople,~J.~A. The performance of a family of density functional methods. \emph{The Journal of Chemical Physics} \textbf{1993}, \emph{98}, 5612--5626\relax
\mciteBstWouldAddEndPuncttrue
\mciteSetBstMidEndSepPunct{\mcitedefaultmidpunct}
{\mcitedefaultendpunct}{\mcitedefaultseppunct}\relax
\EndOfBibitem
\bibitem[Asadchev and Valeev(2023)Asadchev, and Valeev]{libintx_up_to_iiii}
Asadchev,~A.; Valeev,~E.~F. High-performance evaluation of high angular momentum 4-center Gaussian integrals on modern accelerated processors. \emph{The Journal of Physical Chemistry A} \textbf{2023}, \emph{127}, 10889--10895\relax
\mciteBstWouldAddEndPuncttrue
\mciteSetBstMidEndSepPunct{\mcitedefaultmidpunct}
{\mcitedefaultendpunct}{\mcitedefaultseppunct}\relax
\EndOfBibitem
\bibitem[Tornai \latin{et~al.}(2019)Tornai, Ladj{\'a}nszki, R{\'a}k, Kis, and Cserey]{brianqc}
Tornai,~G.~J.; Ladj{\'a}nszki,~I.; R{\'a}k,~{\'A}.; Kis,~G.; Cserey,~G. Calculation of quantum chemical two-electron integrals by applying compiler technology on GPU. \emph{Journal of Chemical Theory and Computation} \textbf{2019}, \emph{15}, 5319--5331\relax
\mciteBstWouldAddEndPuncttrue
\mciteSetBstMidEndSepPunct{\mcitedefaultmidpunct}
{\mcitedefaultendpunct}{\mcitedefaultseppunct}\relax
\EndOfBibitem
\bibitem[Zhang \latin{et~al.}(2015)Zhang, El-Roz, Frei, Mendoza-Cortes, Head-Gordon, Lacy, and Peters]{tetraaza_complex}
Zhang,~M.; El-Roz,~M.; Frei,~H.; Mendoza-Cortes,~J.~L.; Head-Gordon,~M.; Lacy,~D.~C.; Peters,~J.~C. Visible light sensitized CO2 activation by the tetraaza [CoIIN4H (MeCN)] 2+ complex investigated by FT-IR spectroscopy and DFT calculations. \emph{The Journal of Physical Chemistry C} \textbf{2015}, \emph{119}, 4645--4654\relax
\mciteBstWouldAddEndPuncttrue
\mciteSetBstMidEndSepPunct{\mcitedefaultmidpunct}
{\mcitedefaultendpunct}{\mcitedefaultseppunct}\relax
\EndOfBibitem
\bibitem[Ziegler \latin{et~al.}(2018)Ziegler, Torquato, Levine, Nicolay, Celik, and Tilley]{dicopper_complex}
Ziegler,~M.~S.; Torquato,~N.~A.; Levine,~D.~S.; Nicolay,~A.; Celik,~H.; Tilley,~T.~D. Dicopper alkyl complexes: synthesis, structure, and unexpected persistence. \emph{Organometallics} \textbf{2018}, \emph{37}, 2807--2823\relax
\mciteBstWouldAddEndPuncttrue
\mciteSetBstMidEndSepPunct{\mcitedefaultmidpunct}
{\mcitedefaultendpunct}{\mcitedefaultseppunct}\relax
\EndOfBibitem
\bibitem[Li \latin{et~al.}(1999)Li, Eddaoudi, O'Keeffe, and Yaghi]{mof5_original}
Li,~H.; Eddaoudi,~M.; O'Keeffe,~M.; Yaghi,~O.~M. Design and synthesis of an exceptionally stable and highly porous metal-organic framework. \emph{Nature} \textbf{1999}, \emph{402}, 276--279\relax
\mciteBstWouldAddEndPuncttrue
\mciteSetBstMidEndSepPunct{\mcitedefaultmidpunct}
{\mcitedefaultendpunct}{\mcitedefaultseppunct}\relax
\EndOfBibitem
\bibitem[Spicher \latin{et~al.}(2020)Spicher, Bursch, and Grimme]{mof5_geometry}
Spicher,~S.; Bursch,~M.; Grimme,~S. Efficient calculation of small molecule binding in metal--organic frameworks and porous organic cages. \emph{The Journal of Physical Chemistry C} \textbf{2020}, \emph{124}, 27529--27541\relax
\mciteBstWouldAddEndPuncttrue
\mciteSetBstMidEndSepPunct{\mcitedefaultmidpunct}
{\mcitedefaultendpunct}{\mcitedefaultseppunct}\relax
\EndOfBibitem
\end{mcitethebibliography}
\end{document}

\maketitle

\newpage

\section{Derivatives of Integrals with Respect to Atom Positions}
The derivatives of the integrals with respect to the atom positions are essential for the computation of forces required for geometry optimizations, \textit{ab initio} molecular dynamics and (finite-difference) frequency calculations. Here, we provide a brief overview for computing the derivatives of integrals vs atom positions. Since we use pGTO pairs as our data structure for integral calculations, we start from the derivative of pairs $\mu(\vec{r})\nu(\vec{r})$ with respect to atomic center locations $\vec{A}$ and $\vec{B}$.\cite{pair_gradient, terachem_gpu_3} In the main text we have shown that the pair can be represented as a summation over Hermite Gaussians centered at $\vec{P} = \dfrac{a\vec{A} + b\vec{B}}{a+b}$:
\begin{align}
\mu(\vec{r})\nu(\vec{r}) = C_\mu C_\nu \sum^{i_x + j_x}_{t_x = 0} E^{i_x, j_x}_{t_x, x} \left( \frac{\partial}{\partial P_x} \right)^{t_x} \sum^{i_y + j_y}_{t_y = 0} E^{i_y, j_y}_{t_y, y} \left( \frac{\partial}{\partial P_y} \right)^{t_y} \sum^{i_z + j_z}_{t_z = 0} E^{i_z, j_z}_{t_z, z} \left( \frac{\partial}{\partial P_z} \right)^{t_z} e^{-p \left| \vec{r} - \vec{P} \right|^2}
\label{eq:si_pair_hermite}
\end{align}
In order to take a derivative of a pair under McMurchie-Davidson scheme, we define $\vec{\Delta} = \vec{A} - \vec{B}$ and perform the following change of variables:
\begin{align}
\vec{P} & = \frac{a\vec{A} + b\vec{B}}{a+b} & & \vec{A} = \vec{P} + \frac{b}{a+b} \vec{\Delta} \notag \\
\vec{\Delta} & = \vec{A} - \vec{B} & & \vec{B} = \vec{P} - \frac{a}{a+b} \vec{\Delta}
\label{eq:si_AB_to_PDelta}
\end{align}
This analogous to tranforming the coordinates of a two body problem into the center of mass coordinate (here $\vec{P}$) and interbody separation (here $ \vec{\Delta}$) that can instead be treated as independent variables. The corresponding derivative transformations therefore are:
\begin{align}
\frac{\partial}{\partial P_\tau} & = \frac{\partial}{\partial A_\tau} + \frac{\partial}{\partial B_\tau} & & \frac{\partial}{\partial A_\tau} = \frac{a}{a+b} \frac{\partial}{\partial P_\tau} + \frac{\partial}{\partial \Delta_{\tau}} \notag \\
\frac{\partial}{\partial \Delta_{\tau}} & = \frac{b}{a+b} \frac{\partial}{\partial A_\tau} - \frac{a}{a+b} \frac{\partial}{\partial B_\tau} & & \frac{\partial}{\partial B_\tau} = \frac{b}{a+b} \frac{\partial}{\partial P_\tau} - \frac{\partial}{\partial \Delta_{\tau}}
\label{eq:si_AB_to_PDelta_derivative}
\end{align}
where $\tau$ is one of the Cartesian directions ($x$, $y$ or $z$).

In equation \ref{eq:si_pair_hermite}, the Hermite Gaussian part $\left(\left( \dfrac{\partial}{\partial P_x} \right)^{t_x} \left( \dfrac{\partial}{\partial P_y} \right)^{t_y} \left( \dfrac{\partial}{\partial P_z} \right)^{t_z} e^{-p \left| \vec{r} - \vec{P} \right|^2}\right)$ clearly depends only on $\vec{P}$ and not on $\vec{\Delta}$. The Cartesian Gaussian to Hermite Gaussian transformation coefficients ($E^{i_x, j_x}_{t_x, x} E^{i_y, j_y}_{t_y, y} E^{i_z, j_z}_{t_z, z}$), on the other hand, depend only on $\vec{\Delta}$ and not on $\vec{P}$. This can be seen from the recursion relationship (equation \ref{eq:E_recursion}), noticing that although $\vec{P} - \vec{A} = -\dfrac{b}{a+b} \vec{\Delta}$ and $\vec{P} - \vec{B} = \dfrac{a}{a+b} \vec{\Delta}$ looks like they explicitly depend on $\vec{P}$, they actually only depend on $ \vec{\Delta}$.

In order to get the derivative of $E^{i_\tau, j_\tau}_{t_\tau, \tau}$ (labeled $E^{i_\tau, j_\tau, 1}_{t_\tau, \tau} = \frac{\partial}{\partial \Delta_\tau} E^{i_\tau, j_\tau}_{t_\tau, \tau}$), we differentiate both sides of the recursion relationship on $E^{i_\tau, j_\tau}_{t_\tau, \tau}$ (equation \ref{eq:E_recursion}) with respect to $\Delta_\tau$, resulting in the following recursion relationship for $E^{i_\tau, j_\tau, 1}_{t_\tau, \tau}$:
\begin{align}
E^{i_\tau + 1, j_\tau, 1}_{t_\tau, \tau} &= \frac{1}{2p} E^{i_\tau, j_\tau, 1}_{t_\tau - 1, \tau} - \frac{b}{a+b} E^{i_\tau, j_\tau}_{t_\tau, \tau} + (P_\tau - A_\tau) E^{i_\tau, j_\tau, 1}_{t_\tau, \tau} + (t_\tau + 1) E^{i_\tau, j_\tau, 1}_{t_\tau + 1, \tau} \\
E^{i_\tau, j_\tau + 1, 1}_{t_\tau, \tau} &= \frac{1}{2p} E^{i_\tau, j_\tau, 1}_{t_\tau - 1, \tau} + \frac{a}{a+b} E^{i_\tau, j_\tau}_{t_\tau, \tau} + (P_\tau - B_\tau) E^{i_\tau, j_\tau, 1}_{t_\tau, \tau} + (t_\tau + 1)  E^{i_\tau, j_\tau, 1}_{t_\tau + 1, \tau} \\
E^{0,0,1}_{0,\tau} &= 2a (P_\tau - A_\tau) e^{-\frac{ab}{a+b} (A_\tau - B_\tau)^2} \\
E^{i_\tau, j_\tau, 1}_{t_\tau, \tau} &= 0 \qquad \text{ if } t_\tau < 0 \text{ or } t_\tau > i_\tau + j_\tau
\label{eq:si_dE_recursion}
\end{align}
The choice of the form of the base case is not unique, since $a(P_\tau - A_\tau) = -b(P_\tau - B_\tau)$.

Given the variable transformation (equation \ref{eq:si_AB_to_PDelta_derivative}) and derivative of $E^{i_\tau, j_\tau}_{t_\tau, \tau}$, the derivative of $\mu(\vec{r})\nu(\vec{r})$ with respect to $\vec{A}$ and $\vec{B}$ can be expressed as:
\begin{align}
\frac{\partial}{\partial A_x} \mu(\vec{r})\nu(\vec{r}) = & C_\mu C_\nu \sum^{i_x + j_x + 1}_{t_x = 0} \left( \frac{a}{a+b} E^{i_x, j_x}_{t_x - 1, x} + E^{i_x, j_x, 1}_{t_x, x} \right) \left( \frac{\partial}{\partial P_x} \right)^{t_x} \notag \\
    & \sum^{i_y + j_y}_{t_y = 0} E^{i_y, j_y}_{t_y, y} \left( \frac{\partial}{\partial P_y} \right)^{t_y} \sum^{i_z + j_z}_{t_z = 0} E^{i_z, j_z}_{t_z, z} \left( \frac{\partial}{\partial P_z} \right)^{t_z} e^{-p \left| \vec{r} - \vec{P} \right|^2} \\
\frac{\partial}{\partial B_x} \mu(\vec{r})\nu(\vec{r}) = & C_\mu C_\nu \sum^{i_x + j_x + 1}_{t_x = 0} \left( \frac{b}{a+b} E^{i_x, j_x}_{t_x - 1, x} - E^{i_x, j_x, 1}_{t_x, x} \right) \left( \frac{\partial}{\partial P_x} \right)^{t_x} \notag \\
    & \sum^{i_y + j_y}_{t_y = 0} E^{i_y, j_y}_{t_y, y} \left( \frac{\partial}{\partial P_y} \right)^{t_y} \sum^{i_z + j_z}_{t_z = 0} E^{i_z, j_z}_{t_z, z} \left( \frac{\partial}{\partial P_z} \right)^{t_z} e^{-p \left| \vec{r} - \vec{P} \right|^2}
\label{eq:si_pair_derivative}
\end{align}
It is important to notice that the number of terms in the differentiation coordinate direction is increased by one. 

From here on we will show the derivative of all the integrals mentioned in the main text, as well as the actual routines implemented in TeraChem, which contract the integral derivative tensors with the density matrix for force computations. Whenever applicable, we will only provide the derivative along x direction, as the derivative along y and z direction can be obtained by permutation.
 
\subsection{Overlap Integral}

Similar to the overlap integral itself, only the $\vec{t} = \vec{0}$ case remains in the summation, so
\begin{align}
\frac{\partial}{\partial A_x} S_{\mu\nu} &= C_\mu C_\nu E^{i_x, j_x, 1}_{0, x} E^{i_y, j_y}_{0, y} E^{i_z, j_z}_{0, z} \left( \frac{\pi}{p} \right)^{3/2} \\
\frac{\partial}{\partial B_x} S_{\mu\nu} &= - C_\mu C_\nu E^{i_x, j_x, 1}_{0, x} E^{i_y, j_y}_{0, y} E^{i_z, j_z}_{0, z} \left( \frac{\pi}{p} \right)^{3/2}
\label{eq:si_overlap_derivative}
\end{align}

The overlap derivative tensor is usually contracted with the energy-weighted density matrix $W_{\mu\nu}= \displaystyle\sum\limits_{i}^{n_{occ}} C_{\mu i}C_{\nu i}\epsilon_i$ ($i$ running over all occupied orbitals) to form the force, so TeraChem provides the routine for computing $\displaystyle\sum\limits_{\mu\nu}^{n_{AO}} W_{\mu\nu} \dfrac{\partial}{\partial R_\tau} S_{\mu\nu}$, where $\vec{R}$ goes through all atomic centers and $\tau$ goes through all three Cartesian directions.

\subsection{Kinetic Energy Integral}

Since $\dfrac{\partial}{\partial A_\tau} E^{i_\tau, j_\tau}_{t_\tau, \tau} = - \dfrac{\partial}{\partial B_\tau} E^{i_\tau, j_\tau}_{t_\tau, \tau} = \dfrac{\partial}{\partial \Delta_\tau} E^{i_\tau, j_\tau}_{t_\tau, \tau} = E^{i_\tau, j_\tau, 1}_{t_\tau, \tau}$, the derivative of kinetic energy integral can be obtained by differentiating the integral formula (equation \ref{eq:kinetic_formula}):
\begin{align}
\frac{\partial}{\partial A_x} T_{\mu\nu} = C_\mu C_\nu & \left( \left( -\frac{j_x (j_x-1)}{2} E^{i_x, j_x-2, 1}_{0,x} + (2j_x + 1) b E^{i_x, j_x, 1}_{0,x} - 2b^2 E^{i_x, j_x+2, 1}_{0,x} \right) E^{i_y, j_y}_{0,y} E^{i_z, j_z}_{0,z} \right. \notag \\
    & + E^{i_x, j_x, 1}_{0,x} \left( -\frac{j_y (j_y-1)}{2} E^{i_y, j_y-2}_{0,y} + (2j_y + 1) b E^{i_y, j_y}_{0,y} - 2b^2 E^{i_y, j_y+2}_{0,y} \right) E^{i_z, j_z}_{0,z} \notag \\
    & \left. + E^{i_x, j_x, 1}_{0,x} E^{i_y, j_y}_{0,y} \left( -\frac{j_z (j_z-1)}{2} E^{i_z, j_z-2}_{0,z} + (2j_z + 1) b E^{i_z, j_z}_{0,z} - 2b^2 E^{i_z, j_z+2}_{0,z} \right) \right) \left( \frac{\pi}{p} \right)^{3/2} \\
\frac{\partial}{\partial B_x} T_{\mu\nu} = -C_\mu C_\nu & \left( \left( -\frac{j_x (j_x-1)}{2} E^{i_x, j_x-2, 1}_{0,x} + (2j_x + 1) b E^{i_x, j_x, 1}_{0,x} - 2b^2 E^{i_x, j_x+2, 1}_{0,x} \right) E^{i_y, j_y}_{0,y} E^{i_z, j_z}_{0,z} \right. \notag \\
    & + E^{i_x, j_x, 1}_{0,x} \left( -\frac{j_y (j_y-1)}{2} E^{i_y, j_y-2}_{0,y} + (2j_y + 1) b E^{i_y, j_y}_{0,y} - 2b^2 E^{i_y, j_y+2}_{0,y} \right) E^{i_z, j_z}_{0,z} \notag \\
    & \left. + E^{i_x, j_x, 1}_{0,x} E^{i_y, j_y}_{0,y} \left( -\frac{j_z (j_z-1)}{2} E^{i_z, j_z-2}_{0,z} + (2j_z + 1) b E^{i_z, j_z}_{0,z} - 2b^2 E^{i_z, j_z+2}_{0,z} \right) \right) \left( \frac{\pi}{p} \right)^{3/2}
\label{eq:si_kinetic_derivative}
\end{align}

As part of the core Hamiltonian, the kinetic energy integral derivative tensor is usually contracted with the density matrix $D_{\mu\nu}$, so TeraChem provides the routine for computing $\displaystyle\sum\limits_{\mu\nu}^{n_{AO}} D_{\mu\nu} \dfrac{\partial}{\partial R_\tau}  T_{\mu\nu}$.

\subsection{Nuclear Attraction Integral}

The derivative of nuclear attraction integral with respect to GTO center location $\vec{A}$ and $\vec{B}$ can be obtained by applying the pair derivative formula (equation \ref{eq:si_pair_derivative}) to the Hermite Gaussian representation of the integral (equation \ref{eq:v1e_general_step1}). When $\vec{A}\ne \vec{C}$ and $\vec{B}\ne \vec{C}$ (i.e. all atomic centers are distinct):
\begin{align}
\frac{\partial}{\partial A_x} V_{\mu\nu C} = & q_C C_\mu C_\nu \frac{2\pi}{p} \sum^{i_x + j_x + 1}_{t_x = 0} \left( \frac{a}{a+b} E^{i_x, j_x}_{t_x - 1, x} + E^{i_x, j_x, 1}_{t_x, x} \right) \sum^{i_y + j_y}_{t_y = 0} E^{i_y, j_y}_{t_y, y} \sum^{i_z + j_z}_{t_z = 0} E^{i_z, j_z}_{t_z, z} R_{t_x,t_y,t_z}^0 \left( p, \vec{P} - \vec{C} \right) \label{eq:si_v1e_derivative_A}\\
\frac{\partial}{\partial B_x} V_{\mu\nu C} = & q_C C_\mu C_\nu \frac{2\pi}{p} \sum^{i_x + j_x + 1}_{t_x = 0} \left( \frac{b}{a+b} E^{i_x, j_x}_{t_x - 1, x} - E^{i_x, j_x, 1}_{t_x, x} \right) \sum^{i_y + j_y}_{t_y = 0} E^{i_y, j_y}_{t_y, y} \sum^{i_z + j_z}_{t_z = 0} E^{i_z, j_z}_{t_z, z} R_{t_x,t_y,t_z}^0 \left( p, \vec{P} - \vec{C} \right)
\label{eq:si_v1e_derivative_B}
\end{align}

Additionally, we need to take derivatives with respect to the point charge center location $\vec{C}$ as well. It is clear that $E^{i_\tau, j_\tau}_{t_\tau, \tau}$ does not depend on $\vec{C}$, and the derivative of auxiliary integral $R_{t_x,t_y,t_z}^m$ with respect to $\vec{C}$ is still in auxiliary integral form, with a larger index ($t_x\to t_{x}+1$):
\begin{align}
\frac{\partial}{\partial C_x} R_{t_x,t_y,t_z}^m\left( p, \vec{P} - \vec{C} \right) &= \left( \frac{\partial}{\partial P_x} \right)^{t_x} \left( \frac{\partial}{\partial P_y} \right)^{t_y} \left( \frac{\partial}{\partial P_z} \right)^{t_z} \left( \frac{\partial}{\partial C_x} \right) \left( (-2p)^m F_m\left( p\left| \vec{P} - \vec{C} \right|^2 \right) \right) \notag \\
    &= -\left( \frac{\partial}{\partial P_x} \right)^{t_x + 1} \left( \frac{\partial}{\partial P_y} \right)^{t_y} \left( \frac{\partial}{\partial P_z} \right)^{t_z} \left( (-2p)^m F_m\left( p\left| \vec{P} - \vec{C} \right|^2 \right) \right) \notag \\
    &= -R_{t_x + 1,t_y,t_z}^m\left( p, \vec{P} - \vec{C} \right)
\label{eq:si_R_derivative_C}
\end{align}
As a result,
\begin{align}
\frac{\partial}{\partial C_x} V_{\mu\nu C} = -q_C C_\mu C_\nu \frac{2\pi}{p} \sum^{i_x + j_x}_{t_x = 0} E^{i_x, j_x}_{t_x, x} \sum^{i_y + j_y}_{t_y = 0} E^{i_y, j_y}_{t_y, y} \sum^{i_z + j_z}_{t_z = 0} E^{i_z, j_z}_{t_z, z} R_{t_x + 1,t_y,t_z}^0 \left( p, \vec{P} - \vec{C} \right)
\label{eq:si_v1e_derivative_C}
\end{align}
When $\vec{A} \ne \vec{C}$ and $\vec{B}\ne \vec{C}$. From the chain rule, the general case of $\vec{A}= \vec{C}$ or $\vec{B}= \vec{C}$ can be obtained by adding equations \ref{eq:si_v1e_derivative_A}/\ref{eq:si_v1e_derivative_B} with equation \ref{eq:si_v1e_derivative_C}.

As another part of the core Hamiltonian, the nuclear attraction integral derivative tensor is usually contracted with the density matrix $D_{\mu\nu}$ as well, so TeraChem provides one routine for computing both $\displaystyle\sum\limits_{\mu\nu}^{n_{AO}} \displaystyle\sum\limits_{C}^{N_{point-charge}} D_{\mu\nu} \dfrac{\partial}{\partial A_\tau} V_{\mu\nu C}$ and $\displaystyle\sum\limits_{\mu\nu}^{n_{AO}} D_{\mu\nu} \dfrac{\partial}{\partial C_\tau} V_{\mu\nu C}$, where $A_\tau$ goes through all GTO centers, and $C_\tau$ goes through all point charge locations. If neither ghost atoms (more GTO centers than number of atoms) nor QM/MM or similar embedding methods (more point charges than number of atoms) is used, then the indices $\vec{A}$ and $\vec{C}$ eventually covers the same list of atoms, and the two contributions can be summed up. However, in order to get a more general interface, TeraChem integral routine separates the two derivatives into two output vectors, and thus separate the GTO pair derivatives from the electrostatic potential term derivative.

\subsection{Electron Repulsion Integrals (ERIs)}
There are two pieces of Hartree-Fock (HF) or DFT energy contributions that require ERI: the Coulomb ($E_J$) and HF exchange ($E_K$) contributions. They have the following expression:
\begin{align}
E_J &= \frac{1}{2} \sum_{\mu\nu}^{n_{AO}} D_{\mu\nu} \sum_{\lambda\sigma}^{n_{AO}} D_{\lambda\sigma} (\mu\nu|\lambda\sigma) \\
E_K &= -\frac{1}{4} \sum_{\mu\lambda}^{n_{AO}} D_{\mu\lambda} \sum_{\nu\sigma}^{n_{AO}} D_{\nu\sigma} (\mu\nu|\lambda\sigma)
\label{eq:si_E_J_K}
\end{align}

Before diving into the ERI derivatives, we differentiate the contractions $E_J$ and $E_K$ and thereby show that the derivative of ERIs with respect to ket-side GTO center locations ($\vec{C}$ and $\vec{D}$) is not necessary. We then provide the formula for the derivative of ERI with respect to bra-side GTO center locations ($\vec{A}$ and $\vec{B}$).

The derivative of $E_J$ and $E_K$ with respect to any GTO center location $R_\tau$ has the form
\begin{align}
\frac{\partial E_J}{\partial R_\tau} &= \frac{1}{2} \left( \sum_{\mu\nu}^{n_{AO}} D_{\mu\nu} \sum_{\lambda\sigma}^{n_{AO}} D_{\lambda\sigma} \left( \frac{\partial \mu\nu}{\partial R_\tau} \middle| \lambda\sigma \right) + \sum_{\lambda\sigma}^{n_{AO}} D_{\lambda\sigma} \sum_{\mu\nu}^{n_{AO}} D_{\mu\nu} \left( \mu\nu \middle| \frac{\partial \lambda\sigma}{\partial R_\tau} \right) \right) + f_J\left(\frac{\partial D_{\mu\nu}}{\partial R_\tau}\right) \label{eq:si_dE_J_step1} \\
\frac{\partial E_K}{\partial R_\tau} &= -\frac{1}{4} \left( \sum_{\mu\lambda}^{n_{AO}} D_{\mu\lambda} \sum_{\nu\sigma}^{n_{AO}} D_{\nu\sigma} \left( \frac{\partial \mu\nu}{\partial R_\tau} \middle| \lambda\sigma \right) + \sum_{\nu\sigma}^{n_{AO}} D_{\nu\sigma} \sum_{\mu\lambda}^{n_{AO}} D_{\mu\lambda} \left( \mu\nu \middle| \frac{\partial \lambda\sigma}{\partial R_\tau} \right) \right) + f_K\left(\frac{\partial D_{\mu\nu}}{\partial R_\tau}\right) \label{eq:si_dE_K_step1}
\end{align}
Here $f_J(x)$ and $f_K(x)$ are some functions of the density matrix derivatives with respect to $R_\tau$. Since they are unrelated to integral derivatives, we do not care about their actual form here (and they ultimately cancel in the total energy derivative on account of self-consistency of HF/DFT orbitals).

When we contract ERI with density matrices, we already implicitly use the 8-fold symmetry relationship of ERIs with real basis functions (\ref{eq:8_fold_symmetry}). The derivative version of 8-fold symmetry relationship is:
\begin{align}
\left( \frac{\partial \mu\nu}{\partial R_\tau} \middle| \lambda\sigma \right) &= \left( \frac{\partial \nu\mu}{\partial R_\tau} \middle| \lambda\sigma \right) \notag \\
\left( \frac{\partial \mu\nu}{\partial R_\tau} \middle| \lambda\sigma \right) &= \left( \frac{\partial \mu\nu}{\partial R_\tau} \middle| \sigma\lambda \right) \notag \\
\left( \frac{\partial \mu\nu}{\partial R_\tau} \middle| \lambda\sigma \right) &= \left( \lambda\sigma \middle| \frac{\partial \mu\nu}{\partial R_\tau} \right)
\label{eq:si_8_fold_symmetry_derivative}
\end{align}

By applying these symmetry and change of summation variable names, equation \ref{eq:si_dE_J_step1} and \ref{eq:si_dE_K_step1} can be simplified to
\begin{align}
\frac{\partial E_J}{\partial R_\tau} &= \sum_{\mu\nu}^{n_{AO}} D_{\mu\nu} \sum_{\lambda\sigma}^{n_{AO}} D_{\lambda\sigma} \left( \frac{\partial \mu\nu}{\partial R_\tau} \middle| \lambda\sigma \right) + f_J\left(\frac{\partial D_{\mu\nu}}{\partial R_\tau}\right) \label{eq:si_dE_J} \\
\frac{\partial E_K}{\partial R_\tau} &= -\frac{1}{2} \sum_{\mu\lambda}^{n_{AO}} D_{\mu\lambda} \sum_{\nu\sigma}^{n_{AO}} D_{\nu\sigma} \left( \frac{\partial \mu\nu}{\partial R_\tau} \middle| \lambda\sigma \right) + f_K\left(\frac{\partial D_{\mu\nu}}{\partial R_\tau}\right) \label{eq:si_dE_K}
\end{align}

From equation \ref{eq:si_dE_J} and \ref{eq:si_dE_K} it is clear that $\left( \dfrac{\partial \mu\nu}{\partial R_\tau} \middle| \lambda\sigma \right)$ is sufficient and $\left( \mu\nu \middle| \dfrac{\partial \lambda\sigma}{\partial R_\tau} \right)$ is not needed. This simplifies the derivative implementation. TeraChem provides routines to compute the first term in \ref{eq:si_dE_J} and \ref{eq:si_dE_K} respectively.

The ERI derivative can be obtained similarly to the nuclear attraction integral:
\begin{align}
\left( \frac{\partial \mu\nu}{\partial A_x} \middle| \lambda\sigma \right) = & C_\mu C_\nu C_\lambda C_\sigma \notag \\
    & \sum^{i_x + j_x + 1}_{t_x = 0} \left( \frac{a}{a+b} E^{i_x, j_x}_{t_x - 1, x}(A_x, B_x, p) + E^{i_x, j_x, 1}_{t_x, x}(A_x, B_x, p) \right) \notag \\
    & \sum^{i_y + j_y}_{t_y = 0} E^{i_y, j_y}_{t_y, y}(A_y, B_y, p) \sum^{i_z + j_z}_{t_z = 0} E^{i_z, j_z}_{t_z, z}(A_z, B_z, p) \notag \\
    & \sum^{k_x + l_x}_{s_x = 0} E^{k_x, l_x}_{s_x, x}(C_x, D_x, p) \sum^{k_y + l_y}_{s_y = 0} E^{k_y, l_y}_{s_y, y}(C_y, D_y, p) \sum^{k_z + l_z}_{s_z = 0} E^{k_z, l_z}_{s_z, z}(C_z, D_z, p) \notag \\
    & (-1)^{s_x + s_y + s_z} \frac{2\pi^{5/2}}{pq\sqrt{p+q}}  R_{t_x + s_x, t_y + s_y, t_z + s_z}^0\left( \frac{pq}{p+q}, \vec{P} - \vec{Q} \right) \label{eq:si_eri_derivative_A} \\
\left( \frac{\partial \mu\nu}{\partial B_x} \middle| \lambda\sigma \right) = & C_\mu C_\nu C_\lambda C_\sigma \notag \\
    & \sum^{i_x + j_x + 1}_{t_x = 0} \left( \frac{b}{a+b} E^{i_x, j_x}_{t_x - 1, x}(A_x, B_x, p) - E^{i_x, j_x, 1}_{t_x, x}(A_x, B_x, p) \right) \notag \\
    & \sum^{i_y + j_y}_{t_y = 0} E^{i_y, j_y}_{t_y, y}(A_y, B_y, p) \sum^{i_z + j_z}_{t_z = 0} E^{i_z, j_z}_{t_z, z}(A_z, B_z, p) \notag \\
    & \sum^{k_x + l_x}_{s_x = 0} E^{k_x, l_x}_{s_x, x}(C_x, D_x, p) \sum^{k_y + l_y}_{s_y = 0} E^{k_y, l_y}_{s_y, y}(C_y, D_y, p) \sum^{k_z + l_z}_{s_z = 0} E^{k_z, l_z}_{s_z, z}(C_z, D_z, p) \notag \\
    & (-1)^{s_x + s_y + s_z} \frac{2\pi^{5/2}}{pq\sqrt{p+q}}  R_{t_x + s_x, t_y + s_y, t_z + s_z}^0\left( \frac{pq}{p+q}, \vec{P} - \vec{Q} \right) \label{eq:si_eri_derivative_B}
\end{align}

\subsection{Exchange-Correlation Integral}

The exchange-correlation contribution of the DFT energy, under the generalized gradient approximation, has the form:
\begin{align}
E_{XC} = \iiint_\infty d\vec{r} \ \varepsilon_{XC}(\rho(\vec{r}), \vec{\nabla} \rho(\vec{r}))
\label{eq:si_xc_energy}
\end{align}
It is evaluated numerically as a weighted sum of grid-point evaluation of the integrands (similar to equation \ref{eq:xc_grid_sum_0} and \ref{eq:xc_grid_sum_1}). Since atom-centered standard quadrature grids and Becke partitioning is used in TeraChem for grid construction, both grid point locations $\vec{r}_g$ and weights $w_g$ depend on the atom center locations, which needs to be handled when taking derivatives. We first expand the grid-point summation in a more detailed form:
\begin{align}
E_{XC} &\approx \sum_g^{N_{grid}} w_g \varepsilon_{XC}(\rho(\vec{r}_g), \vec{\nabla} \rho(\vec{r}_g))
\label{eq:si_xc_energy_grid}\\
&= \sum_{C}^{N_{atom}} \sum_g^{N_{grid}(C)} w_{gC}^{Becke} w_{gC}^{quadrature} \varepsilon_{XC}\left( \rho(\vec{r}_{gC}^{\ quadrature} + \vec{C}), \vec{\nabla} \rho(\vec{r}_{gC}^{\ quadrature} + \vec{C}) \right)
\label{eq:si_xc_energy_grid_detailed}
\end{align}
where for each atom $C$, a spherical grid around $\vec{C}$ is constructed, with quadrature position offset $\vec{r}_{gC}^{\ quadrature}$ and quadrature weight $w_{gC}^{quadrature}$. The quadrature position offset and weight are usually pre-defined and do not vary with atomic center locations. The Becke weight $w_{gC}^{Becke}$ provides a smooth way of assigning each grid point to individual atom by positional vicinity, and Becke weight of each grid point depends on atomic center locations of all atoms. To simplify the notation, we will use $w_{gC} = w_{gC}^{Becke} w_{gC}^{quadrature}$ and $\vec{r}_{gC} = \vec{r}_{gC}^{\ quadrature} + \vec{C}$ wherever applicable.

The derivative of $E_{XC}$ with respect to any atomic center location $R_\tau$ is
\begin{align}
\frac{\partial E_{XC}}{\partial R_\tau} \approx & \sum_{C}^{N_{atom}} \sum_g^{N_{grid}(C)} \frac{\partial w_{gC}^{Becke}}{\partial R_\tau} w_{gC}^{quadrature} \varepsilon_{XC}\left( \rho(\vec{r}_{gC}), \vec{\nabla} \rho(\vec{r}_{gC}) \right) \notag \\
    & + \sum_{C}^{N_{atom}} \sum_g^{N_{grid}(C)} w_{gC} \left. \frac{\partial \varepsilon_{XC}}{\partial \rho} \right|_{\rho = \rho(\vec{r}_{gC})} \sum_{\mu\nu}^{n_{AO}} D_{\mu\nu} \left. \frac{\partial \left( \mu\nu \right)}{\partial R_\tau} \right|_{\vec{r} = \vec{r}_{gC}} \notag \\
    & + \sum_{C}^{N_{atom}} \sum_g^{N_{grid}(C)} w_{gC} \left. \frac{\partial \varepsilon_{XC}}{\partial \rho} \right|_{\rho = \rho(\vec{r}_{gC})} \sum_{\mu\nu}^{n_{AO}} D_{\mu\nu} \left. \frac{\partial \left( \mu\nu \right)}{\partial \tau} \right|_{\vec{r} = \vec{r}_{gC}} \frac{\partial C_\tau}{\partial R_\tau} \notag \\
    & + \sum_{C}^{N_{atom}} \sum_g^{N_{grid}(C)} w_{gC}  \left. \frac{\partial \varepsilon_{XC}}{\partial \vec{\nabla} \rho} \right|_{ \rho = \rho(\vec{r}_{gC}), \vec{\nabla} \rho = \vec{\nabla} \rho(\vec{r}_{gC}) } \cdot \sum_{\mu\nu}^{n_{AO}} D_{\mu\nu} \left. \frac{\partial \left( \vec{\nabla}_{r} (\mu\nu) \right) }{\partial R_\tau} \right|_{\vec{r} = \vec{r}_{gC}} \notag \\
    & + \sum_{C}^{N_{atom}} \sum_g^{N_{grid}(C)} w_{gC}  \left. \frac{\partial \varepsilon_{XC}}{\partial \vec{\nabla} \rho} \right|_{ \rho = \rho(\vec{r}_{gC}), \vec{\nabla} \rho = \vec{\nabla} \rho(\vec{r}_{gC}) } \cdot \sum_{\mu\nu}^{n_{AO}} D_{\mu\nu} \left. \frac{\partial \left( \vec{\nabla}_{r} (\mu\nu) \right) }{\partial \tau} \right|_{\vec{r} = \vec{r}_{gC}} \frac{\partial C_\tau}{\partial R_\tau} \notag \\
    & + f_{XC}\left(\frac{\partial D_{\mu\nu}}{\partial R_\tau}\right)
\label{eq:si_xc_derivative_step1}
\end{align}
where the first 5 terms are the partial derivative contributions from Becke weights, pairs in density, grid points for density evaluation, pairs in density gradient, and grid points for density gradient evaluation, respectively. Similar to ERI derivative, we do not care about the density matrix element derivative term $\dfrac{\partial D_{\mu\nu}}{\partial R_\tau}$ as it ultimately cancels out in the total energy gradient expression. \cite{pulay_hf_derivative, pople_hf_derivative}

In order to simplify the grid point derivative expressions, it is easy to note that $\dfrac{\partial C_\tau}{\partial R_\tau} = \delta_{C,R}$. Also since $\mu(\vec{r})$ is just a function of the difference between $\vec{r}$ and atomic center location $\vec{A}$, or in other words $\mu(\vec{r}; \vec{A}) = \mu(\vec{r} - \vec{A})$, it is evident that:
\begin{align}
\frac{\partial \left( \mu(\vec{r})\nu(\vec{r}) \right)}{\partial \tau} = - \frac{\partial \left( \mu(\vec{r})\nu(\vec{r}) \right)}{\partial A_\tau} - \frac{\partial \left( \mu(\vec{r})\nu(\vec{r}) \right)}{\partial B_\tau}
\label{eq:si_xc_derivative_C_to_AB}
\end{align}

So the expression for exchange-correlation energy derivative can be simplified to
\begin{align}
\frac{\partial E_{XC}}{\partial R_\tau} \approx & \sum_{C}^{N_{atom}} \sum_g^{N_{grid}(C)} \frac{\partial w_{gC}^{Becke}}{\partial R_\tau} w_{gC}^{quadrature} \varepsilon_{XC}\left( \rho(\vec{r}_{gC}), \vec{\nabla} \rho(\vec{r}_{gC}) \right) \notag \\
    & + \sum_{C}^{N_{atom}} \sum_g^{N_{grid}(C)} w_{gC} \left. \frac{\partial \varepsilon_{XC}}{\partial \rho} \right|_{\rho = \rho(\vec{r}_{gC})} \sum_{\mu\nu}^{n_{AO}} D_{\mu\nu} \left. \frac{\partial \left( \mu\nu \right)}{\partial R_\tau} \right|_{\vec{r} = \vec{r}_{gC}} \notag \\
    & - \sum_g^{N_{grid}(R)} w_{gR} \left. \frac{\partial \varepsilon_{XC}}{\partial \rho} \right|_{\rho = \rho(\vec{r}_{gR})} \sum_{\mu\nu}^{n_{AO}} D_{\mu\nu} \left( \left. \frac{\partial \left( \mu\nu \right)}{\partial A_\tau} \right|_{\vec{r} = \vec{r}_{gR}} +  \left. \frac{\partial \left( \mu\nu \right)}{\partial B_\tau} \right|_{\vec{r} = \vec{r}_{gR}} \right) \notag \\
    & + \sum_{C}^{N_{atom}} \sum_g^{N_{grid}(C)} w_{gC}  \left. \frac{\partial \varepsilon_{XC}}{\partial \vec{\nabla} \rho} \right|_{ \rho = \rho(\vec{r}_{gC}), \vec{\nabla} \rho = \vec{\nabla} \rho(\vec{r}_{gC}) } \cdot \sum_{\mu\nu}^{n_{AO}} D_{\mu\nu} \left. \frac{\partial \left( \vec{\nabla}_{r} (\mu\nu) \right) }{\partial R_\tau} \right|_{\vec{r} = \vec{r}_{gC}} \notag \\
    & - \sum_g^{N_{grid}(R)} w_{gR}  \left. \frac{\partial \varepsilon_{XC}}{\partial \vec{\nabla} \rho} \right|_{ \rho = \rho(\vec{r}_{gR}), \vec{\nabla} \rho = \vec{\nabla} \rho(\vec{r}_{gR}) } \cdot \sum_{\mu\nu}^{n_{AO}} D_{\mu\nu} \left( \left. \frac{\partial \left( \vec{\nabla}_{r} (\mu\nu) \right) }{\partial A_\tau} \right|_{\vec{r} = \vec{r}_{gR}} + \left. \frac{\partial \left( \vec{\nabla}_{r} (\mu\nu) \right) }{\partial B_\tau} \right|_{\vec{r} = \vec{r}_{gR}} \right) \notag \\
    & + f_{XC}\left(\frac{\partial D_{\mu\nu}}{\partial R_\tau}\right)
\label{eq:si_xc_derivative}
\end{align}

Details about how to obtain Becke weights $w_{gC}^{Becke}$ and its derivatives $\dfrac{\partial w_{gC}^{Becke}}{\partial R_\tau}$ can be found in the appendix B of Johnson, Gill and Pople's work.\cite{dft_gradient}

The new terms in equation \ref{eq:si_xc_derivative} are all derivatives of pair values:
\begin{align}
\left. \frac{\partial (\mu\nu)}{\partial A_x} \right|_{\vec{r}} = & C_\mu C_\nu \left( -i_x (x - A_x)^{i_x-1} (x - B_x)^{j_x} + 2a (x - A_x)^{i_x+1} (x - B_x)^{j_x} \right)  \notag \\
    & (y - A_y)^{i_y} (y - B_y)^{j_y} (z - A_z)^{i_z} (z - B_z)^{j_z} e^{-\frac{ab}{a+b} \left| \vec{A} - \vec{B} \right|^2} e^{-p \left| \vec{r} - \vec{P} \right|^2} \\
\left. \frac{\partial (\mu\nu)}{\partial B_x} \right|_{\vec{r}} = & C_\mu C_\nu \left( -j_x (x - A_x)^{i_x} (x - B_x)^{j_x-1} + 2b (x - A_x)^{i_x} (x - B_x)^{j_x+1} \right)  \notag \\
    & (y - A_y)^{i_y} (y - B_y)^{j_y} (z - A_z)^{i_z} (z - B_z)^{j_z} e^{-\frac{ab}{a+b} \left| \vec{A} - \vec{B} \right|^2} e^{-p \left| \vec{r} - \vec{P} \right|^2}
\label{eq:si_xc_dpair_dAB}
\end{align}
\begin{align}
\left. \frac{\partial \left( \vec{\nabla}_{r} (\mu\nu) \right)}{\partial A_x} \right|_{\vec{r}} = \left( \left.\frac{\partial^2 (\mu\nu)}{\partial A_x \partial x} \right|_{\vec{r}}, \left.\frac{\partial^2 (\mu\nu)}{\partial A_x \partial y} \right|_{\vec{r}}, \left.\frac{\partial^2 (\mu\nu)}{\partial A_x \partial z} \right|_{\vec{r}} \right) \\
\left. \frac{\partial \left( \vec{\nabla}_{r} (\mu\nu) \right)}{\partial B_x} \right|_{\vec{r}} = \left( \left.\frac{\partial^2 (\mu\nu)}{\partial B_x \partial x} \right|_{\vec{r}}, \left.\frac{\partial^2 (\mu\nu)}{\partial B_x \partial y} \right|_{\vec{r}}, \left.\frac{\partial^2 (\mu\nu)}{\partial B_x \partial z} \right|_{\vec{r}} \right)
\label{eq:si_xc_dpairgrad_dAB}
\end{align}
\begin{align}
\left.\frac{\partial^2 (\mu\nu)}{\partial A_x \partial x} \right|_{\vec{r}} = & C_\mu C_\nu \left( 2a(2i_x+1) (x - A_x)^{i_x} (x - B_x)^{j_x} \right. \notag \\
    & + 2bi_x (x - A_x)^{i_x-1} (x - B_x)^{j_x+1} + 2aj_x (x - A_x)^{i_x+1} (x - B_x)^{j_x-1} \notag \\
    & - i_x(i_x-1) (x - A_x)^{i_x-2} (x - B_x)^{j_x} - i_xj_x (x - A_x)^{i_x-1} (x - B_x)^{j_x-1} \notag \\
    &\left. - 4a^2 (x - A_x)^{i_x+2} (x - B_x)^{j_x} - 4ab (x - A_x)^{i_x+1} (x - B_x)^{j_x+1} \right) \notag \\
    & (y - A_y)^{i_y} (y - B_y)^{j_y} (z - A_z)^{i_z} (z - B_z)^{j_z} e^{-\frac{ab}{a+b} \left| \vec{A} - \vec{B} \right|^2} e^{-p \left| \vec{r} - \vec{P} \right|^2} \\
\left.\frac{\partial^2 (\mu\nu)}{\partial A_x \partial y} \right|_{\vec{r}} = & C_\mu C_\nu \left( -i_x (x - A_x)^{i_x-1} (x - B_x)^{j_x} + 2a (x - A_x)^{i_x+1} (x - B_x)^{j_x} \right)  \notag \\
    & \left( -2a (y - A_y)^{i_y+1} (y - B_y)^{j_y} - 2b (y - A_y)^{i_y} (y - B_y)^{j_y+1} \right. \notag \\
    & \left. + i_y (y - A_y)^{i_y-1} (y - B_y)^{j_y} + j_y (y - A_y)^{i_y} (y - B_y)^{j_y-1} \right) \notag \\
    & (z - A_z)^{i_z} (z - B_z)^{j_z} e^{-\frac{ab}{a+b} \left| \vec{A} - \vec{B} \right|^2} e^{-p \left| \vec{r} - \vec{P} \right|^2} \\
\left.\frac{\partial^2 (\mu\nu)}{\partial A_x \partial z} \right|_{\vec{r}} = & C_\mu C_\nu \left( -i_x (x - A_x)^{i_x-1} (x - B_x)^{j_x} + 2a (x - A_x)^{i_x+1} (x - B_x)^{j_x} \right)  \notag \\
    & (y - A_y)^{i_y} (y - B_y)^{j_y} \\
    & \left( -2a (z - A_z)^{i_z+1} (z - B_z)^{j_z} - 2b (z - A_z)^{i_z} (z - B_z)^{j_z+1} \right. \notag \\
    & \left. + i_z (z - A_z)^{i_z-1} (z - B_z)^{j_z} + j_z (z - A_z)^{i_z} (z - B_z)^{j_z-1} \right) e^{-\frac{ab}{a+b} \left| \vec{A} - \vec{B} \right|^2} e^{-p \left| \vec{r} - \vec{P} \right|^2}
\label{eq:si_xc_dpair_dA_dxyz}
\end{align}
\begin{align}
\left.\frac{\partial^2 (\mu\nu)}{\partial B_x \partial x} \right|_{\vec{r}} = & C_\mu C_\nu \left( 2b(2j_x+1) (x - A_x)^{i_x} (x - B_x)^{j_x} \right. \notag \\
    & + 2bi_x (x - A_x)^{i_x-1} (x - B_x)^{j_x+1} + 2aj_x (x - A_x)^{i_x+1} (x - B_x)^{j_x-1} \notag \\
    & - j_x(j_x-1) (x - A_x)^{i_x} (x - B_x)^{j_x-2} - i_xj_x (x - A_x)^{i_x-1} (x - B_x)^{j_x-1} \notag \\
    &\left. - 4b^2 (x - A_x)^{i_x} (x - B_x)^{j_x+2} - 4ab (x - A_x)^{i_x+1} (x - B_x)^{j_x+1} \right) \notag \\
    & (y - A_y)^{i_y} (y - B_y)^{j_y} (z - A_z)^{i_z} (z - B_z)^{j_z} e^{-\frac{ab}{a+b} \left| \vec{A} - \vec{B} \right|^2} e^{-p \left| \vec{r} - \vec{P} \right|^2} \\
\left.\frac{\partial^2 (\mu\nu)}{\partial B_x \partial y} \right|_{\vec{r}} = & C_\mu C_\nu \left( -j_x (x - A_x)^{i_x} (x - B_x)^{j_x-1} + 2b (x - A_x)^{i_x} (x - B_x)^{j_x+1} \right)  \notag \\
    & \left( -2a (y - A_y)^{i_y+1} (y - B_y)^{j_y} - 2b (y - A_y)^{i_y} (y - B_y)^{j_y+1} \right. \notag \\
    & \left. + i_y (y - A_y)^{i_y-1} (y - B_y)^{j_y} + j_y (y - A_y)^{i_y} (y - B_y)^{j_y-1} \right) \notag \\
    & (z - A_z)^{i_z} (z - B_z)^{j_z} e^{-\frac{ab}{a+b} \left| \vec{A} - \vec{B} \right|^2} e^{-p \left| \vec{r} - \vec{P} \right|^2} \\
\left.\frac{\partial^2 (\mu\nu)}{\partial B_x \partial z} \right|_{\vec{r}} = & C_\mu C_\nu \left( -j_x (x - A_x)^{i_x} (x - B_x)^{j_x-1} + 2b (x - A_x)^{i_x} (x - B_x)^{j_x+1} \right) \notag \\
    & (y - A_y)^{i_y} (y - B_y)^{j_y} \\
    & \left( -2a (z - A_z)^{i_z+1} (z - B_z)^{j_z} - 2b (z - A_z)^{i_z} (z - B_z)^{j_z+1} \right. \notag \\
    & \left. + i_z (z - A_z)^{i_z-1} (z - B_z)^{j_z} + j_z (z - A_z)^{i_z} (z - B_z)^{j_z-1} \right) e^{-\frac{ab}{a+b} \left| \vec{A} - \vec{B} \right|^2} e^{-p \left| \vec{r} - \vec{P} \right|^2}
\label{eq:si_xc_dpair_dB_dxyz}
\end{align}

\section{Integral Derivative Implementation}

Similar to the Fock term, the implementations of the overlap and kinetic energy integral derivatives are also not GPU accelerated. The algorithm first allocates and zeroes the space of size $3 N_{atom}$ for the gradient vector. Then for each angular pair, we iterate through all primitive shell pairs, and for each pair, we fetch its corresponding density matrix values (there can be more than one value, like in PP kernel, each primitive shell pair maps to 9 density matrix values), compute the overlap or kinetic energy integral derivative with respect to the $\mu$ orbital center $\vec{A}$ and $\nu$ orbital center $\vec{B}$, multiply the derivative with density matrix value, and sum the derivative result into the gradient vector according to atomic indices of $A$ and $B$. Since the operation is sequential, we do not worry about atomic addition. After iterating through all primitive shell pair, we finish the computation of $\displaystyle\sum\limits_{\mu\nu}^{n_{AO}} W_{\mu\nu} \dfrac{\partial}{\partial R_\tau} S_{\mu\nu}$ or $\displaystyle\sum\limits_{\mu\nu}^{n_{AO}} D_{\mu\nu} \frac{\partial}{\partial R_\tau} T_{\mu\nu}$.

Implementing the derivatives of the nuclear repulsion integral is complicated because we also need to take care of derivative with respect to point charge location. In order to avoid atomic write operation on GPU, we design the following GPU algorithm: we assign each GPU thread a primitive shell pair $\mu\nu$, and allocate a memory space of 6 numbers for the derivative of $\displaystyle\sum\limits_{C}^{N_{point-charge}} V_{\mu\nu C} D_{\mu\nu}$ with respect to shell-pair centers $\vec{A}$ and $\vec{B}$ (the $D_{\mu\nu}$ dependency on atom position is not considered here). In addition, for each thread block, we allocate a memory space of $3 N_{point-charge}$ to store the derivative of $\displaystyle\sum\limits_{\mu\nu \in block} V_{\mu\nu C}D_{\mu\nu}$ with respect to each point charge location $\vec{C}$. In each thread block we will perform an internal summation of $D_{\mu\nu} \dfrac{\partial}{\partial C_\tau} V_{\mu\nu C}$ among each pair associated to each thread, which is what $(\mu\nu \in block)$ in the summation index means. The purpose of this reduction operation is to save GPU memory, as we do not want to allocate $3 N_{point-charge}$ of space for each thread. With the GPU memory properly allocated, a thread will iterate through the list of point charges, and at each iteration, it computes the pair-center position derivatives $D_{\mu\nu} \dfrac{\partial}{\partial A_\tau} V_{\mu\nu C}$ and $D_{\mu\nu} \dfrac{\partial}{\partial B_\tau} V_{\mu\nu C}$ and accumulates them to its local registers, as well as the point charge location derivatives $D_{\mu\nu} \dfrac{\partial}{\partial C_\tau} V_{\mu\nu C}$. The whole thread block will then synchronize and sum the $D_{\mu\nu} \dfrac{\partial}{\partial C_\tau} V_{\mu\nu C}$, and once the summation is done, the result is written back to GPU memory. The synchronization and write-back happens every iteration. After all GPU kernel calls have returned, we sum up the derivative values with respect to $\vec{C}$ from all thread blocks to obtain $\displaystyle\sum\limits_{\mu\nu}^{n_{AO}} D_{\mu\nu} \dfrac{\partial}{\partial C_\tau} V_{\mu\nu C}$.

The Coulomb matrix implementation is greatly simplified due to the separation of transformation coefficients $E^{i_\tau, j_\tau}_{t_\tau, \tau}$ and auxiliary integrals $R_{t_x,t_y,t_z}^0$, and this applies to the derivative of Coulomb energy as well:
\begin{align}
\sum_{\mu\nu}^{n_{AO}} D_{\mu\nu} \sum_{\lambda\sigma}^{n_{AO}} D_{\lambda\sigma} \left( \frac{\partial \mu\nu}{\partial A_\tau} \middle| \lambda\sigma \right) &= \sum_{P+1} D_{\mu\nu} \frac{\partial E_P^{\mu\nu}}{\partial A_\tau} \sum_Q R_{PQ} \sum_{\lambda\sigma} E_Q^{\lambda\sigma} D_{\lambda\sigma} \notag \\
\sum_{\mu\nu}^{n_{AO}} D_{\mu\nu} \sum_{\lambda\sigma}^{n_{AO}} D_{\lambda\sigma} \left( \frac{\partial \mu\nu}{\partial B_\tau} \middle| \lambda\sigma \right) &= \sum_{P+1} D_{\mu\nu} \frac{\partial E_P^{\mu\nu}}{\partial B_\tau} \sum_Q R_{PQ} \sum_{\lambda\sigma} E_Q^{\lambda\sigma} D_{\lambda\sigma}
\label{eq:si_dj_formula}
\end{align}
where $\dfrac{\partial E_P^{\mu\nu}}{\partial A_\tau}$ and $\dfrac{\partial E_P^{\mu\nu}}{\partial B_\tau}$ represent the derivative of $E^{i_\tau, j_\tau}_{t_\tau, \tau}$ terms shown in equation \ref{eq:si_eri_derivative_A} and \ref{eq:si_eri_derivative_B}, and $P+1$ in the summation index emphasizes the +1 in the Hermite Gaussian index summation ($t_\tau \in [0, i_\tau + j_\tau + 1]$) in the Cartesian direction of the differentiation. Since the last two summations have the exact same form as in Coulomb matrix calculation (equation \ref{eq:j_formula}), we can reuse the Hermite density computation code and GPU-accelerated auxiliary integral code developed for that purpose. In order to handle the additional summation index in $P$ (or precisely, $\vec{t}$), we call the auxiliary integral kernel with one angular momentum higher in $\mu$ or $\nu$. If the angular momentum of $\mu$ ($L_\mu$) is lower than the angular momentum of $\nu$ ($L_\nu$), then we call the kernel of type $((L_\mu + 1) L_\nu |L_\lambda L_\sigma)$, otherwise we call the kernel of type $( L_\mu (L_\nu + 1) | L_\lambda L_\sigma)$. For example, when computing the derivative of $(ff|ff)$ type integral, we call the $(fg|ff)$ auxiliary integral kernel. This will provide us necessary $R_{t_x,t_y,t_z}^0$ terms for derivative calculation. We therefore had to implement 10 more auxiliary integral kernel functions of type $(fg|L_\lambda L_\sigma)$ to support the derivatives of all $f$ orbital based ERI integrals (not accounting for single/double precision). Once the GPU auxiliary integral GPU kernel returns, we obtain the $\mathbf{J}$ matrix elements in Hermite Gaussian basis with one additional $P$ index. Then, on CPU, we compute $\dfrac{\partial E_P^{\mu\nu}}{\partial A_\tau}$ and $\dfrac{\partial E_P^{\mu\nu}}{\partial B_\tau}$ to transform the Hermite $\mathbf{J}$ to derivatives of $\mathbf{J}$ matrix elements in the Cartesian Gaussian basis, and contract it with density matrix elements to form the derivative of $E_J$.

The derivative implementation of the HF exchange energy is also ``simplified" by the loss of permutational symmetry. In the derivative form in equation \ref{eq:si_dE_K}, almost all 8-fold symmetry of ERI is lost, except for one, where we interchange $\mu$ and $\nu$ as well as $\lambda$ and $\sigma$ at the same time:
\begin{align}
\sum_{\mu\lambda}^{n_{AO}} D_{\mu\lambda} \sum_{\nu\sigma}^{n_{AO}} D_{\nu\sigma} \left( \frac{\partial \mu\nu}{\partial R_\tau} \middle| \lambda\sigma \right) = \sum_{\nu\sigma}^{n_{AO}} D_{\nu\sigma} \sum_{\mu\lambda}^{n_{AO}} D_{\mu\lambda} \left( \frac{\partial \nu\mu}{\partial R_\tau} \middle| \sigma\lambda \right)
\label{eq:si_dE_K_symmetry}
\end{align}
As a consequence, there are 160 different angular momentum combinations, and we need to generate the same number of GPU kernel functions (not accounting for single/double precision). In each kernel, in order to avoid the atomic write of derivative result on GPU, we assign each thread a bra pair, and let each thread block loop through all ket pairs, similar to the logic in the Coulomb matrix auxiliary integrals. Consequently, at each ket pair iteration, each thread will get a new set of $\lambda$ and $\sigma$ indices, and will have to access both density matrices $D_{\mu\lambda}$ and $D_{\nu\sigma}$ in an uncoalesced memory access fashion. The threads will accumulate into their local registers the value of $D_{\mu\lambda} D_{\nu\sigma} \dfrac{\partial \mu\nu}{\partial A_\tau}$ and $D_{\mu\lambda} D_{\nu\sigma} \dfrac{\partial \mu\nu}{\partial B_\tau}$, and eventually perform an internal summation on the ket direction within a thread block, again very similar to the Coulomb matrix auxiliary integral logic. The result is copied back to CPU and sequentially summed into the gradient vector according to the atomic index of $A$ and $B$ for each pair. As an important note, we place the $E^{i_\tau, j_\tau}_{t_\tau, \tau}$ term computation into the HF exchange derivatives as well, which indeed provides us with run time improvement. However it also makes the HF exchange derivative GPU kernels a severe compilation bottleneck, which makes development painful and results in gigabytes of CUDA source code. We are presently investigating good ways to split kernel functions into small relocatable device functions, with the objective of reducing compile time without hurting the run time performance.\cite{libintx_up_to_iiii}

For the derivative implementation of exchange-correlation energy, the density and exchange-correlation term evaluation on each grid point is unchanged. However, when performing the summation of exchange-correlation term over grid points, extra care needs to be taken about the ``source" of each grid point. We redesign the box data structure mentioned in the main text as follows: we first group all grid points according to the associated atom, and then for every atom, we construct a bounding box and split them into smaller boxes. In the GPU kernel, each thread will hold a primitive shell pair, and only iterate through all boxes (and thus all grid points) belonging to a particular atom. That means we need to set up $N_{pair} \times N_{atom}$ threads. Each thread will compute $D_{\mu\nu} \displaystyle\sum\limits_g^{N_{grid}(C)} w_{gC} \left. \dfrac{\partial \varepsilon_{XC}}{\partial \rho} \right|_{\rho = \rho(\vec{r}_{gC})} \left. \dfrac{\partial \left( \mu\nu \right)}{\partial A_\tau} \right|_{\vec{r} = \vec{r}_{gC}}$ and $D_{\mu\nu} \displaystyle\sum\limits_g^{N_{grid}(C)} w_{gC} \left. \dfrac{\partial \varepsilon_{XC}}{\partial \rho} \right|_{\rho = \rho(\vec{r}_{gC})} \left. \dfrac{\partial \left( \mu\nu \right)}{\partial B_\tau} \right|_{\vec{r} = \vec{r}_{gC}}$ and the corresponding GGA terms, where $C$ is the atom that these grid points belong to. The result is copied back to host memory and distributed into the final gradient vector according to the pattern in equation \ref{eq:si_xc_derivative}.

\newpage

\section{Performance on Different GPUs for Organic Molecules}

\begin{figure}[htb!]
    \centering
    \includegraphics[width=0.6\linewidth]{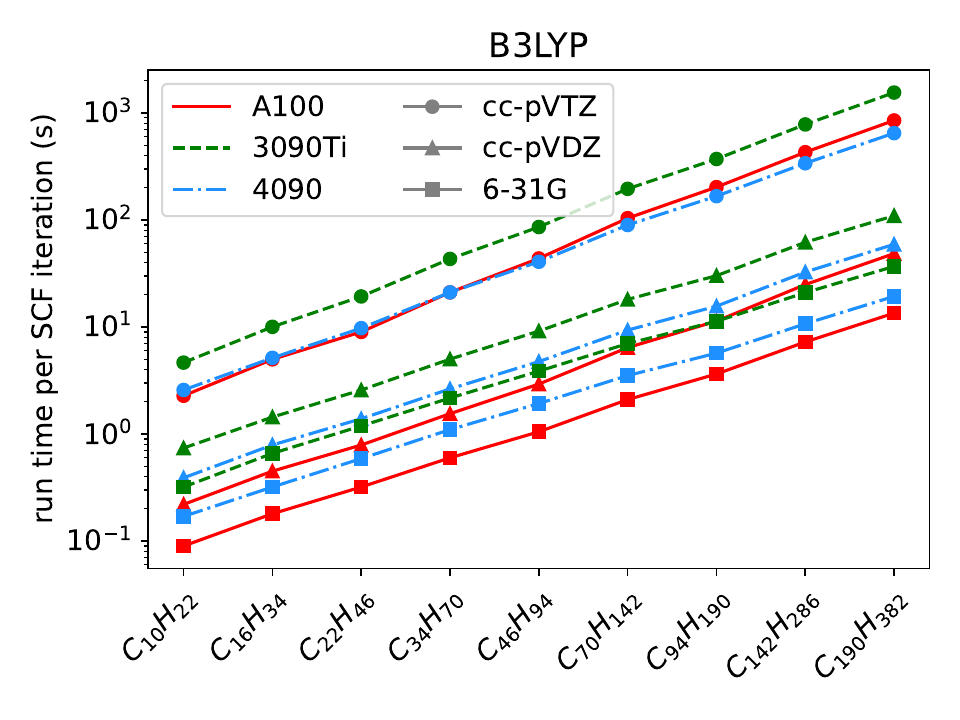}
\caption{Run time per SCF iteration (averaged over all cycles) for TeraChem B3LYP calculations on branched alkanes, compared between three types of GPUs: NVIDIA A100, NVIDIA RTX 4090, and NVIDIA RTX 3090Ti. A single GPU was used for all calculations. 
}
\label{fig:si_alkane_overall_runtime_gpus}
\end{figure}

\begin{figure}[htb!]
    \centering
    \includegraphics[width=0.6\linewidth]{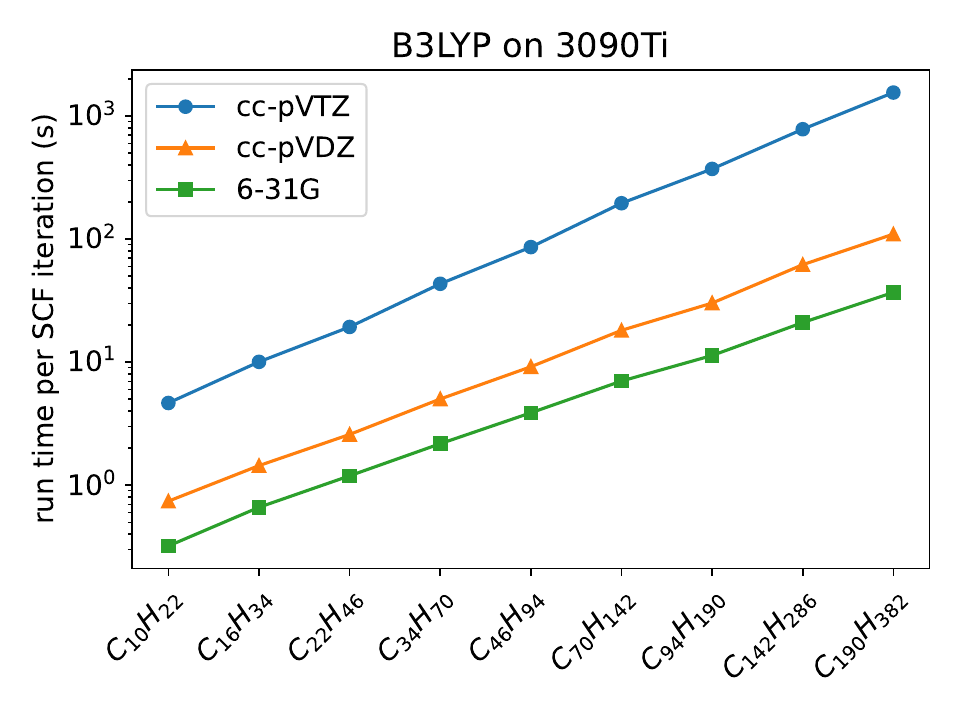}
\caption{Run time per SCF iteration (averaged over all cycles) for TeraChem B3LYP calculations on branched alkanes, using one NVIDIA RTX 3090Ti GPU. We use the 6-31G, cc-pVDZ and cc-pVTZ basis sets. The scaling of run time with respect to the number of basis functions ($n_{AO}$) is $O(n_{AO}^{2.0})$ for the B3LYP/cc-pVTZ calculation, $O(n_{AO}^{1.7})$ for the B3LYP/cc-pVDZ calculation and $O(n_{AO}^{1.6})$ for the B3LYP/6-31G calculation.
}
\label{fig:si_alkane_overall_runtime_3090ti}
\end{figure}

\begin{figure}[htb!]
    \centering
    \includegraphics[width=0.45\linewidth]{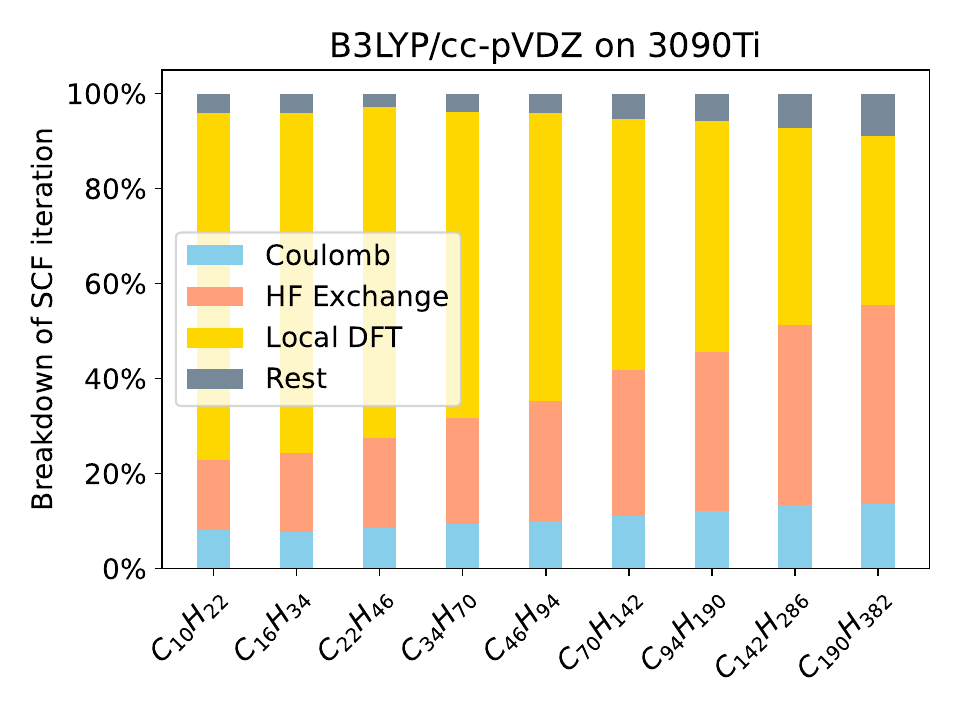}
    \includegraphics[width=0.45\linewidth]{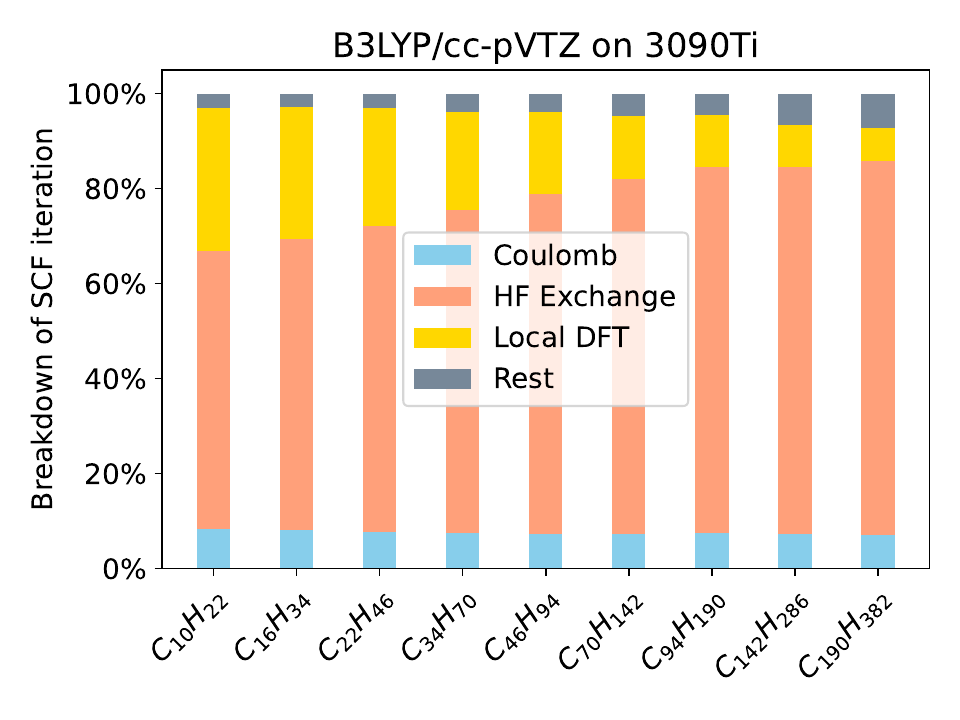}
\caption{Distribution of SCF iteration run time components (averaged over all cycles) for B3LYP/cc-pVDZ (left) and B3LYP/cc-pVTZ (right) calculations on branched alkanes, using one NVIDIA RTX 3090Ti GPU.
}
\label{fig:si_alkane_runtime_component_3090ti}
\end{figure}

\begin{figure}[htb!]
    \centering
    \includegraphics[width=0.6\linewidth]{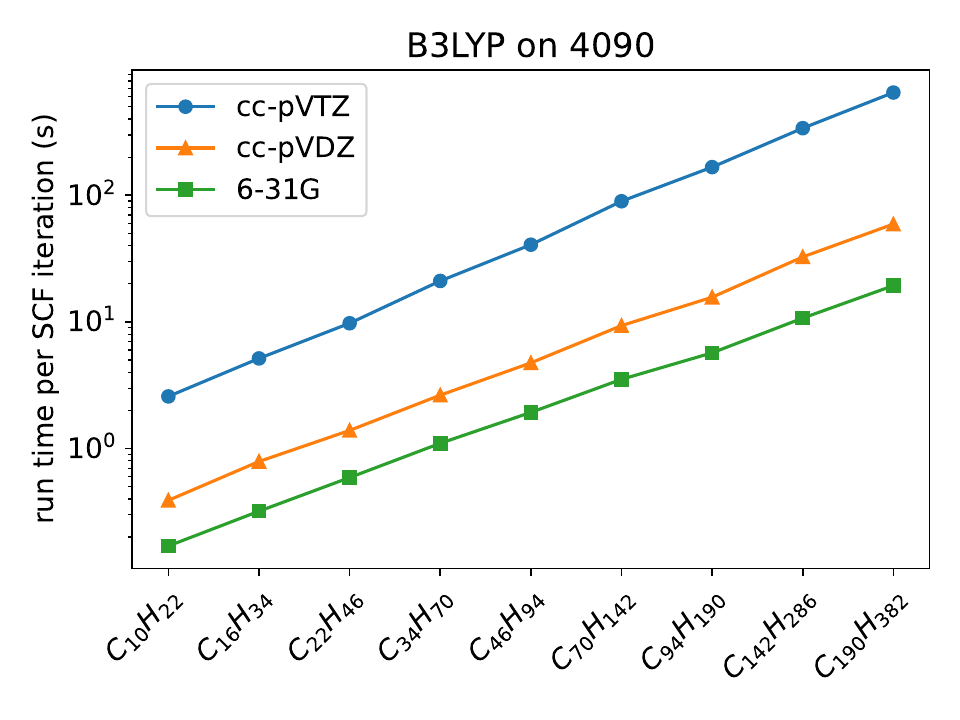}
\caption{Run time per SCF iteration (averaged over all cycles) for TeraChem B3LYP calculations on branched alkanes, using one NVIDIA RTX 4090 GPU. We use the 6-31G, cc-pVDZ and cc-pVTZ basis sets. The scaling of run time with respect to the number of basis functions ($n_{AO}$) is $O(n_{AO}^{1.9})$ for the B3LYP/cc-pVTZ calculation, $O(n_{AO}^{1.7})$ for the B3LYP/cc-pVDZ calculation and $O(n_{AO}^{1.6})$ for the B3LYP/6-31G calculation.
}
\label{fig:si_alkane_overall_runtime_4090}
\end{figure}

In this section we provide TeraChem timings for branched alkanes and other organic molecules on the NVIDIA RTX 3090Ti and RTX 4090 GPUs, using mixed precision. The RTX 4090 GPU generally leads to the smallest run times (even in comparison to the A100 GPU results reported in the main text, as shown in Table \ref{tab:si_timing_comparison_gpu} and Fig. \ref{fig:si_alkane_overall_runtime_gpus}). The TeraChem computation time grows quadratically (or slower) with the size of the branched alkanes (as shown in Figs. \ref{fig:si_alkane_overall_runtime_3090ti} and \ref{fig:si_alkane_overall_runtime_4090}). The cc-pVTZ run times are furthermore dominated by the evaluation of HF exchange (as shown in Figs. \ref{fig:si_alkane_runtime_component_3090ti} and \ref{fig:si_alkane_overall_runtime_4090}). Comparison to BrianQC is also made for the RTX 3090Ti GPU (as shown in Table \ref{tab:si_comparison_to_brianqc_3090ti}).

\newpage

\begin{table}[htb!]
\begin{tabular}{c | c | r | r | r | r} 
 \hline
 Molecule & Basis Set & $n_{AO}$ & A100 (s) & 4090 (s) & 3090Ti (s) \\
 \hline
 \multirow{2}{8em}{\ce{C46H94}} & cc-pVDZ & 1160 & 3 & 5 & 9 \\
 & cc-pVTZ & 3020 & 44 & 41 & 84 \\
 \hline
\multirow{2}{8em}{\ce{C70H142}} & cc-pVDZ & 1760 & 6 & 9 & 18 \\
 & cc-pVTZ & 4580 & 104 & 90 & 192 \\
 \hline
 \multirow{2}{8em}{\ce{C94H190}} & cc-pVDZ & 2360 & 11 & 16 & 29 \\
 & cc-pVTZ & 6140 & 203 & 167 & 364 \\
 \hline
 \multirow{2}{8em}{fullerene (\ce{C60})} & cc-pVDZ & 900 & 2 & 3 & 6 \\
 & cc-pVTZ & 2100 & 45 & 37 & 80 \\
 \hline
 \multirow{2}{10em}{taxol (\ce{C47H51NO14})}& cc-pVDZ & 1185 & 2 & 3 & 6 \\
 & cc-pVTZ & 2935 & 35 & 30 & 66 \\
 \hline
\multirow{2}{12em}{valinomycin (\ce{C54H90N6O18})} & cc-pVDZ & 1620 & 4 & 5 & 9 \\
 & cc-pVTZ & 4080 & 65 & 57 & 119 \\
 \hline
\end{tabular}
\caption{Run time per SCF iteration (averaged over all cycles) for TeraChem B3LYP/cc-pVDZ and B3LYP/cc-pVTZ calculations on several organic molecules, performed on three different types of GPUs: NVIDIA A100, NVIDIA RTX 4090, and NVIDIA RTX 3090Ti. A single GPU was used for all calculations. }
\label{tab:si_timing_comparison_gpu}
\end{table}

\newpage

\newpage

\begin{figure}[htb!]
    \centering
    \includegraphics[width=0.45\linewidth]{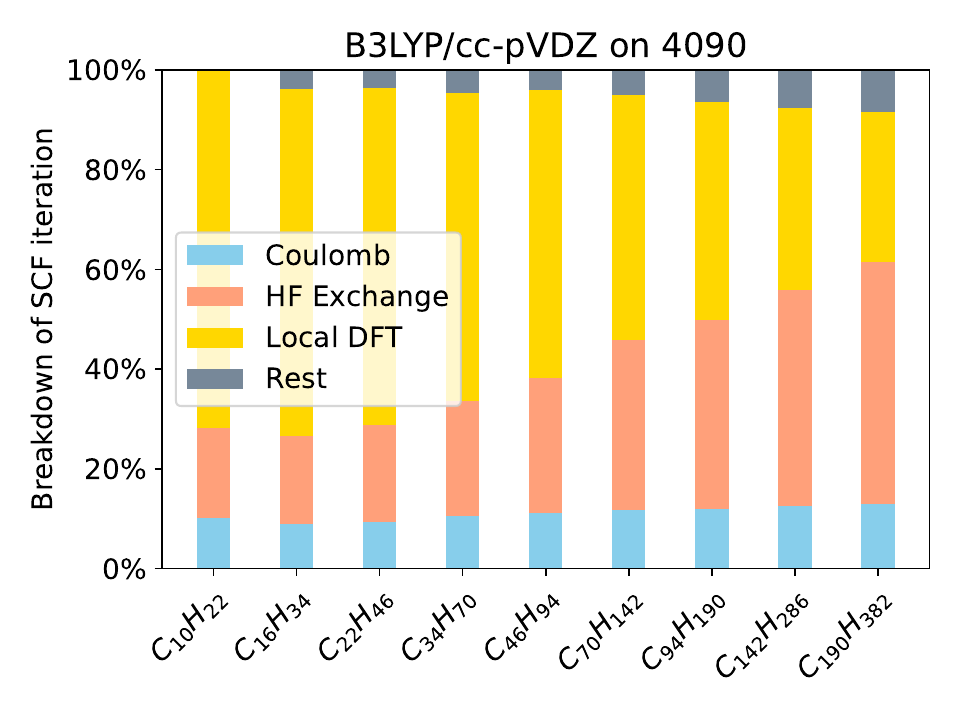}
    \includegraphics[width=0.45\linewidth]{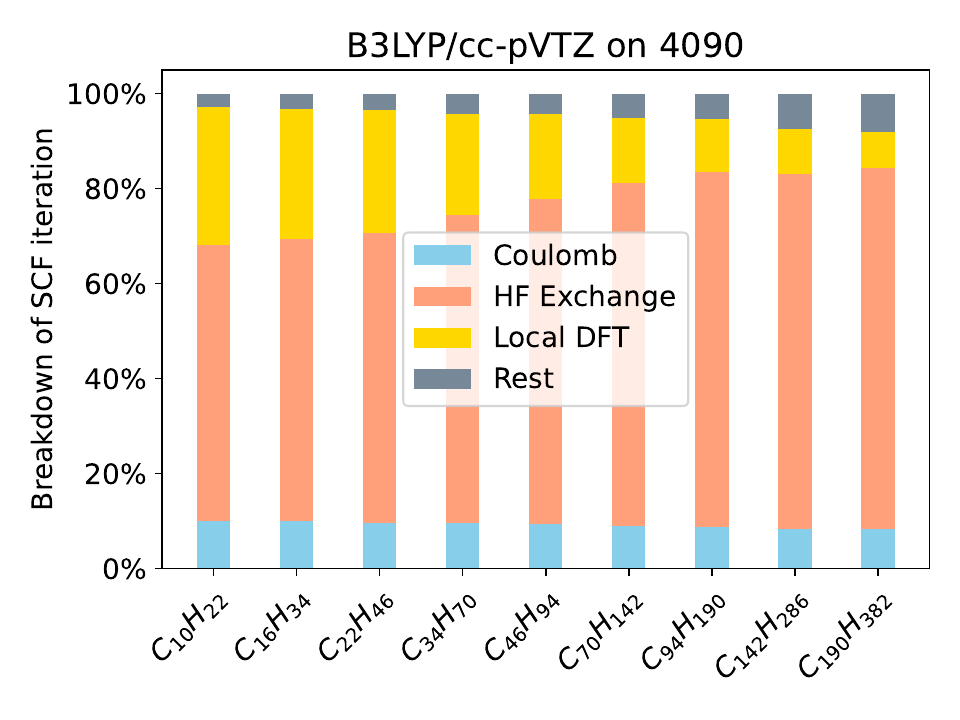}
\caption{Distribution of SCF iteration run time components (averaged over all cycles) for B3LYP/cc-pVDZ (left) and B3LYP/cc-pVTZ (right) calculations on branched alkanes, using one NVIDIA RTX 4090 GPU.
}
\label{fig:si_alkane_runtime_component_4090}
\end{figure}

\newpage

\begin{table}[htb!]
\begin{tabular}{c | c | r | r | r} 
 \hline
 Molecule & Basis Set & $n_{AO}$ & TeraChem (s) & BrianQC (s) \\
 \hline
 \multirow{2}{4em}{\ce{C46H94}} & cc-pVDZ & 1160 & 9 & 14 \\
 & cc-pVTZ & 3020 & 84 & 250 \\
 \hline
\multirow{2}{4em}{\ce{C70H142}} & cc-pVDZ & 1760 & 18 & 34 \\
 & cc-pVTZ & 4580 & 192 & 733 \\
 \hline
 \multirow{2}{4em}{\ce{C94H190}} & cc-pVDZ & 2360 & 29 & 72 \\
 & cc-pVTZ & 6140 & 364 & 1633 \\
 \hline
\multirow{2}{8em}{fullerene (\ce{C60})} & cc-pVDZ & 900 & 6 & 12 \\
 & cc-pVTZ & 2100 & 80 & 156 \\
 \hline
 \multirow{2}{10em}{taxol (\ce{C47H51NO14})} & cc-pVDZ & 1185 & 6 & 10 \\
 & cc-pVTZ & 2935 & 66 & 156 \\
 \hline
\multirow{2}{12em}{valinomycin (\ce{C54H90N6O18})} & cc-pVDZ & 1620 & 9 & 19 \\
 & cc-pVTZ & 4080 & 119 & 331 \\
 \hline
\end{tabular}
\caption{Run time per SCF iteration (averaged over all cycles) for B3LYP/cc-pVDZ and B3LYP/cc-pVTZ calculations on several organic molecules, compared between the present TeraChem implementation and BrianQC \cite{brianqc}. A single NVIDIA RTX 3090Ti GPU was used for all calculations. For BrianQC we only report the sum of the Coulomb, HF exchange, and exchange-correlation computation times, which the program reports rounded to the nearest second. The matrix linear algebra time per iteration, and other miscellaneous components is therefore not included in the BrianQC timing.}
\label{tab:si_comparison_to_brianqc_3090ti}
\end{table}

\newpage
\section{Performance on Different GPUs for 
Transition Metal Containing Molecules}

In this section we provide TeraChem timings for some transition metal containing species on the NVIDIA RTX 3090Ti and RTX 4090 GPUs with mixed precision, in Table \ref{tab:si_timing_comparison_gpu_metal}. Comparison to BrianQC is also made for the RTX 3090Ti GPU (as shown in Table \ref{tab:si_comparison_to_brianqc_3090ti_metal}).  

\begin{table}[htb!]
\begin{tabular}{c | c | r| r | r | r} 
 \hline
 Molecule & Basis Set & $n_{AO}$ & A100 (s) & 4090 (s) & 3090Ti(s)\\
 \hline
 \multirow{2}{18em}{Tetra-aza Co(II) complex\cite{tetraaza_complex} +\ce{CO2} \ce{[CoC16H22N4O2]+}} & def2-SVP & 476 & 1 & 2 & 3\\
 & def2-TZVP & 976 & 6 & 7 &12\\
 \hline
 \multirow{2}{18em}{$\mu$-alkyl dicopper(I) complex\cite{dicopper_complex} \ce{[Cu2C35H31N6F2]+}} & def2-SVP & 872 & 2 & 4 & 6\\
 & def2-TZVP & 1838 & 27 & 29 & 56\\
 \hline
 \multirow{2}{18em}{MOF-5\cite{mof5_original}  Cluster model\cite{mof5_geometry}(0.5 pore) \ce{Zn8C28H34O26}} & def2-SVP & 1268 & 3 & 4 & 7\\
 & def2-TZVP & 2588 & 24 & 24 & 47\\
 \hline
  \multirow{2}{18em}{MOF-5\cite{mof5_original}  Cluster model\cite{mof5_geometry} (1 pore) \ce{Zn16C64H64O52}} & def2-SVP & 2636 & 9 & 11 & 21 \\
 & def2-TZVP & 5440 & 96 & 83 & 178\\
 \hline
\end{tabular}
\caption{Run time per SCF iteration (averaged over all cycles) for TeraChem B3LYP/def2-SVP and B3LYP/def2-TZVP calculations on several metal containing molecules, performed on three different types of GPUs: NVIDIA A100, NVIDIA RTX 4090, and NVIDIA RTX 3090Ti. A single GPU was used for all calculations.}
\label{tab:si_timing_comparison_gpu_metal}
\end{table}

\begin{table}[htb!]
\begin{tabular}{c | c | r| r | r} 
 \hline
 Molecule & Basis Set & $n_{AO}$ & TeraChem (s) & BrianQC (s) \\
 \hline
 \multirow{2}{18em}{Tetra-aza Co(II) complex\cite{tetraaza_complex} +\ce{CO2} \ce{[CoC16H22N4O2]+}} & def2-SVP & 476 & 3 & 4 \\
 & def2-TZVP & 976 & 12 & 18 \\
 \hline
 \multirow{2}{18em}{$\mu$-alkyl dicopper(I) complex\cite{dicopper_complex} \ce{[Cu2C35H31N6F2]+}} & def2-SVP & 872 & 6 & 8 \\
 & def2-TZVP & 1838 & 56 & 69 \\
 \hline
 \multirow{2}{18em}{MOF-5\cite{mof5_original}  Cluster model\cite{mof5_geometry}(0.5 pore) \ce{Zn8C28H34O26}} & def2-SVP & 1268 & 7 & 11 \\
 & def2-TZVP & 2588 & 47 & 94 \\
 \hline
  \multirow{2}{18em}{MOF-5\cite{mof5_original}  Cluster model\cite{mof5_geometry} (1 pore) \ce{Zn16C64H64O52}} & def2-SVP & 2636 & 21 & 39 \\
 & def2-TZVP & 5440 & 178 & 469 \\
 \hline
\end{tabular}
\caption{Run time per SCF iteration (averaged over all cycles) for B3LYP/def2-SVP and B3LYP/def2-TZVP calculations on several metal containing molecules, compared between the present TeraChem implementation and BrianQC \cite{brianqc}. A single NVIDIA RTX 3090Ti GPU was used for all calculations. For BrianQC we only report the sum of the Coulomb, HF exchange, and exchange-correlation computation times, which the program reports rounded to the nearest second. The matrix linear algebra time per iteration, and other miscellaneous components is therefore not included in the BrianQC timing.}
\label{tab:si_comparison_to_brianqc_3090ti_metal}
\end{table}

\newpage

\section{Profiling of Compute Times over Kernels}
We show the distribution of compute times over kernels of different angular momentum for the NVIDIA RTX 4090 GPU, as well as an additional system (a cluster carved out from the ZnS Wurtzite crystal with the def2-SVP basis).

\begin{figure}[htb!]
    \centering
    \includegraphics[width=0.6\linewidth]{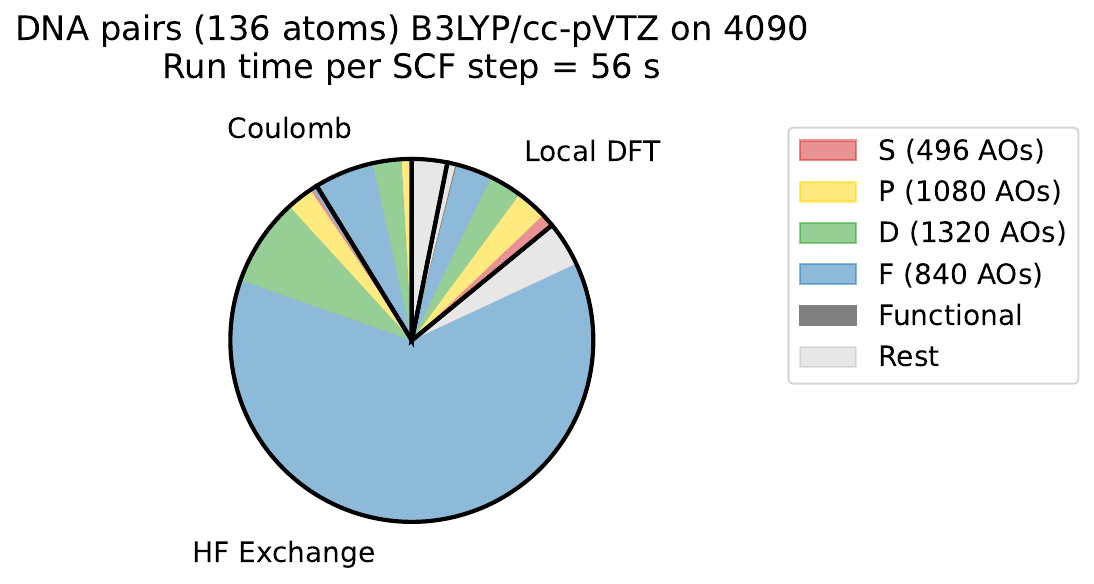}
\caption{Kernel run time distributions across SCF iterations for a model organic (two DNA CG base pairs, B3LYP/cc-pVTZ) system on one NVIDIA RTX 4090 GPU.}
\label{fig:si_angular_component_organic_4090}
\end{figure}

\begin{figure}[htb!]
    \centering
    \includegraphics[width=0.6\linewidth]{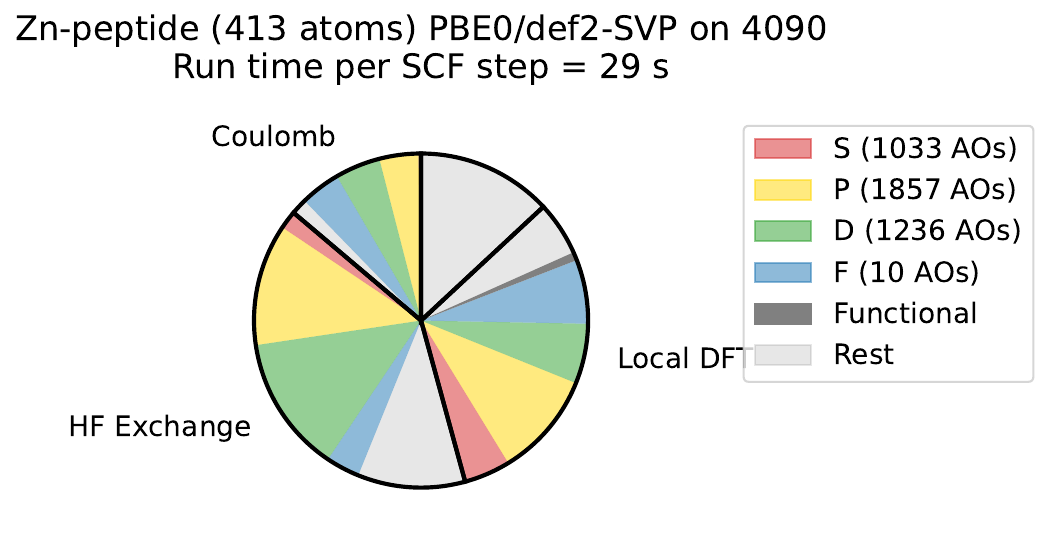}
\caption{Kernel run time distributions across SCF iterations for a model bioinorganic system (protein 6UFA, PBE0/def2-SVP) system on one NVIDIA RTX 4090 GPU.}
\label{fig:si_angular_component_metalorganic_4090}
\end{figure}

\newpage

\begin{figure}[htb!]
    \centering
    \includegraphics[width=0.6\linewidth]{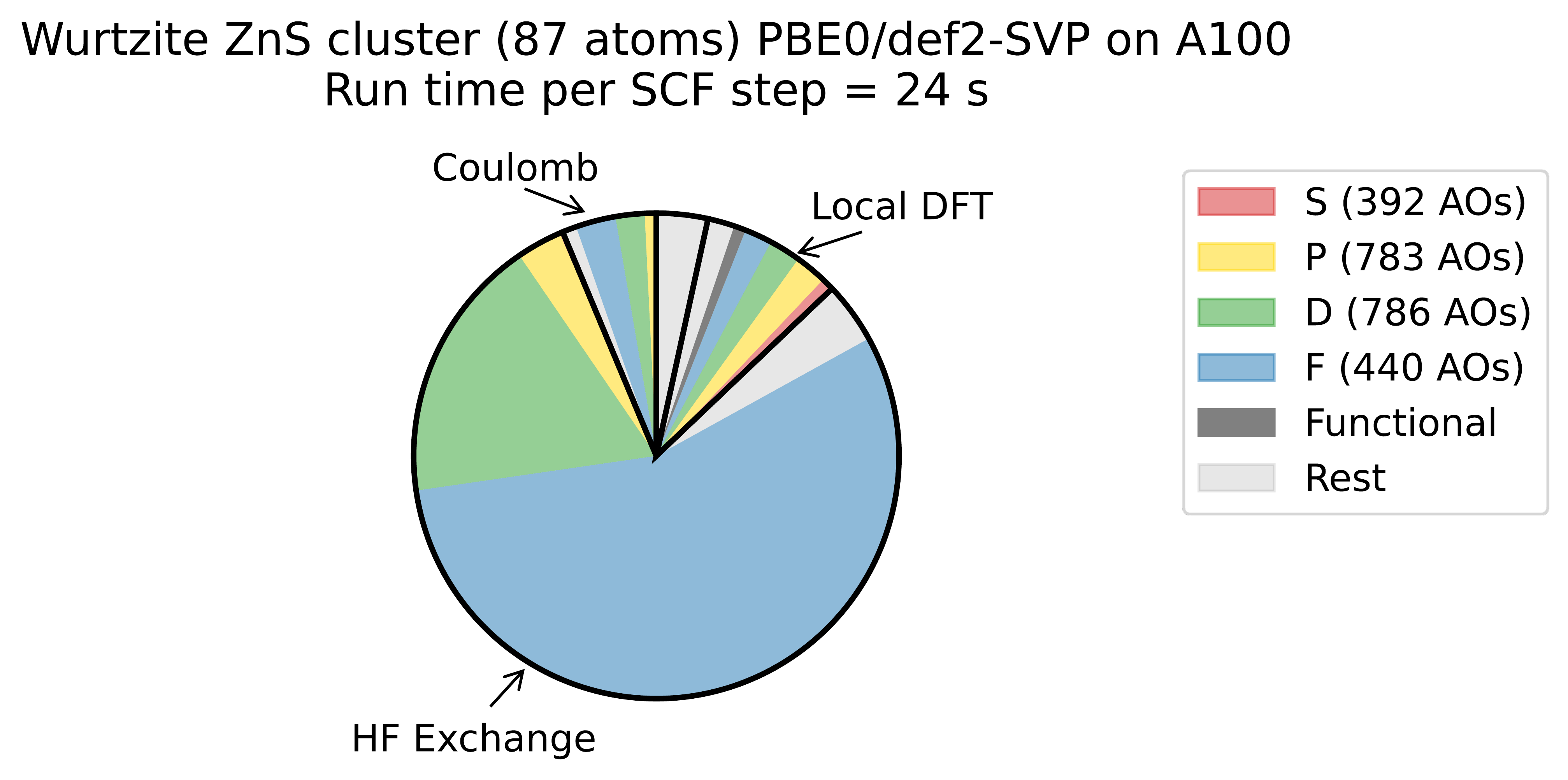}
\caption{Kernel run time distributions across SCF iterations for a model inorganic crystal cluster (Wurtzite zinc sulfide, PBE0/def2-SVP) system on one NVIDIA A100 GPU.}
\label{fig:si_angular_component_cluster_a100}
\end{figure}

\begin{figure}[htb!]
    \centering
    \includegraphics[width=0.6\linewidth]{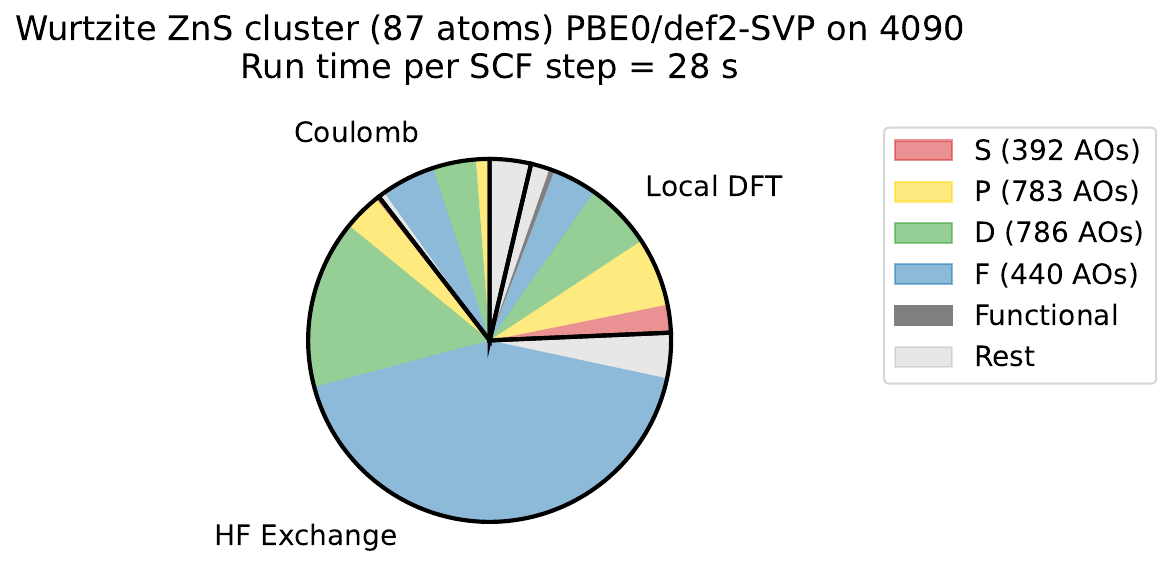}
\caption{Kernel run time distributions across SCF iterations for a model inorganic crystal cluster (Wurtzite zinc sulfide, PBE0/def2-SVP) system on one NVIDIA RTX 4090 GPU.}
\label{fig:si_angular_component_cluster_4090}
\end{figure}

\newpage

\section{Performance of HF Exchange Matrix Kernels against Kernel Splitting Strategies}

\begin{figure}[htb!]
\begin{minipage}{0.48\textwidth}
    \centering
    \includegraphics[width=\linewidth] {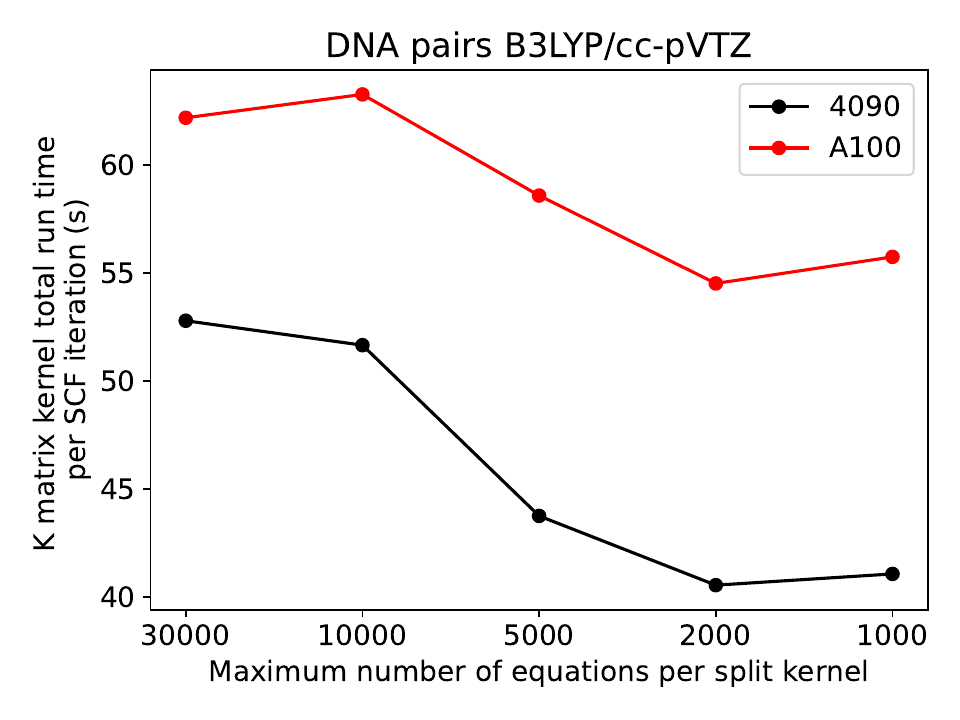}
\end{minipage}
 \begin{minipage}{0.48\textwidth}
    \centering
    \includegraphics[width=\linewidth] {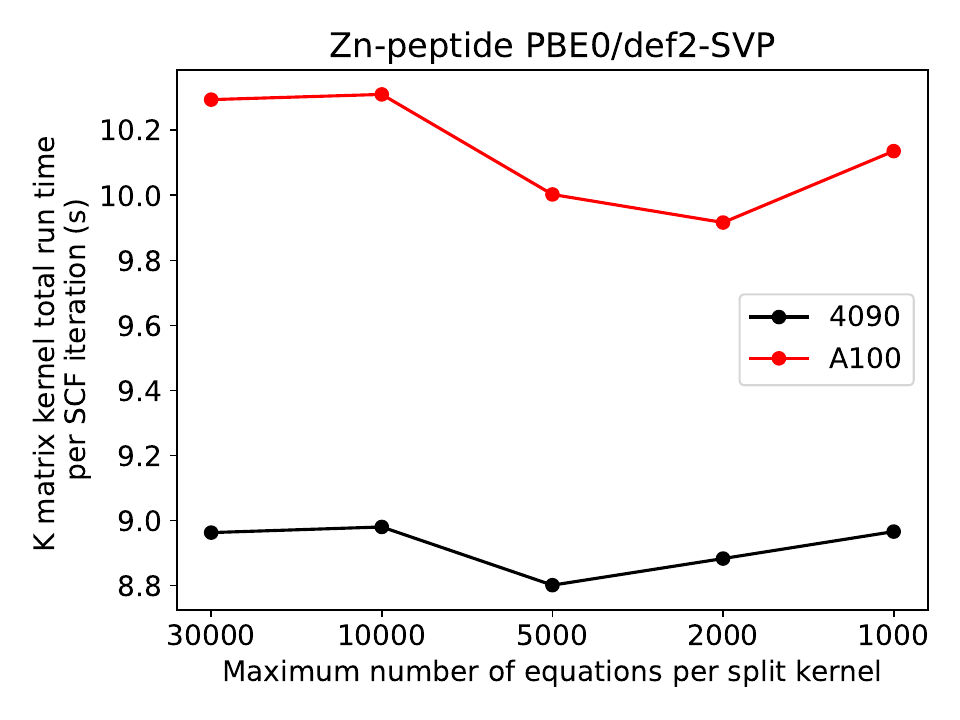}
\end{minipage}
  \begin{minipage}{0.48\textwidth}
    \centering
    \includegraphics[width=\linewidth] {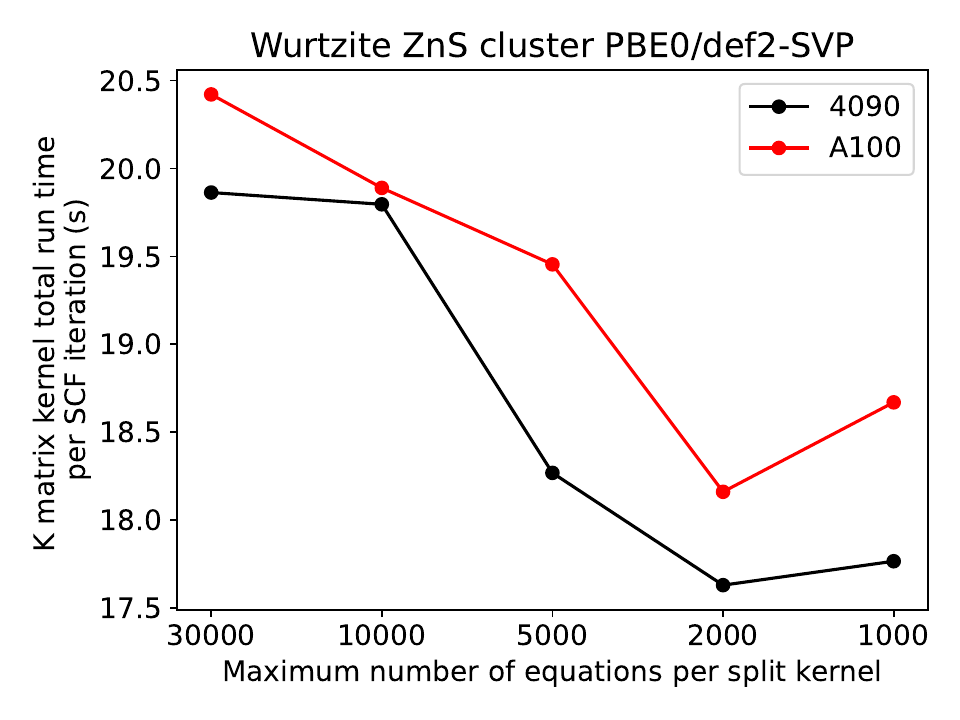}
\end{minipage}      
\caption{The total HF exchange kernel run time per SCF iteration with different kernel splitting strategies for the three model systems. A larger value for  ``maximum number of equations per kernel" leads to  fewer split kernels.}
\label{fig:si_runtime_vs_kernel_split_all}
\end{figure}

HF exchange matrix kernels involving high angular momentum (such as $(ff|ff)$) are quite complex, and can often not be compiled as a single entity with the 2 TB memory machines available to us. Splitting such kernels into smaller pieces eases compilation, but the effect on run times is harder to gauge in advance. Indeed, it is very hard to predict the best splitting strategy  as a bunch different factors are changed when a kernel is split, including kernel launch overhead, amount of re-computation, and amount of register variables moving between register and GPU global memory. We consider different splitting strategies based on the maximum number of equations in each split kernel (a larger value means fewer split kernels are needed) for the three model systems reported above: organic (two DNA CG base pairs, B3LYP/cc-pVTZ), bioinorganic (protein 6UFA, PBE0/def2-SVP) and inorganic cluster (Wurtzite zinc sulfide, PBE0/def2-SVP). The results for the A100 and 4090 GPUs are shown in shown in Fig. \ref{fig:si_runtime_vs_kernel_split_all}, which show that kernel splitting can actually perceptibly reduce run times but the behavior is not monotonic. It is however clear that very large kernels lead to the worst performance and thus some kernel splitting strategy ought to be used for better performance. Our current implementation utilizes a maximum of 4000 equations per kernel, which appears to be reasonable based on the results for the model systems reported in Fig. \ref{fig:si_runtime_vs_kernel_split_all}.

\begin{figure}[htb!]
    \centering
    \includegraphics[width=0.45\linewidth] {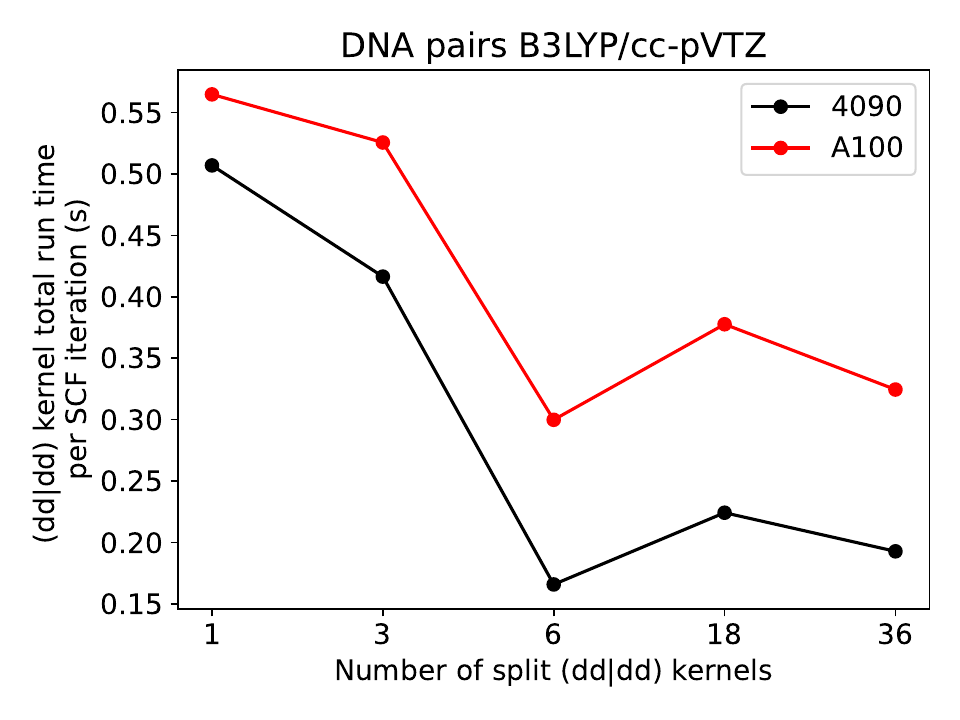}
    \includegraphics[width=0.45\linewidth] {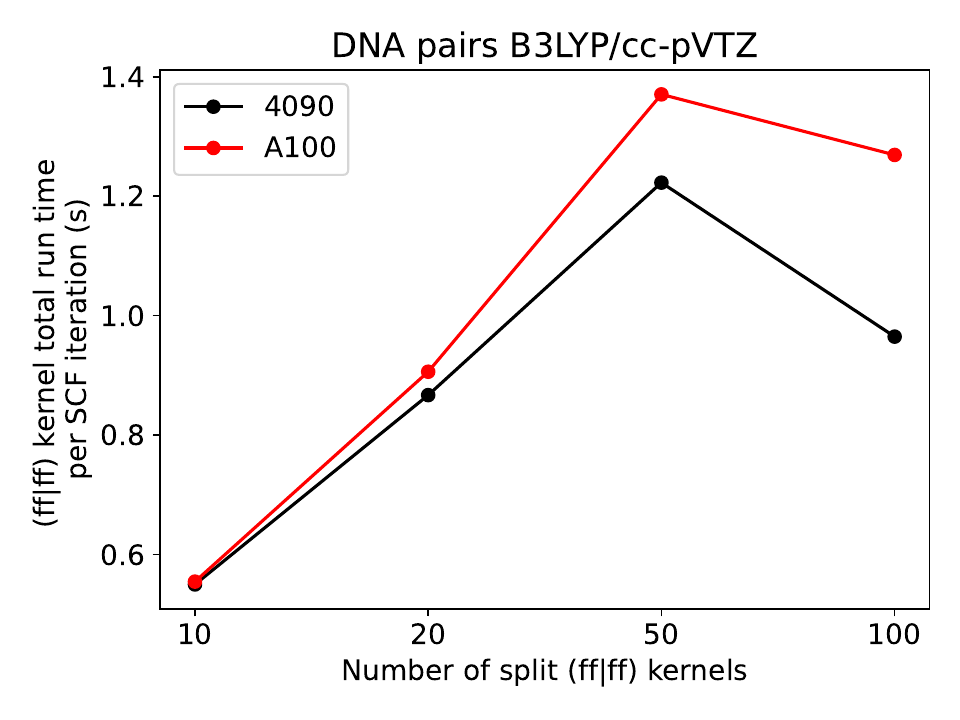}
    \includegraphics[width=0.45\linewidth] {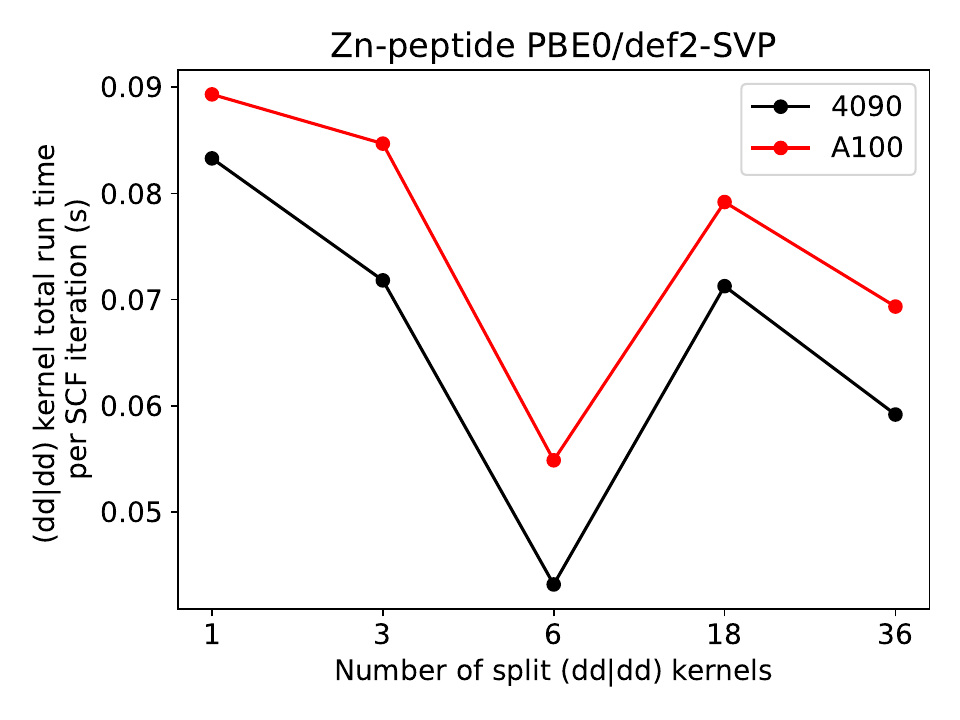}
    \includegraphics[width=0.45\linewidth] {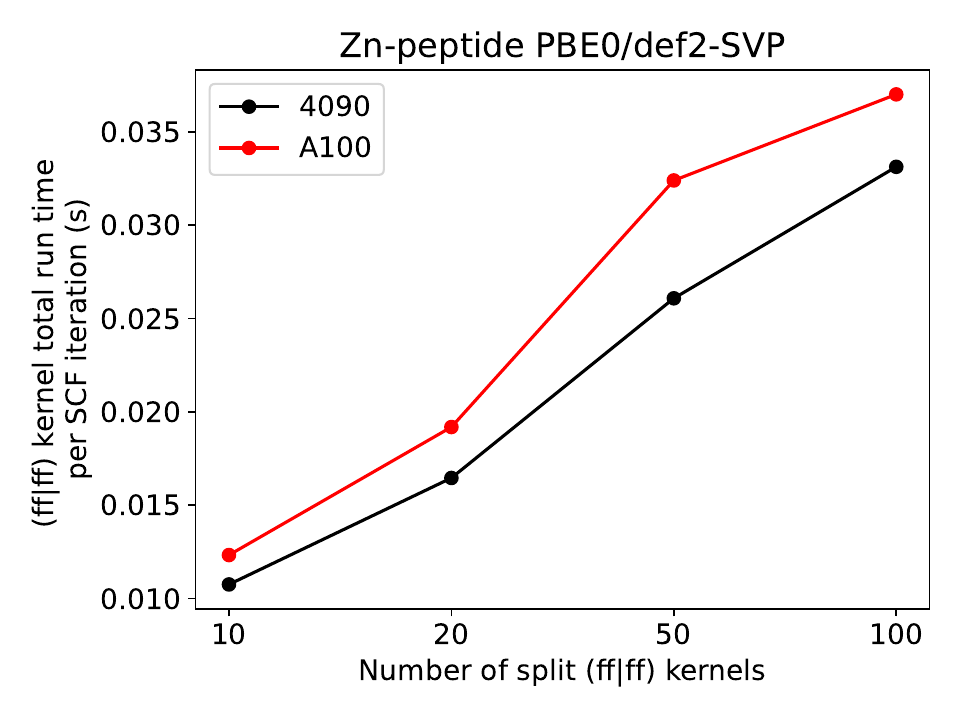}
    \includegraphics[width=0.45\linewidth] {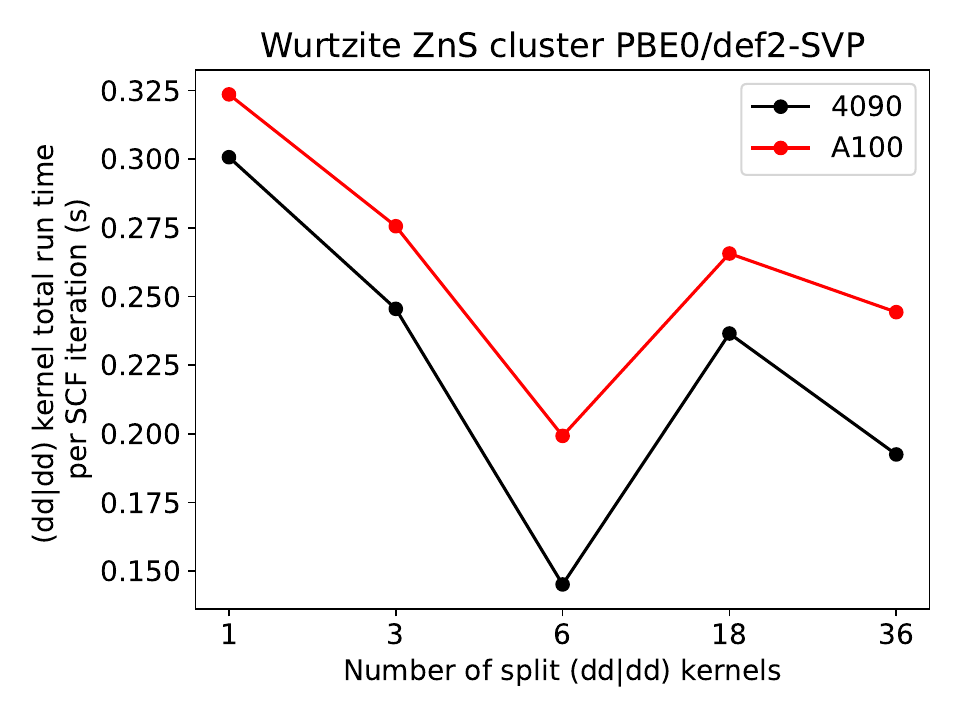}
    \includegraphics[width=0.45\linewidth] {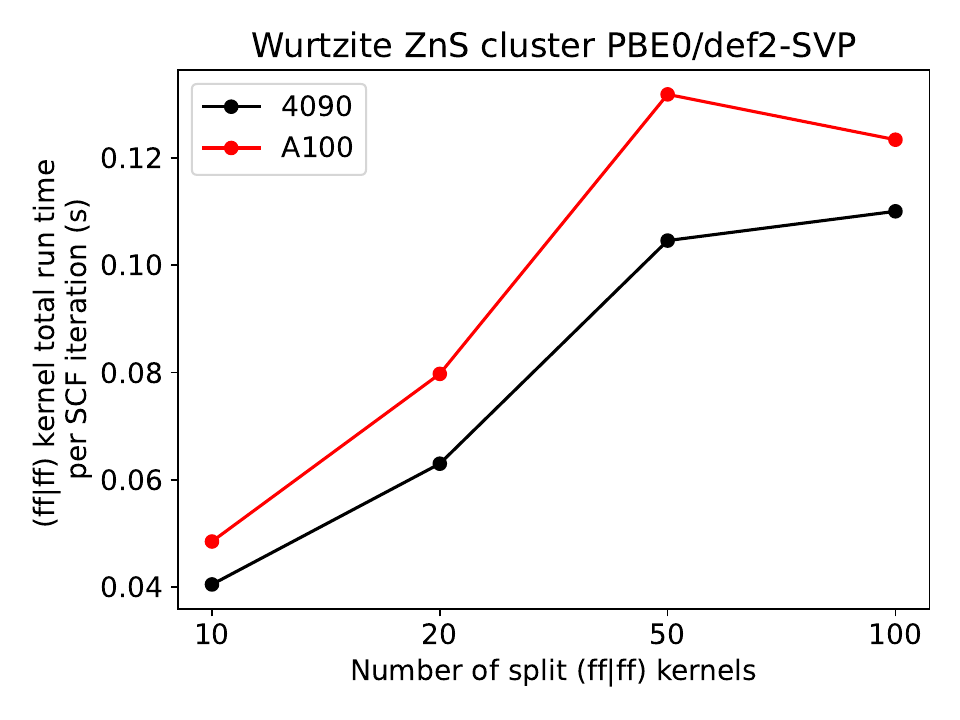}
\caption{The $(dd|dd)$ (left column) and $(ff|ff)$ (right column) HF exchange kernel run time with different kernel splitting strategies for the three model systems. We are not able to compile the whole unsplit $(ff|ff)$ kernel (number of split kernels = 1) due to memory limitations.}
\label{fig:si_runtime_vs_kernel_split}
\end{figure}

We also investigate the behavior of the $(dd|dd)$ and $(ff|ff)$ kernels for the model systems, with the results shown in Fig. \ref{fig:si_runtime_vs_kernel_split}. The $(ff|ff)$ results are less straightforward in that splitting into 100 pieces increases the runtime vs splitting into 10 (although intermediate behavior is not always monotonic). Such effects were not evident in the total run times shown in Fig. \ref{fig:si_runtime_vs_kernel_split_all}, as the $(ff|ff)$ kernel was not a significant bottleneck (contributing to $<5\%$ of the total run time). A six fold splitting on the other hand, appears to be quite effective for $(dd|dd)$ kernels. It thus appears that there is no obvious `one size fits all' strategy for total run time improvement via kernel splitting. It would therefore be necessary to device optimal splitting strategies individually for each kernel (with appropriate hardware considerations) for further improvements to performance. This will be explored in future work. 

\clearpage

\section{Performance of Single Precision Nuclear Attraction Integral Kernels against Number of Intermediates}

We also include the nuclear attraction integral run time vs number of intermediates in the equations for single precision calculations. However, we note that TeraChem always uses double precision nuclear attraction integral kernels for \textbf{all} calculations (including those described as mixed precision). These single precision nuclear attraction integral kernels therefore were only implemented for the profiling results shown here. These computations reveal a slight increase in kernel run time for the FF type integrals at the low number of multiplication operations limit, as discussed in the main text.

\begin{figure}[htb!]
    \centering
    \includegraphics[width=0.45\linewidth] {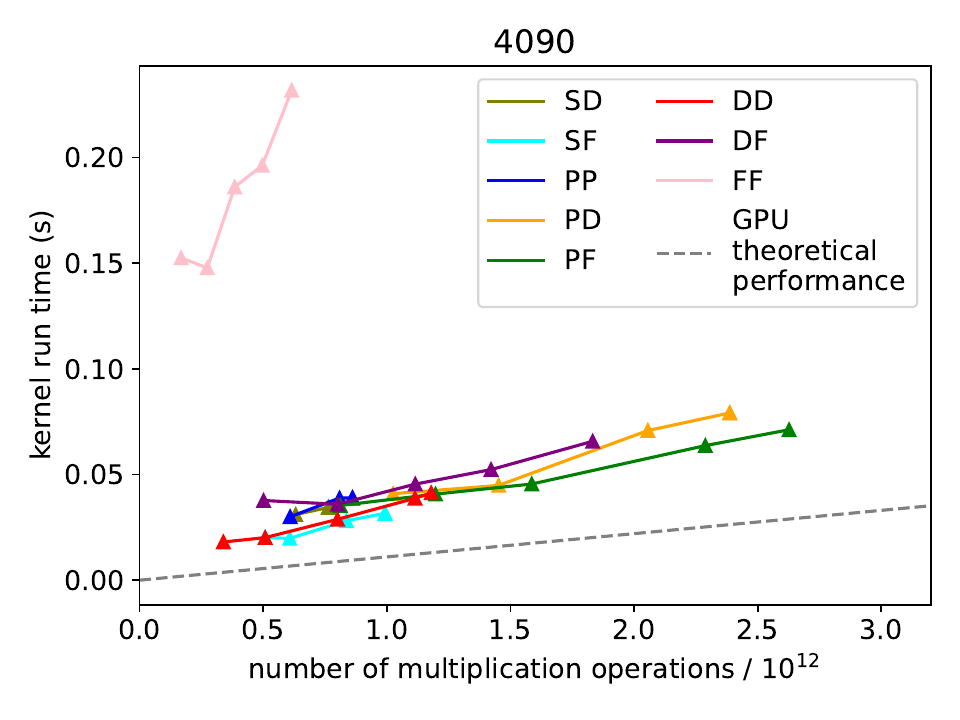}
    \includegraphics[width=0.45\linewidth] {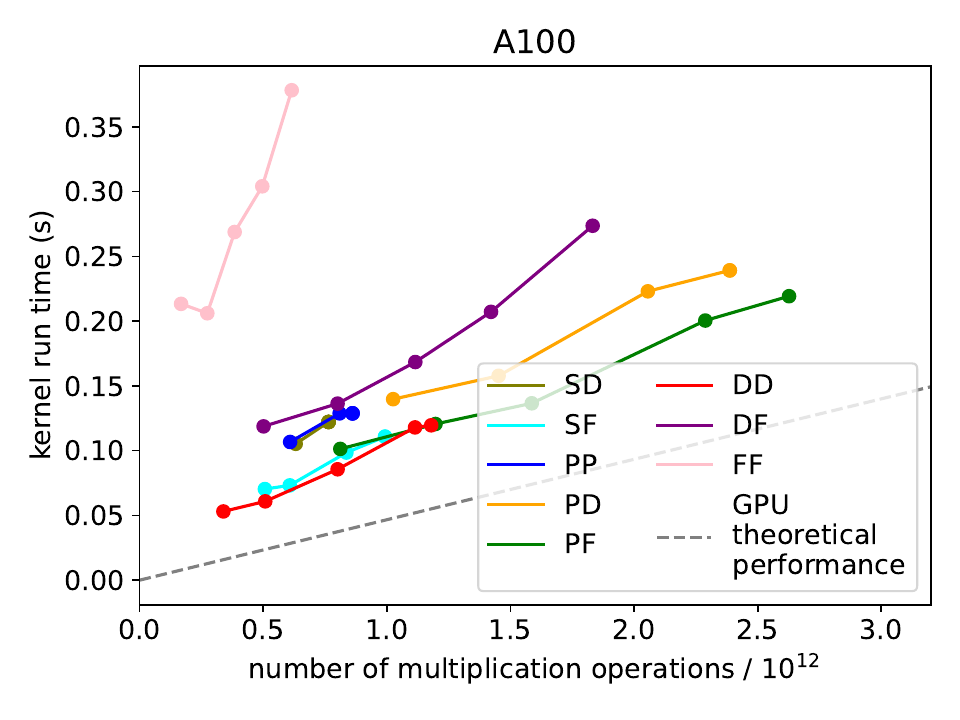}
\caption{The nuclear attraction integral evaluation time for PYP QM region 5 with cc-pVTZ basis set. Full single precision is used in nuclear attraction integral computation.}
\label{fig:si_runtime_vs_nvariables_single}
\end{figure}

\newpage

\section{Electronic and Nuclear Energy Contributions for Nickel Catalyzed N-N Coupling}
\begin{table}[htb!]
\begin{tabular}{l|r|r|r|r|r}
\hline 
Relative free energy (in kcal/mol)    & Reactant & TS1 & Intermediate & TS2 & Product  \\  \hline 
pcseg-1 with f functions   &                   &              &                       &              &                  \\ \hline 
Electronic             & 0.00              & 19.12        & -3.63                 & 32.45        & -1.37            \\
Nuclear            & 0.00              & -2.68        & 0.18                  & -1.12        & 0.48             \\
Total              & 0.00              & 16.44        & -3.45                 & 31.32        & -0.89            \\
\hline 
pcseg-1 without f functions   &                   &              &                       &              &                  \\ \hline 
Electronic             & 0.00              & 19.07        & -3.63                 & 32.14        & -1.56            \\
Nuclear            & 0.00              & -1.95        & 0.05                  & 0.52         & 1.00             \\
Total             & 0.00              & 17.12        & -3.58                 & 32.66        & -0.56      \\     \hline 
\end{tabular}
\caption{Free energies at 300 K (in kcal/mol, relative to the reactant) for the process shown in Fig \ref{fig:Stolz_step} of the main text, separated into electronic contributions (from $\omega$B97X-D3(BJ)/pcseg-1 on the electronic ground state) and nuclear free energy (from the rigid rotor harmonic oscillator approximation at the $\omega$B97X-D3(BJ)/pcseg-1 optimized geometries and associated normal mode harmonic frequencies). Results both with and without $f$ functions in the pcseg-1 basis are shown.}
\end{table}

\newpage

\bibliography{references}